\documentclass[a4paper,11pt]{article}
\usepackage[active]{srcltx}

\usepackage{jheppub}
\usepackage{amsmath}
\usepackage{amsfonts}
\usepackage{amssymb}

\usepackage{graphicx}
\usepackage{dcolumn}
\usepackage{bm}
\usepackage{epsfig}
\newcommand{\dhd}{{\textstyle d}
	\lower.03ex\hbox{\kern-0.38em$^{\scriptstyle-}$}\kern-0.05em{}}
\newcommand{\dbar}{{\textstyle \delta}
	\lower.03ex\hbox{\kern-0.38em$^{\scriptstyle-}$}\kern-0.05em{}}
\newcommand{\half}{{1\over 2}}

\newcommand{\calb}{{\cal B}}

\newcommand{\calo}{{\cal O}}

\newcommand{\hatp}{{\hat p}}

\newcommand \ket [1] {|{#1}\rangle}
\newcommand \bra [1] {\langle {#1}|}

\newcommand{\ketx}{\ket{x}}
\newcommand{\kety}{\ket{y}}
\newcommand{\ketz}{\ket{z}}

\newcommand{\brax}{\bra{x}}

\newcommand{\braz}{\bra{z}}
\newcommand{\brazi}{\bra{z'}}

\newcommand{\ketxp}{\ket{x_\perp}}
\newcommand{\ketyp}{\ket{y_\perp}}
\newcommand{\ketzp}{\ket{z_\perp}}
\newcommand{\ketzip}{\ket{z'_\perp}}

\newcommand{\braxp}{\bra{x_\perp}}
\newcommand{\brayp}{\bra{y_\perp}}
\newcommand{\brazp}{\bra{z_\perp}}
\newcommand{\brazip}{\bra{z'_\perp}}

\newcommand \sslash [1] {\slash\hspace{-0.2cm}{#1}}
\newcommand{\ssp}{\sslash{p}}
\newcommand{\ssx}{\sslash{x}}
\newcommand{\ssy}{\sslash{y}}

\newcommand{\slashd}{\sslash{\partial}}

\newcommand \Slash [1] {\slash\hspace{-0.23cm}{#1}}
\newcommand{\Sa}{\Slash{A}}
\newcommand{\Sp}{\Slash{P}}

\begin{document}
	
	\title{Sub-eikonal corrections to scattering amplitudes at high energy}

	\author[]{Giovanni Antonio Chirilli}
	\affiliation[]{Institut f\"ur Theoretische Physik, Universit\"at Regensburg,\\ D-93040 Regensburg, Germany}
	\emailAdd{giovanni.chirilli@ur.de}
	
\abstract{
Most of the progress in high-energy Quantum Chromodynamics has been obtained within the eikonal approximation and infinite Wilson-line operators. Evolution equations of Wilson lines with respect to the rapidity parameter encode the dynamics of the hadronic processes at high energy. However, even at high energy many interesting aspects of hadron dynamics are not accessible within the eikonal approximation, the
spin physics being an obvious example. The higher precision reached by the experiments and the possibility to probe spin dynamics
at future Electron Ion Colliders make the study of deviations from eikonal approximation especially timely.  In this paper, I derive the sub-eikonal quark and gluon propagators which can serve as a starting point of studies of these effects.  
}

	\maketitle

\section{Introduction}
It is well known that high-energy behavior of QCD amplitudes can be described by the evolution of 
relevant Wilson-line operators. The typical example is the deep inelastic scattering (DIS) at low 
Bjorken $x_B$,
where the T-product of two electromagnetic currents can
be approximated by a perturbative expansion in terms of coefficient functions (photon impact factors) and matrix 
elements of Wilson-line operators evaluated in the proton or nucleus state. The evolution equation of the Wilson-line operators
with respect to the rapidity parameter provides the energy dependence of the cross section. 
This procedure takes the name of high-energy Operator Product Expansion (OPE)
(see ref. \cite{balitskyreview} for a review on Wilson-line formalism in high-energy QCD).

The operator describing the
DIS scattering amplitude is a trace of two Wilson lines.
Its evolution equation with respect to the rapidity
of the fields generates a hierarchy, the Balitsky-hierarchy \cite{Balitsky:1995ub}, of evolution equations. 
The Balitsky-hierarchy is equivalent to the 
Jalilian-Marian--Iancu--McLerran--Weigert--Leonidov--Kovner
(JIMWLK)
 \cite{Jalilian-Marian:1997dw,Jalilian-Marian:1997gr,Iancu:2001ad,Iancu:2000hn} 
 evolution equation  thus, they are referred to as B--JIMWLK equation.
The Balitsky-Kovchegov (BK) equation \cite{Kovchegov:1999yj, Kovchegov:1999ua} is the first of the Balitsky hierarchy and 
it is obtained in the mean field approximation. Its linearization coincides with the BFKL \cite{Kuraev:1977fs, Balitsky:1978ic}
equation (for a review, see ref. \cite{Kovchegov:2012mbw}).

The OPE in terms of composite Wilson-line operators \cite{Balitsky:2009yp} 
has been applied up to NLO accuracy in processes like DIS
\cite{Balitsky:2010ze, Balitsky:2012bs}
and proton-nucleus collisions \cite{Chirilli:2011km, Chirilli:2012jd}.
In ref.\cite{Chirilli:2014dcb} the analytic expression of the $\gamma^*\gamma^*$ cross section has been computed
using the NLO OPE in linearized (BFKL) form.  

A relevant part of the program of the proposed Electron Ion Collider \cite{Boer:2011fh, Accardi:2012qut, Aschenauer:2015ata} 
is dedicated to the study of DIS with spin. 
Since the high-energy OPE formalism developed so far is suitable only for unpolarized scattering processes, the extension
of such formalism to high-energy spin-dynamics for Transverse Momentum Distribution (TMD) functions
and $g_1$ structure function is in strong demand. 

The high-energy OPE in terms of infinite Wilson lines is based on a semi-classical description 
of the process. In eikonal-approximation, the quantum free wave function of a particle propagating in a classical external field
is modified by a simple phase factor in QED or gravity 
\cite{Bjorken:1970ah, tHooft:1987vrq}, and by a path-ordered exponential in QCD \cite{Collins:1985gm, Cheng:1987ga, Nachtmann:1991ua}.
In high-energy QCD the rapidity parameter serves as a discriminator between
classical and quantum fields and the propagation of fast moving particles in an external filed is described 
by eikonal interactions. Unfortunately, eikonal interactions are insensitive to the 
spin content of the process. So, in order to bring in spin information, it is necessary to include
sub-eikonal contributions. The subject of this paper is the derivation of these corrections to 
the quark and gluon propagator in the background of a shock-wave. These results represent the first necessary step 
to include spin dynamics in the high-energy OPE formalism. 

In refs. \cite{Balitsky:2015qba, Balitsky:2016dgz}
 sub-eikonal corrections to scalar and gluon propagators have been calculated in order to construct
a formalism that provides evolution equations of gluon Transverse Momentum Distribution (TMD) from low to moderate
$x_B$. In this paper, I calculate the same result for the scalar and gluon propagators but including also the contributions 
coming from the transverse gauge fields neglected in refs. \cite{Balitsky:2015qba, Balitsky:2016dgz}. 
In addition, I compute the sub-eikonal corrections to the quark propagator.
The results obtained in this paper can be used to study, for example, quark-TMDs with spin and to check results obtained in
refs. \cite{Kovchegov:2015pbl, Kovchegov:2016zex, Kovchegov:2017jxc, 
	Kovchegov:2017lsr, Bartels:1995iu, Bartels:1996wc, Gorshkov:1966ht}.

Corrections to the eikonal formalism have been considered also in the contest of spin asymmetries in proton-nucleus collision. 
In ref. \cite{Altinoluk:2014oxa}, for example,
sub-eikonal corrections for the retarded gluon propagator have been calculated. 

Sub-eikonal corrections have also been considered in the contest of high-energy QCD at fixed angle 
and resummation of threshold logarithms \cite{Laenen:2008gt, Laenen:2010uz}.

The paper is structured as follows. In section \ref{sec: eikoscapropsection} and in Appendix \ref{sec: notation}
we introduce the formalism and derive the 
leading-eikonal scalar and quark propagator. The sub-eikonal corrections to the scalar propagator
are derived in section \ref{sec: subeiko-scalar}. This result is a necessary step to derive the sub-eikonal corrections to 
the quark propagator, in section \ref{sec: subeiko-quark}, and to the gluon propagator in light-cone gauge, in section \ref{sec: gluonpro}. 
In the Appendix we consider a modification of the shock-wave picture, that is, we consider the
case in which the particle starts or ends its propagation inside the external field. In the Appendix we also include an 
alternative derivation of the eikonal quark propagator and we calculate the sub-eikonal corrections to the gluon propagator
in background-Feynman gauge.

\section{Eikonal approximation for scalar and quark propagators}

We start our analysis with the derivation of the scalar and quark propagator in the background of
a shock-wave in eikonal approximation. We will use this preliminary step in order to introduce the 
formalism that will be used through out the paper. 

\subsection{Scalar propagator in the eikonal approximation}
\label{sec: eikoscapropsection}

The idea of the shock-wave formalism is based on the observation that 
high-energy regime can be reached not only rescaling the longitudinal momenta of the projectile-particle by a large parameter,
but also performing a longitudinal boost of the fields generated by the target-particle. 
We will study the propagation of a particle in the background of a highly boosted gluon filed.

Let $p_1^\mu$ and $p_2^\mu$ be two light-cone vectors such that $p_{1\mu}p_2^\mu = p_1\cdot p_2 = {s\over 2}$.
We assume that the projectile-particle is propagating along $p_1$ direction, while the target particle is moving along $p_2$
direction. Using the two light-cone vectors we can perform a Sudakov 
decomposition of the momentum $p^\mu = \alpha p_1^\mu + \beta p_2^\mu + p_\perp^\mu$.
We also define the light-cone components $x_\bullet = x_\mu p_1^\mu = \sqrt{s\over 2}x^-$
and $x_* = x_\mu p_2^\mu=\sqrt{s\over 2}x^+$ with $x^\pm = {x^0 \pm x^3\over \sqrt{2}}$.
We refer the reader to the Appendix \ref{sec: notation} for further details on the notation that will be used.

The gauge field $A^\mu$ generated by the target, under a boost, gets rescaled by a large parameter $\lambda$ as follows
\begin{eqnarray}
&&A_\bullet(x_\bullet, x_*, x_\perp) \to \lambda\, A_\bullet(\lambda^{-1}x_\bullet, \lambda\, x_*, x_\perp)\,,\nonumber\\
&&A_*(x_\bullet, x_*, x_\perp) \to  \lambda^{-1}A_*(\lambda^{-1}x_\bullet, \lambda\, x_*, x_\perp)\,,
\label{boost}\\
&&A_\perp(x_\bullet, x_*, x_\perp)  \to  A_\perp(\lambda^{-1}x_\bullet, \lambda\, x_*, x_\perp)\,.\nonumber
\end{eqnarray}
and the field strength tensor as
\begin{eqnarray}
&&F_{i\bullet}(x_\bullet, x_*, x_\perp) \to  \lambda\, F_{i\bullet}(\lambda^{-1}x_\bullet, \lambda\, x_*, x_\perp)\,,\nonumber\\
&&F_{i *}(x_\bullet, x_*, x_\perp) \to  \lambda^{-1}F_{i *}(\lambda^{-1}x_\bullet, \lambda\, x_*, x_\perp)\,,
\label{Fboost}\nonumber
\\
&&F_{\bullet *}(x_\bullet, x_*, x_\perp)  \to  F_{\bullet *}(\lambda^{-1}x_\bullet, \lambda\, x_*, x_\perp)\,,
\nonumber\\
&&F_{ij}(x_\bullet, x_*, x_\perp)  \to  F_{ij}(\lambda^{-1}x_\bullet, \lambda\, x_*, x_\perp)\,.
\end{eqnarray}

In Schwinger representation the free scalar propagator is
\begin{eqnarray}
\brax {i\over p^2 + i\epsilon}\kety = i\int\!\dhd^4 k \,{e^{-ik\cdot(x-y)}\over k^2 + i\epsilon}\,,
\label{schwrep}
\end{eqnarray}
with $\langle k\ketx = e^{ix\cdot k}$. In (\ref{schwrep}) we 
used the $\hbar$-inspired notation $\dhd^4 k \equiv {d^4k\over (2\pi)^4}$ and
$\dbar^{(4)}(k) = (2\pi)^4\delta^{(4)}(k)$ so that, $\int\!\dhd^4 k \,\dbar^{(4)}(k) = 1$.

Because of the infinite boost, in first approximation, we can assume that the field operator $\hat{A}_\mu$ commutes with the 
$\hat{\alpha} = i{\partial\over \partial x_\bullet}$ operator where, as we already mentioned above,
$\alpha$ is the longitudinal component along the light-cone vector $p_1$. 
The eikonal approximation is based on the observation that,
the only component surviving the boost is $A_\bullet$ so, 
$\hat{P}^2 \simeq \hat{p}^2 + 2\alpha g \hat{A}_\bullet$. If $A_\mu(x)$ is small
in comparison with the typical distance $(x-y)$, we can represent the propagator as a series
\begin{eqnarray}
\hspace{-1cm}\brax {i\over \hat{P}^2+i\epsilon}\kety\nonumber
\simeq\!\!\!&&  \brax {i\over \hat{p}^2 + i\epsilon}\kety 
\nonumber\\
&&+g\int d^4z \brax {i\over \hatp^2 + i\epsilon}\ketz 2\,i\,\alpha A_\bullet(z_*,z_\perp)
\braz {i\over \hatp^2 + i\epsilon}\kety + \dots\,.
\label{expand}
\end{eqnarray}
Alternatively, expansion (\ref{expand}) can also be expressed through Schwinger's proper time integral
\begin{eqnarray}
\hspace{-1cm}\brax {i\over \hat{P}^2+i\epsilon}\kety 
= - \!\! &&\! i\int_0^{+\infty}\!\!\!dt\,\brax e^{(\hatp^2 + 2 \alpha \hat{A}_\bullet+i\epsilon)t}\kety
= -i\int_0^{+\infty}\!\!\!dt\bigg[\brax e^{i(\hat{p}^2+i\epsilon)t}
\nonumber\\
+ \!\!&&\! i \int_0^t\!\!dt'\,\brax e^{i(t-t')(\hatp^2+i\epsilon)}\,2\,\alpha\, g \hat{A}_\bullet\,e^{i(\hatp^2+i\epsilon)t}\kety
+ \dots\bigg]\,.
\label{schexpand}
\end{eqnarray}
Using either (\ref{expand}) or (\ref{schexpand}), the expansion of the scalar propagator reduces to
\begin{eqnarray}
\hspace{-2cm}\brax {i\over \hat{P}^2+i\epsilon}\kety 
=\!\!\!&&
\left[\int_0^{+\infty}\!\!{\dhd \alpha\over 2\alpha}\theta(x_*-y_*) - 
\int_{-\infty}^0\!\!{\dhd\alpha\over 2\alpha}\theta(y_*-x_*) \right]e^{-i\alpha(x_\bullet - y_\bullet)} 
\nonumber\\
&&\times\bigg[\braxp e^{-i{\hatp^2_\perp\over \alpha s}(x_*-y_*)}\ketyp 
\nonumber\\
&&+ 
\int_{y_*}^{x_*}\!\!\!dz_*\,\braxp e^{-i{\hatp^2\over\alpha s}(x_*-z_*)}\,i\,{2\over s} g \hat{A}_\bullet (z_*)
\, e^{-i{\hatp^2_\perp\over \alpha s}(z_*-y_*)}\ketyp + \dots\bigg]\,.
\label{scalexpan}
\end{eqnarray}

\begin{figure}
	\begin{center}
		\includegraphics[width=3.1in]{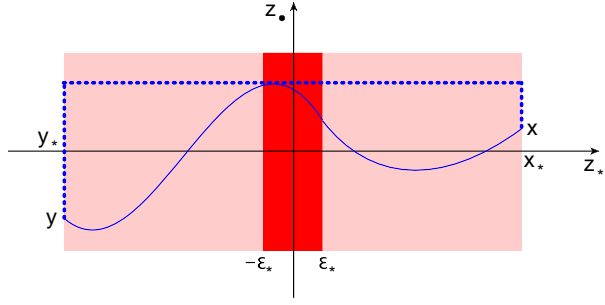}
		\caption{Particle propagating in an external gluonic field. The red strip represents the
			shock-wave defined in the infinitesimal interval $[-\epsilon_*,\epsilon_*]$ in which $F_{\mu\nu}\ne 0$. 
			The light-red areas to the left and to the right of the shock-wave represent the background filed made of pure gage field.
			The curvy line is the particle's path from point $x$ to point $y$. The dotted lines represent the Wilson lines.}
		\label{swprpagation}
	\end{center}
\end{figure}

We are interested in the shock-wave picture relevant for high-energy scattering so, we assume that the particle starts and ends its propagation
outside the interval in which the field strength tensor is different then zero (see figure \ref{swprpagation}).
With this assumption we can rewrite expansion (\ref{scalexpan}) as
\begin{eqnarray}
\hspace{-1cm}\brax {i\over \hat{P}^2 + i\epsilon}\kety
=\!\!\!&& \left[\int_0^{+\infty}\!\!{\dhd \alpha\over 2\alpha}\theta(x_*-y_*) - 
\int_{-\infty}^0\!\!{\dhd\alpha\over 2\alpha}\theta(y_*-x_*) \right]e^{-i\alpha(x_\bullet - y_\bullet)}
\nonumber\\
&&\times\int d^2z d^2 z'
\braxp\, e^{-i{\hatp^2_\perp\over \alpha s}x_*}\ketzp
\nonumber\\
&&\times
\brazp{\rm Pexp} \left\{ig\!\! \int_{y_*}^{x_*}d{2\over s}\omega_*\, 
e^{i{\hatp^2_\perp\over \alpha s}\omega_*}A_\bullet(\omega_*)
e^{-i{\hatp^2_\perp\over \alpha s}\omega_*} \right\} 
\ketzip
\nonumber\\
&&\times
\brazip
e^{i{\hatp^2_\perp\over \alpha s}y_*}\ketyp\,.
\label{pathexpscal}
\end{eqnarray}

So far, to arrive at the propagator given in eq. (\ref{pathexpscal}), we have implemented only two consequences of the
longitudinal Lorentz boost of the external field: the commutation relation $[\hat{\alpha},\hat{A}]=0$ 
and the fact that the most dominant component of the gauge external field is $A_\bullet$. Indeed, 
since we are considering an infinite
Lorentz boost, we can perform further approximations. 

Propagator (\ref{pathexpscal}) describes the propagation, following any path from point 
$z_\perp + {2\over s}x_*p_1$ to point $z'_\perp + {2\over s}y_*p_1$ in the external field $A^\mu(x)=(A_\bullet(x_*,x_\perp),0,0)$,
of a spinless particle. Notice that, the particle propagates between points
$(x_\bullet, x_*, x_\perp)$ and $(y_\bullet, y_*, y_\perp)$, while the field strength tensor, because of the longitudinal boost,
is defined within an infinitesimal interval in the longitudinal direction. In other words,
$F^{\mu\nu}(\omega_*,\omega_\perp) \neq 0$ for $\omega_* \in [-\epsilon_*, \epsilon_*]$ with $0<\epsilon_*\ll 1$
and, since we are in the shock-wave case, $ x_*,y_* \notin[-\epsilon_*,\epsilon_*]$.

Since we are boosting the coordinates, the longitudinal distance 
traveled by the particle in the external filed is rescaled under a boost
as $\omega_*\to {1\over \lambda}\omega_*$ while the gauge field is rescaled as $A_\bullet \to \lambda A_\bullet$
with $\lambda\gg 1$ the boost parameter. So, we can make a further approximation 
in eq. (\ref{pathexpscal}) and write
\begin{eqnarray}
e^{i{\hatp^2_\perp\over \alpha s}\omega_*}A_\bullet(\omega_*)
e^{-i{\hatp^2_\perp\over \alpha s}\omega_*} = A_\bullet(\omega_*) + O(\lambda^0)\,.
\label{commutexpa}
\end{eqnarray}
Making use of (\ref{commutexpa}) in propagator (\ref{pathexpscal}), we obtain
\begin{eqnarray}
\brax {i\over \hat{P}^2+i\epsilon}\kety 
= \!\!\!&&\left[\int_0^{+\infty}\!\!{\dhd \alpha\over 2\alpha}\theta(x_*-y_*) - 
\int_{-\infty}^0\!\!{\dhd\alpha\over 2\alpha}\theta(y_*-x_*) \right]e^{-i\alpha(x_\bullet - y_\bullet)}
\nonumber\\
&&\times\!\int d^2z \braxp\, e^{-i{\hatp^2_\perp\over \alpha s}x_*}\ketzp[x_*,y_*]_z
\brazp e^{i{\hatp^2_\perp\over \alpha s}y_*}\ketyp\,,
\label{leadingscapro}
\end{eqnarray}
where we have defined the gauge link at fixed transverse position $z_\perp$ as
\begin{eqnarray}
[x_*,y_*]_z = {\rm Pexp} \left\{ig{2\over s}\!\! \int_{y_*}^{x_*}\!\!\!d\omega_*\, 
A_\bullet(\omega_*,z_\perp)\right\}\,.
\end{eqnarray}
Propagator (\ref{leadingscapro}) describes the propagation of the particle in the external field along a straight-line. Deviation 
from the straight-line propagation are taken into account by the higher order terms neglected in eq. (\ref{commutexpa}).
We will consider them in the next section.

In the shock wave approximation we can trade the finite gauge link with the infinite Wilson line because, under the infinite boost,
the dominant component of the field strength tensor, $F_{\bullet i}$, has an infinitesimal thin support in $x_*$ coordinate.
We assumed that $F_{\bullet i}$ is peaked at the origin and
outside the infinitesimal interval $[-\epsilon_*,\epsilon_*]$, $F_{\bullet i} = 0$ (see figure \ref{swprpagation}). 

In the gauge rotated field $A^\Omega$ the gauge field outside the external field is zero 
so, we can trade the gauge link $[\epsilon_*,-\epsilon_*]$ with the infinite Wilson line $[\infty p_1,-\infty p_1]$.
In this gauge we can then write
\begin{eqnarray}
\hspace{-1cm}\brax {i\over P^2+i\epsilon}\kety 
=\!\!\!&& \left[\int_0^{+\infty}\!\!{\dhd \alpha\over 2\alpha}\theta(x_*-y_*) - 
\int_{-\infty}^0\!\!{\dhd\alpha\over 2\alpha}\theta(y_*-x_*) \right]e^{-i\alpha(x_\bullet - y_\bullet)}
\nonumber\\
&&\times
\!\int\! d^2z\braxp\, e^{-i{\hatp^2_\perp\over \alpha s}x_*}
\ketzp U_z\brazp
e^{i{\hatp^2_\perp\over \alpha s}y_*}\ketyp\,,
\label{leadingscapro-sw}
\end{eqnarray}
where we have defined the infinite Wilson line $U_z$ at fixed transverse position $z_\perp$ as
\begin{eqnarray}
U_z \equiv [p_1\infty, - p_1\infty]_z = {\rm Pexp}\left\{ig{2\over s}\int_{-\infty}^{+\infty}\!\!dz_* A_{\bullet}({2\over s}p_1 z_* + z_\perp)\right\}\,.
\end{eqnarray}
Notice that, because of the infinite boost, we have set $A^\mu_\perp$ component to zero, 
so, since it is a pure gauge, it can be restored, for example, as a transverse gauge link (see figure \ref{swprpagation}).

\subsection{Quark propagator in eikonal approximation}
\label{sec: leadinqpro}

We consider now the quark propagator in the eikonal approximation. 
From now on we work in the gauge rotated field $A^\Omega$. This allows us to set to
zero the transverse fields at the edges of the gauge fields, i.e., $A_i(x_*) = A_i(y_*) = 0$.

We proceed in the same way as in the scalar propagator case. After boosting the fields, we consider only the dominant
component of the gauge fields and write
\begin{eqnarray}
\hspace{-1cm}\brax{i\over \hat{\Sp} +i\epsilon}\kety \simeq\!\!\!&&
\brax\,\hat{\Sp}{i\over \hat{P}^2 +  ig{2\over s}\gamma_\perp^\rho \ssp_2F_{\rho\bullet}+i\epsilon}\kety
\nonumber\\
 =\!\!\!&&  \Big(i\sslash{\partial}^x + g{2\over s}\ssp_2 A_\bullet(x_*,x_\perp)\Big)
 \nonumber\\
 &&\times\brax\, \left[{i\over \hat{P}^2+i\epsilon}  - 
 {i\over \hat{P}^2+i\epsilon}g{2\over s}\gamma_\perp^\rho \ssp_2F_{\rho\bullet}{i\over \hat{P}^2+i\epsilon}\right]\kety\,,
\label{exactexpa}
\end{eqnarray}
where $P^2 \simeq p^2 + 2g\alpha A_\bullet$.
Notice that, because $\ssp_2\ssp_2 = 0$, there is only one term surviving the expansion in (\ref{exactexpa}).
In appendix \ref{sec: 2type-exp} we will consider an alternative way of expanding the quark propagator in the external field.
However, to study the sub-eikonal corrections the expansion used in this section is more convenient.

In (\ref{exactexpa}), for each factor ${1\over P^2+i\epsilon}$  we use the scalar propagator (\ref{leadingscapro}), and using
\begin{eqnarray}
&&\left(i\slashd^x + g{2\over s}\ssp_2 A_\bullet(x_*,x_\perp) \right)e^{-i\alpha(x_\bullet - y_\bullet)}
\braxp\, e^{-i{\hatp^2_\perp\over \alpha s}x_*}\ketzp \nonumber\\
=&&
e^{-i\alpha(x_\bullet - y_\bullet)}\braxp\, e^{-i{\hatp^2_\perp\over \alpha s}x_*}\left({1\over \alpha s}\hat{\ssp}\ssp_2\hat{\ssp}
+{2\over s}\ssp_2 iD^x_\bullet \right)\ketzp\,,
\label{usefla1}
\end{eqnarray}
together with the identity 
\begin{eqnarray}
 iD^x_\bullet[x_*,y_*] = \Big(i\partial^x_\bullet + g A_\bullet(x_*) \Big) [x_*,y_*]= 0
\end{eqnarray}
we arrive at
\begin{eqnarray}
\brax {i\over \Sp + i\epsilon} \kety
= \!\!\!&&
\left[\int_0^{+\infty}\!{\dhd\alpha\over 2\alpha}\,\theta(x_* - y_*) - 
\int^0_{-\infty}\!{\dhd\alpha\over 2\alpha}\theta(y_* - x_*)\right]e^{-i\alpha(x_\bullet-y_\bullet)}
\nonumber\\
&&\times
{1\over \alpha s}\braxp\, e^{-i{\hatp^2_\perp\over \alpha s}x_*}\,\hat{\ssp}\ssp_2
\nonumber\\
&&\times\left( \hat{\ssp}[x_*,y_*]
- g{2\over s}\gamma_\perp^\rho\int_{y_*}^{x_*}\!\!\!d\omega_*\,[x_*,\omega_*]F_{\bullet \rho}[\omega_*,y_*]
\right)e^{i{\hatp^2_\perp\over\alpha s}y_*}\ketyp\,.
\label{qpropa-gcv}
\end{eqnarray}
Note that, in using eq. (\ref{usefla1}) we have neglected the delta function $\delta(x_*-y_*)$ coming from the differentiation
of the Theta functions $\theta(x_* - y_*)$ and $\theta(y_* - x_*)$ of the scalar propagator eq. (\ref{leadingscapro}).
This is because we are assuming that $x_*\neq y_*$.

Quark propagator (\ref{qpropa-gcv}) is in covariant gauge form. 
An equivalent expression of the quark propagator (\ref{qpropa-gcv}) is
\begin{eqnarray}
\brax {i\over \Sp + i\epsilon} \kety
=\!\!\!&&  {1\over s}\left[\int_0^{+\infty}\!{\dhd\alpha\over 2\alpha^2}\,\theta(x_* - y_*) - 
\int^0_{-\infty}\!{\dhd\alpha\over 2\alpha^2}\theta(y_* - x_*)\right]e^{-i\alpha(x_\bullet-y_\bullet)}
\nonumber\\
&&
\times\braxp\,e^{-i{\hatp^2_\perp\over \alpha s}x_*}\Bigg( \hat{\ssp}\,
\ssp_2[x_*,y_*]\hat{\ssp}
- \,\hat{\ssp}\,\ssp_2\Big[g\,\Sa_\perp(x_*)[x_*,y_*]
- [x_*,y_*] g\,\Sa_\perp(y_*)\Big] \Bigg)
\nonumber\\
&&\times e^{i{\hatp^2_\perp\over \alpha s}y_*}\ketyp\,.
\label{qpropa-nogcv}
\end{eqnarray}
To arrive at eq. (\ref{qpropa-nogcv}) from eq. (\ref{qpropa-gcv}) 
we have carried one of the $\hat{\ssp}$ operator to the right of the gauge-link.
The quark propagator in eq. (\ref{qpropa-nogcv}) is not in gauge covariant form but it is suitable for the shock-wave picture.
Indeed, in the shock-wave approximation, $\Sa_\perp(x_*)$ and 
$\Sa_\perp(y_*)$ are zero in the gauge rotated field $A^\Omega$ we are working in. Moreover, the gauge-link $[x_*,y_*]$,
reduces to $[\epsilon_*, -\epsilon_*]$. At this point we can extend the limit of integration $\epsilon_*\to \infty$
and $-\epsilon_*\to -\infty$ thus, (\ref{qpropa-nogcv}) becomes
\begin{eqnarray}
\brax {i\over \Sp + i\epsilon} \kety
=\!\!\!&&  {1\over s}\left[\int_0^{+\infty}\!{\dhd\alpha\over 2\alpha^2}\,\theta(x_* - y_*) - 
\int^0_{-\infty}\!{\dhd\alpha\over 2\alpha^2}\theta(y_* - x_*)\right]e^{-i\alpha(x_\bullet-y_\bullet)}
\nonumber\\
&&
\times\braxp\,e^{-i{\hatp^2_\perp\over \alpha s}x_*} \ssp\,\ketzp
\ssp_2U_z\brazp\ssp\,
e^{i{\hatp^2_\perp\over \alpha s}y_*}\ketyp\,.
\label{qpropa-sw}
\end{eqnarray}
In (\ref{qpropa-sw}) the shock-wave picture is now evident: we have free propagation from point $x$ to
the point of interaction $z$ with the shock wave, 
eikonal interaction with the shock wave, which is represented by the infinite Wilson line multiplied by $\ssp_2$, 
and then again free propagation from point $z$ to point $y$ (see figure 1).
Propagator (\ref{qpropa-sw}) is the one that has been used so far in all the results that have been
obtained in the high-energy Wilson-lines formalism. In this paper we are interested in the 
sub-eikonal corrections to (\ref{qpropa-sw}).

\section{Sub-eikonal corrections to the scalar propagator}
\label{sec: subeiko-scalar}

In the previous section, to obtain the quark propagator in the eikonal approximation,
we have used the scalar propagator as intermediate step. 
For the sub-eikonal corrections we proceed in a similar way: we first derive the 
sub-eikonal correction to the scalar propagator, and then use this result as a useful intermediate step
in order to get the sub-eikonal corrections to the quark propagator.

We consider a background gauge field with all components different then zero: 
$A^{cl}_\mu(x_*,x_\perp)=(A^{cl}_*(x_*,x_\perp), A^{cl}_\bullet(x_*,x_\perp), A^{cl}_\perp(x_*,x_\perp))$. 
From now on we shall omit the superscript ``cl'' from the classical field.
The shock-wave now has a finite width, so we have to identify the sub-dominant components of the momentum operator 
$\hat{P}^2= \{\hatp^\mu_\perp, \hat{A}^\perp_\mu\} + \{{2\over s}\hat{P}_\bullet, \hat{A}_*\}  - g\hat{A}^2_\perp$. 
First, let us notice that we can get rid of the term $\{\hat{P}_\bullet, A_*\}$ observing that
\begin{eqnarray}
&&\brax{1\over \hatp^2 + 2\alpha g A_\bullet + i\epsilon} A_*\hat{P}_\bullet{1\over \hatp^2 + 2\alpha g A_\bullet + i\epsilon}\kety
\nonumber\\
 && =
\left[-\int_0^{+\infty}\!\!{\dhd \alpha\over 4\alpha^2}\theta(x_*-y_*) 
+ \int_{-\infty}^0\!\!{\dhd\alpha\over 4\alpha^2}\theta(y_*-x_*) \right]e^{-i\alpha(x_\bullet - y_\bullet)}
\nonumber\\
&&~~\times
\int^{x_*}_{y_*}\!\!\!d{2\over s}z_{*}\braxp e^{-i{\hatp^2_\perp\over \alpha s}x_*}
[x_*,z_*]\big(A_* + {iz_*\over \alpha s}[\hatp^2_\perp,A_*]\big)
\nonumber\\
&&~~\times
\left({iz_*\over\alpha s}[\hatp^2_\perp,A_\bullet] + {\hatp^2_\perp\over 2\alpha} + iD^z_\bullet\right)
[z_*,y_*]e^{i{\hatp^2_\perp\over \alpha s}y_*}\ketyp 
= O(\lambda^{-2})\,,
\end{eqnarray}
where in the first step we inserted $\int\!d^4z \,\ketz\braz=1$ and used 
\begin{eqnarray}
e^{i{\hatp^2_\perp\over \alpha s}z_*}A_*e^{-i{\hatp^2_\perp\over \alpha s}z_*} \simeq
A_* + {iz_*\over \alpha s}[\hatp^2_\perp,A_*] 
\end{eqnarray} 
and
\begin{eqnarray}
&&\big(i\partial^z_\bullet+ g A_\bullet(z_*,z_\perp)\big)\brazp e^{-i{\hatp^2_\perp\over \alpha s}z_*}
[z_*,y_*]e^{i{\hatp^2_\perp\over \alpha s}y_*}\ketyp 
\nonumber\\
=&& 
\brazp e^{-i{\hatp^2_\perp\over \alpha s}z_*}
\left({iz_*\over\alpha s}[\hatp^2_\perp,A_\bullet] + {\hatp^2_\perp\over 2\alpha} + iD^z_\bullet\right)
[z_*,y_*]e^{i{\hatp^2_\perp\over \alpha s}y_*}\ketyp\,,
\end{eqnarray} 
while in the last step we used $\big(i\partial^x_\bullet + gA_\bullet(x_*)\big)[x_*,y_*] = 0$.
Similarly, we can drop the term $\hat{P}_\bullet A_*$. In Appendix \ref{sec: astarterm}, following 
an alternative procedure, we will show that $\big\{\hat{P}_\bullet, A_*\big\}$ contributes only as sub-sub-eikonal correction.

Let us define $\hat{O} \equiv \{\hatp^\mu_\perp, \hat{A}^\perp_\mu\}  - g\hat{A}^2_\perp$ so, 
$\hat{P}^2 = \hatp^2 + 2g\alpha A_\bullet + g\hat{O}$.
The scalar propagator in the boosted external filed can now be written as	
\begin{eqnarray}
\hspace{-0.5cm}\brax {i\over \hat{P}^2+i\epsilon}\kety =\!\!\! && \brax {i\over \hat{p}^2 + 2\alpha g \hat{A}_\bullet + g \hat{O}+i\epsilon}\kety
\nonumber\\
=\!\!\!&& \left[\int_0^{+\infty}\!\!{\dhd \alpha\over 2\alpha}\theta(x_*-y_*) - 
\int_{-\infty}^0\!\!{\dhd\alpha\over 2\alpha}\theta(y_*-x_*) \right]e^{-i\alpha(x_\bullet - y_\bullet)}
\nonumber\\
&&\times
\braxp\, e^{-i{\hatp^2_\perp\over \alpha s}x_*}
{\rm Pexp} \left\{ig\!\! \int_{y_*}^{x_*}d{2\over s}\omega_*\, 
e^{i{\hatp^2_\perp\over \alpha s}\omega_*}\left(\hat{A}_\bullet(\omega_*) + {\hat{O}(\omega_*)\over 2\alpha}\right)
e^{-i{\hatp^2_\perp\over \alpha s}\omega_*} \right\} 
\nonumber\\
&&\times
e^{i{\hatp^2_\perp\over \alpha s}y_*}\ketyp\,.
\label{scamidpoint}
\end{eqnarray}

We are interested in corrections to the eikonal propagator that go like ${1\over \lambda}$, so in the following expansion it 
is sufficient to consider the first sub-dominant contribution which goes like $\lambda^0$
\begin{eqnarray}
&&e^{i{\hatp^2_\perp\over \alpha s}\omega_*}\left(\hat{A}_\bullet + {\hat{O}\over 2\alpha}\right)
e^{-i{\hatp^2_\perp\over \alpha s}\omega_*}
=  \hat{A}_\bullet + {\hat{O}\over 2\alpha} + 
i{\omega_*\over \alpha s}[\hatp^2_\perp, \hat{A}_\bullet ]
+ O(\lambda^{-1})\,.
\label{subscaexp}
\end{eqnarray}
Making use of (\ref{subscaexp}) and of the following identity
\begin{eqnarray}
{\hat{O}\over 2\alpha} + {i\omega_*\over \alpha s}[\hatp^2_\perp,\hat{A}_\bullet] = 
{1\over 2\alpha}\Big(
\{\hat{p}^i, {2\over s}\omega_*F_{i\bullet} + (D_\bullet{2\over s}\omega_* \hat{A}_i)\}
- g\hat{A}^2_\perp\Big)\,,
\end{eqnarray}
after some algebra, we arrive at the desired expression of the scalar propagator with sub-eikonal corrections
\begin{eqnarray}
\hspace{-0.2cm}\brax {i\over \hat{P}^2+i\epsilon}\kety 
=\!\!\!&& \left[\int_0^{+\infty}\!\!{\dhd \alpha\over 2\alpha}\theta(x_*-y_*) - 
\int_{-\infty}^0\!\!{\dhd\alpha\over 2\alpha}\theta(y_*-x_*) \right]e^{-i\alpha(x_\bullet - y_\bullet)}
\nonumber\\
&&\times
\braxp\, e^{-i{\hatp^2_\perp\over \alpha s}x_*}\Bigg\{[x_*,y_*] 
+ {ig\over 2\alpha}\bigg[{2\over s}x_*\Big(\{P_i,A^i(x_*)\} - gA_i(x_*)A^i(x_*)\Big)[x_*,y_*]
\nonumber\\
&&-[x_*,y_*]{2\over s}y_*\Big(\{P_i,A^i(y_*)\} - gA_i(y_*)A^i(y_*)\Big)
\nonumber\\
&&
+ \int^{x_*}_{y_*}\!\!d{2\over s}\omega_*\Big(
\big\{P^i,[x_*,\omega_*]\,{2\over s}\,\omega_*\, F_{i\bullet}(\omega_*)\,[\omega_*,y_*]\big\}
\nonumber\\
&& + g\!\!\int^{x_*}_{\omega_*}\!\!d{2\over s}\,\omega'_*\,{2\over s}\big(\omega_* - \omega'_*\big)
[x_*,\omega'_*]F^i_{~\bullet}[\omega'_*,\omega_*]\,
\,F_{i\bullet}\,[\omega_*,y_*]\Big)\bigg]\Bigg\}
e^{i{\hatp^2_\perp\over \alpha s}y_*}\ketyp
\nonumber\\
&&
+ O(\lambda^{-2})\,.
\label{scalpropa-nogcv}
\end{eqnarray}

In the shock-wave limit, the fields outside the shock-wave are pure gauge 
and the scalar propagator takes the following form
\begin{eqnarray}
\brax {i\over \hat{P}^2+i\epsilon}\kety 
=\!\!\!&& \left[\int_0^{+\infty}\!\!{\dhd \alpha\over 2\alpha}\theta(x_*-y_*) - 
\int_{-\infty}^0\!\!{\dhd\alpha\over 2\alpha}\theta(y_*-x_*) \right]e^{-i\alpha(x_\bullet - y_\bullet)}
\braxp\, e^{-i{\hatp^2_\perp\over \alpha s}x_*}
\label{scalpropa-sw}\\
&&\times
\Bigg\{[\infty p_1 , - \infty p_1] 
+ {ig\over 2\alpha}
\int^{+\infty}_{-\infty}\!\!d{2\over s}\omega_*\bigg(
\big\{\hatp^i,[\infty p_1,\omega_*]\,{2\over s}\,\omega_*
\, F_{i\bullet}(\omega_*)\,[\omega_*, -\infty p_1]\big\}
\nonumber\\
&& + g\!\!\int^{+\infty}_{\omega_*}\!\!d{2\over s}\,\omega'_*\,{2\over s}\big(\omega_* - \omega'_*\big)
[\infty p_1,\omega'_*]F^i_{~\bullet}[\omega'_*,\omega_*]\,
\,F_{i\bullet}\,[\omega_*, -\infty p_1]
\bigg)\Bigg\}e^{i{\hatp^2_\perp\over \alpha s}y_*}\ketyp\,.
\nonumber
\end{eqnarray}
Except for the terms with the transverse fields at the edges (\textit{i.e.} at the points $x_*$ and $y_*$),
propagator (\ref{scalpropa-nogcv}) coincides with the one obtained in Ref. \cite{Balitsky:2015qba}.

In next section we will derive, following a similar procedure, the sub-eikonal corrections to the quark propagator. 
To this end the result obtained in this section, eq. (\ref{scalpropa-nogcv}), will be an essential intermediate step.
It is useful to notice that, the terms with transverse fields at points $x_*$ and $y_*$ in eq. (\ref{scalpropa-nogcv})
cannot be set to zero when we use the scalar propagator as intermediate step because, as we will see in next section,
the points $x_*$ and $y_*$ might just be the points in the middle of the shock-wave.

\section{Sub-eikonal corrections to the quark propagator}
\label{sec: subeiko-quark}

In this section we derive the sub-eikonal correction to the quark propagator in the background of
gluon filed and quark anti-quark field. We will star with the gluon field case first.

\subsection{Quark propagator in the background of gluon field}
As already mentioned before, the sub-eikonal corrections to the scalar propagator, eq. (\ref{scalpropa-nogcv}) 
will be an essential intermediate step. 

Our starting point is
\begin{eqnarray}
\brax{i\over \hat{\Sp}+i\epsilon}\kety
=\!\!\!&&\brax\hat{\Sp}\,
{i\over p^2 + 2\alpha g A_\bullet + gB + i {2\over s}gF_{\bullet i}\,\ssp_2\gamma^i
	+ i\epsilon}\kety\nonumber\\
=\!\!\!&& \left(i\slashd^x + g{2\over s}A_\bullet(x_*,x_\perp)\ssp_2 + g\Sa_\perp(x_*,x_\perp)\right)
\nonumber\\
&&
\times\left[\int_0^{+\infty}\!\!{\dhd\alpha\over 2\alpha}\theta(x_*-y_*) - 
\int_{-\infty}^0\!\!{\dhd\alpha\over 2\alpha}\theta(y_*-x_*)\right]
e^{-i\alpha(x_\bullet - y_\bullet)}
\braxp\,e^{-i{\hatp^2_\perp\over \alpha s}x_*}
\nonumber\\
&&\times
{\rm Pexp}\Bigg\{ig\!\int_{y_*}^{x_*}\!\!\!d{2\over s}\omega_*
\,e^{i{\hatp^2_\perp\over \alpha s}\omega_*}\left(A_\bullet(\omega_*) + {B(\omega_*)\over 2\alpha} +
{i\over 2\alpha} {2\over s}F_{\bullet i}(\omega_*)\,\ssp_2\gamma^i\right)
\nonumber\\
&&\times
e^{-i{\hatp^2_\perp\over \alpha s}\omega_*}\Bigg\}
e^{i{\hatp^2_\perp\over \alpha s}y_*}\ketyp\,,
\end{eqnarray}
where we have defined $B\equiv O + {4\over s^2}gF_{\bullet*}\sigma_{*\bullet} + {g\over 2}F_{ij}\sigma^{ij}$ 
and $\sigma^{\mu\nu} = {i\over 2}(\gamma^\mu\gamma^\nu - \gamma^\nu\gamma^\mu)$.

Expanding again up to $\lambda^0$ contributions, we get
\begin{eqnarray}
&&\hspace{-1cm}e^{i{\hatp^2_\perp\over \alpha s}\omega_*}\left(A_\bullet(\omega_*) + {B(\omega_*)\over 2\alpha}+
{i\over 2\alpha}  {2\over s}F_{\bullet i}(\omega_*)\,\ssp_2\gamma^i\right)\,e^{-i{\hatp^2_\perp\over \alpha s}\omega_*}
\nonumber\\
&&\hspace{-0.6cm}\simeq A_\bullet + {B\over 2\alpha}+
{i\over 2\alpha}  {2\over s}F_{\bullet i}\,\ssp_2\gamma^i
+ i{\omega_*\over \alpha s}[p^2_\perp, A_\bullet] + 
i{\omega_*\over \alpha s}[p^2_\perp, {i\over 2\alpha}  {2\over s}F_{\bullet i}\,\ssp_2\gamma^i]\,.
\label{expan-sub-q}
\end{eqnarray}

Now we have to expand the path ordered exponential of the right-hand-side of (\ref{expan-sub-q}) up to the desired order.
We have
\begin{eqnarray}
&&\hspace{-0.3cm}{\rm Pexp}\left\{ig\!\int_{y_*}^{x_*}\!\!\!d{2\over s}\omega_*
\,e^{i{\hatp^2_\perp\over \alpha s}\omega_*}\left(A_\bullet(\omega_*) + {B(\omega_*)\over 2\alpha}+
{i\over 2\alpha}  {2\over s}F_{\bullet i}(\omega_*)\,\ssp_2\gamma^i\right)\,e^{-i{\hatp^2_\perp\over \alpha s}\omega_*}\right\}
\nonumber\\
&&\hspace{-0.3cm}= [x_*,y_*]
+ ig\!\int_{y_*}^{x_*}\!\!\!d{2\over s}\omega_*
\,[x_*,\omega_*]\left({O(\omega_*)\over 2\alpha} + i{\omega_*\over \alpha s}[p^2_\perp, A_\bullet]\right)[\omega_*,y_*]
\nonumber\\
&&
+  {ig\over 2\alpha}\!\int_{y_*}^{x_*}\!\!\!d{2\over s}\omega_*\,[x_*,\omega_*]
\left({4\over s^2}F_{\bullet*}\sigma_{*\bullet} + {1\over 2}F_{ij}\sigma^{ij}\right)[\omega_*,y_*]
\nonumber\\
&& + ig\!\int_{y_*}^{x_*}\!\!\!d{2\over s}\omega_*\,[x_*,\omega_*]
\left({i\over 2\alpha}  {2\over s}F_{\bullet i}\,\ssp_2\gamma^i 
+ i{\omega_*\over \alpha s}[p^2_\perp, {i\over 2\alpha}  {2\over s}F_{\bullet i}\,\ssp_2\gamma^i]\right)[\omega_*,y_*]
\nonumber\\
&& + (ig)^2\!\int_{y_*}^{x_*}\!\!\!d{2\over s}\omega_*\!\int_{y_*}^{\omega_*}\!\!\!d{2\over s}\omega'_*\,
[x_*,\omega_*]\left({B(\omega_*)\over 2\alpha}
+ i{\omega_*\over \alpha s}[p^2_\perp, A_\bullet] \right)
[\omega_*,\omega'_*]\,
{i\over 2\alpha} {2\over s}F_{\bullet i}(\omega'_*)\,\ssp_2\gamma^i \,[\omega'_*,y_*]
\nonumber\\
&& + (ig)^2\!\int_{y_*}^{x_*}\!\!\!d{2\over s}\omega_*\!\int_{y_*}^{\omega_*}\!\!\!d{2\over s}\omega'_*\,
[x_*,\omega_*]\,{i\over 2\alpha} {2\over s}F_{\bullet i}(\omega_*)\,\ssp_2\gamma^i 
[\omega_*,\omega'_*]\left({B(\omega'_*)\over 2\alpha} + i{\omega'_*\over \alpha s}[p^2_\perp, A_\bullet] \right)[\omega'_*,y_*]
\nonumber\\
&& + O(\lambda^{-2})\,.
\label{4terms-quapro}
\end{eqnarray}
With the help of some algebra (see Appendix \ref{sec: calc-details} for some details of the derivation),
we can rewrite eq. (\ref{4terms-quapro}) in a gauge invariant form as
\begin{eqnarray}
&&\hspace{-0.2cm}{\rm Pexp}\left\{ig\!\int_{y_*}^{x_*}\!\!\!d{2\over s}\omega_*
\,e^{i{\hatp^2_\perp\over \alpha s}\omega_*}\left(A_\bullet(\omega_*) + {B(\omega_*)\over 2\alpha}+
{i\over 2\alpha}  {2\over s}F_{\bullet i}(\omega_*)\,\ssp_2\gamma^i\right)\,e^{-i{\hatp^2_\perp\over \alpha s}\omega_*}\right\}
\nonumber\\
&&\hspace{-0.2cm}=
\Bigg(\big(1 - {1\over 2\alpha}{2\over s}\ssp_2\,i\,\Slash{\mathfrak{D}}_\perp\big)[x_*,y_*] 
+ {ig\over 2\alpha}\!\int_{y_*}^{x_*}\!\!\!d{2\over s}\omega_*\,[x_*,\omega_*]
\calb_1[\omega_*,y_*]\Bigg) 
\nonumber\\
&&\hspace{-0.2cm}
+ {1\over 4\alpha^2} \int_{y_*}^{x_*}\!\!\!d{2\over s}z_*
\Big[i(\Slash{\mathfrak{D}}_\perp {2\over s}\ssp_2 [x_*,z_*])ig\calb_1[z_*,y_*] 
+ [x_*,z_*]ig\calb_1(i\Slash{\mathfrak{D}}_\perp{2\over s}\ssp_2[z_*,y_*])\Big]
\nonumber\\
&&\hspace{-0.2cm}
+ {ig\over 2\alpha}\Bigg[
\int^{x_*}_{y_*}\!\!\!d{2\over s}\omega_*\bigg(
\big\{p^i,[x_*,\omega_*]\,{2\over s}\,\omega_*\, F_{i\bullet}(\omega_*)\,[\omega_*,y_*]\big\}
\nonumber\\
&&\hspace{-0.2cm}
+ g\!\!\int^{x_*}_{\omega_*}\!\!\!d{2\over s}\,\omega'_*\,{2\over s}\big(\omega_* - \omega'_*\big)
[x_*,\omega'_*]F^i_{~\bullet}[\omega'_*,\omega_*]
\,F_{i\bullet}\,[\omega_*,y_*]\bigg)\Bigg]
- {ig\over 4\alpha^2}\gamma^j{2\over s}\ssp_2\int_{y_*}^{x_*}\!\!\!d{2\over s}\omega_*\,{2\over s}\omega_*
\nonumber\\
&&\hspace{-0.2cm}
\times\bigg[
(i\mathfrak{D}^i[x_*,\omega_*])(iD_iF_{j\bullet})[\omega_*,y_*]
- [x_*,\omega_*](iD^iF_{j\bullet})(i\mathfrak{D}_i[\omega_*,y_*]) 
- \{p^i,[x_*,\omega_*] (iD_iF_{j\bullet})[\omega_*,y_*]\}
 \bigg]
\nonumber\\
&&\hspace{-0.2cm} 
+ {g\over 4\alpha^2}{2\over s}\ssp_2\int_{y_*}^{x_*}\!\!\!d{2\over s}\omega_*
\bigg[
{2\over s}\omega_*(\Slash{\mathfrak{D}}_\perp[x_*,\omega_*])F_{i\bullet}(i\mathfrak{D}^i[\omega_*,y_*])
-{2\over s}\omega_*
(i\mathfrak{D}^i[x_*,\omega_*])F_{i\bullet}(\Slash{\mathfrak{D}}_\perp[\omega_*,y_*])
\nonumber\\
&&\hspace{-0.2cm}
+ \{p^i,[x_*,\omega_*]{2\over s}\omega_*F_{i\bullet}(\Slash{\mathfrak{D}}_\perp[\omega_*,y_*])\}
+ \{p^i,(\Slash{\mathfrak{D}}_\perp[x_*,\omega_*]){2\over s}\omega_*F_{i\bullet}[\omega_*,y_*]\}
\nonumber\\
&&\hspace{-0.2cm} + [x_*,\omega_*]{2\over s}\omega_*F_{i\bullet}\,i\mathfrak{D}^i(\Slash{\mathfrak{D}}_\perp[\omega_*,y_*])
-(i\mathfrak{D}^i(\Slash{\mathfrak{D}}_\perp[x_*,\omega_*])){2\over s}\omega_*F_{i\bullet}[\omega_*,y_*]\bigg]
\nonumber\\
&&\hspace{-0.2cm}
+ O(\lambda^{-2})\,,
\label{sum4lines}
\end{eqnarray}
where we have defined $\calb_1\equiv{4\over s^2}F_{\bullet*}\sigma_{*\bullet} + {1\over 2}F_{ij}\sigma^{ij}$.

To continue our analysis we observe that
\begin{eqnarray}
&&\hspace{-1.2cm}\big(i\slashd^x + g{2\over s}\ssp_2 A_\bullet(x_*,x_\perp) + 
g\Sa_\perp(x_*,x_\perp) \big)e^{-i\alpha(x_\bullet - y_\bullet)}
\braxp\, e^{-i{\hatp^2_\perp\over \alpha s}x_*} \ketzp\nonumber\\
&&\hspace{-0.8cm}= 
e^{-i\alpha(x_\bullet - y_\bullet)}\braxp\, e^{-i{\hatp^2_\perp\over \alpha s}x_*}
\nonumber\\
\times&&\!\!\!
\Big({1\over \alpha s}\ssp\ssp_2\ssp + i{2\over s}\ssp_2D^x_\bullet
+ {ix_*\over \alpha s}[p^2_\perp,g{2\over s}\ssp_2 A_\bullet(x_*)]
+ g\Sa_\perp(x_*) \Big)\ketzp\,,
\label{usefla}
\end{eqnarray}
and note that $i[p^2_\perp, gA_\bullet(x_*)] = g\{p^i, F_{i\bullet}(x_*) + D_\bullet A_i(x_*)\}$. 
The field strength tensor $F_{i\bullet}(x_*) = 0$ since $x_*$ is outside the shock-wave (see figure \ref{swprpagation}). Similarly,
we can set all the transverse fields
 at the edges of the gauge-link (outside the shock-wave) to zero since they are pure gauge. So, (\ref{usefla}) reduces to
\begin{eqnarray}
&&\big(i\slashd^x + g{2\over s}\ssp_2 A_\bullet(x_*,x_\perp) + g\Sa_\perp(x_*,x_\perp) \big)e^{-i\alpha(x_\bullet - y_\bullet)}
\braxp\, e^{-i{\hatp^2_\perp\over \alpha s}x_*}\ketzp \nonumber\\
&&~~= 
e^{-i\alpha(x_\bullet - y_\bullet)}\braxp\, e^{-i{\hatp^2_\perp\over \alpha s}x_*}\Big({1\over \alpha s}\ssp\ssp_2\ssp 
+ i{2\over s}\ssp_2D^x_\bullet \Big)\ketzp\,.
\label{usefla2}
\end{eqnarray}

Using (\ref{usefla2}) and the identity $iD^x_\bullet[x_*,y_*] = 0$,
we arrive at the following expression for the quark propagator 
\begin{eqnarray}
&&\hspace{-0.8cm}\brax{i\over \hat{\Sp} +i\epsilon}\kety
\nonumber\\
&&\hspace{-0.8cm}=\left[\int_0^{+\infty}\!\!{\dhd \alpha\over 2\alpha}\theta(x_*-y_*) - 
\int_{-\infty}^0\!\!{\dhd\alpha\over 2\alpha}\theta(y_*-x_*) \right]e^{-i\alpha(x_\bullet - y_\bullet)}{1\over \alpha s}
\braxp\, e^{-i{\hatp^2_\perp\over \alpha s}x_*}
\nonumber\\
\times\!\!\!&&
\Bigg\{\ssp\ssp_2\ssp
\big(1-{1\over 2\alpha}{2\over s}\ssp_2i\Slash{\mathfrak{D}}_\perp\big)[x_*,y_*]
+{ig\over 2\alpha}\int_{y_*}^{x_*}\!\!\!d{2\over s}z_*\ssp\ssp_2\ssp[x_*,z_*]\calb_1[z_*,y_*]
\nonumber\\
&&\hspace{-0.4cm}
+ {1\over 4\alpha^2}{2\over s} \int_{y_*}^{x_*}\!\!\!d{2\over s}z_*\,
\ssp\ssp_2\ssp\Big[(i\,\Slash{\mathfrak{D}}_\perp \ssp_2 [x_*,z_*])ig\calb_1[z_*,y_*] 
+ [x_*,z_*]ig\calb_1(i\,\Slash{\mathfrak{D}}_\perp\ssp_2[z_*,y_*])\Big]
\nonumber\\
&&\hspace{-0.4cm}
+ {ig\over 2\alpha}
\int^{x_*}_{y_*}\!\!d{2\over s}\omega_*\,\ssp\ssp_2\ssp\,\bigg(\big\{P^i,[x_*,\omega_*]\,{2\over s}\,\omega_*\, F_{i\bullet}(\omega_*)\,[\omega_*,y_*]\big\}
\nonumber\\
&&\hspace{-0.4cm}
+ g\!\!\int^{x_*}_{\omega_*}\!\!d{2\over s}\,\omega'_*\,{2\over s}\big(\omega_* - \omega'_*\big)
[x_*,\omega'_*]F^i_{~\bullet}[\omega'_*,\omega_*]\,
\,F_{i\bullet}\,[\omega_*,y_*]\bigg)
\nonumber\\
&&\hspace{-0.4cm}
+ {ig\over 2\alpha}\ssp\gamma^j\ssp_2\int_{y_*}^{x_*}\!\!\!d{2\over s}z_*\,{2\over s}z_*\bigg[
\{P^i,[x_*,z_*] \big(iD_iF_{j\bullet}\big)[z_*,y_*]\} - \big(i\,\mathfrak{D}^i[x_*,z_*]\big)\big(iD_iF_{j\bullet}\big)[z_*,y_*]
\nonumber\\
&&\hspace{-0.4cm}
+ [x_*,z_*]\big(iD_iF_{j\bullet}\big)\big(i\mathfrak{D}^i[z_*,y_*]\big)\bigg]
+ {ig\over 2\alpha}\ssp\ssp_2\int_{y_*}^{x_*}\!\!\!d{2\over s}z_*\,{2\over s}z_*
\bigg[ (i\,\mathfrak{D}^i(i\,\Slash{\mathfrak{D}}_\perp[x_*,z_*]))F_{i\bullet}[z_*,y_*]
\nonumber\\
&&\hspace{-0.4cm}
- [x_*,z_*]\, F_{i\bullet}\big(i\,\mathfrak{D}^i\big(i\,\Slash{\mathfrak{D}}_\perp[z_*,y_*]\big)\big)
+  \big(i\,\mathfrak{D}^i[x_*,z_*]\big)\,F_{i\bullet}\big(i\,\Slash{\mathfrak{D}}_\perp[z_*,y_*]\big)
\nonumber\\
&&\hspace{-0.4cm}
-\{P^i,(i\,\Slash{\mathfrak{D}}_\perp[x_*,z_*])F_{i\bullet}[z_*,y_*]\} - \big\{P^i,[x_*,z_*]\, F_{i\bullet}(i\,\Slash{\mathfrak{D}}_\perp[z_*,y_*])\big\} 
\nonumber\\
&&\hspace{-0.4cm}
- \big(i\,\Slash{\mathfrak{D}}_\perp[x_*,z_*]\big)F^i_{~\bullet}\big(i\,\mathfrak{D}_i[z_*,y_*]\big)\bigg]
\Bigg\}e^{i{\hatp^2_\perp\over \alpha s}y_*}\ketyp\,.
\label{quarksubnoedge}
\end{eqnarray}
In order to have $\hat{\ssp}$ to the left and to the right of the gauge link,
which is suitable for the shock-wave picture, we will perform similar steps we performed in going from
eq. (\ref{qpropa-gcv}) to eq. (\ref{qpropa-nogcv}), and set again to zero the fields $A_i$ at the point $x_*$ and $y_*$.
To this end, we notice that
\begin{eqnarray}
&&\hspace{-0.2cm}
{ig\over 2\alpha}
\int^{x_*}_{y_*}\!\!d{2\over s}\omega_*\,\hat{\ssp}\ssp_2\hat{\ssp}\,\bigg(\big\{P^i,[x_*,\omega_*]\,{2\over s}\,\omega_*\, F_{i\bullet}(\omega_*)\,[\omega_*,y_*]\big\}
\nonumber\\
&&\hspace{-0.2cm}
+ g\!\!\int^{x_*}_{\omega_*}\!\!d{2\over s}\,\omega'_*\,{2\over s}\big(\omega_* - \omega'_*\big)
[x_*,\omega'_*]F^i_{~\bullet}[\omega'_*,\omega_*]\,
\,F_{i\bullet}\,[\omega_*,y_*]\bigg)
\nonumber\\
&&\hspace{-0.2cm}
+ {ig\over 2\alpha}\hat{\ssp}\gamma^j\ssp_2\int_{y_*}^{x_*}\!\!\!d{2\over s}z_*\,{2\over s}z_*\bigg[
\{P^i,[x_*,z_*] \big(iD_iF_{j\bullet}\big)[z_*,y_*]\} - \big(i\,\mathfrak{D}^i[x_*,z_*]\big)i\,\mathfrak{D}_iF_{j\bullet}[z_*,y_*]
\nonumber\\
&&\hspace{-0.2cm}
+ [x_*,z_*]\big(iD_iF_{j\bullet}\big)\big(i\,\mathfrak{D}^i[z_*,y_*]\big)\bigg]
+ {ig\over 2\alpha}\hat{\ssp}\ssp_2\int_{y_*}^{x_*}\!\!\!d{2\over s}z_*\,{2\over s}z_*
\bigg[  (i\,\mathfrak{D}^i(i\,\Slash{\mathfrak{D}}_\perp[x_*,z_*]))F_{i\bullet}[z_*,y_*]
\nonumber\\
&&\hspace{-0.2cm}
- [x_*,z_*]\, F_{i\bullet}\big(i\,\mathfrak{D}^i\big(i\,\Slash{\mathfrak{D}}_\perp[z_*,y_*]\big)\big)
+  \big(i\,\mathfrak{D}^i[x_*,z_*]\big)\,F_{i\bullet}\big(i\,\Slash{\mathfrak{D}}_\perp[z_*,y_*]\big)
- \big(i\,\Slash{\mathfrak{D}}_\perp[x_*,z_*]\big)F^i_{~\bullet}\big(i\,\mathfrak{D}_i[z_*,y_*]\big)
\nonumber\\
&&\hspace{-0.2cm}
-\{P^i,(i\,\Slash{\mathfrak{D}}_\perp[x_*,z_*])F_{i\bullet}[z_*,y_*]\} - \big\{P^i,[x_*,z_*]\, 
F_{i\bullet}(i\,\Slash{\mathfrak{D}}_\perp[z_*,y_*])\big\} \bigg]
\nonumber\\
&&~
={ig\over 2\alpha}
\int^{x_*}_{y_*}\!\!d{2\over s}\omega_*\,\hat{\ssp}\ssp_2\bigg(
{2\over s}\omega_*\,\big\{P_i,[x_*,\omega_*]F^i_{~\bullet}[\omega_*,y_*]\big\}
\nonumber\\
&&
~~+\! g\!\!\int^{x_*}_{\omega_*}\!\!d{2\over s}\,\omega'_*\,{2\over s}\big(\omega_* - \omega'_*\big)
[x_*,\omega'_*]F^i_{~\bullet}[\omega'_*,\omega_*]\,
\,F_{i\bullet}\,[\omega_*,y_*]
\bigg)\hat{\ssp}
\,,
\label{simplification1}
\end{eqnarray}
where we have used $F_{ij}(x_*)=F_{ij}(y_*)=0$, as points $x_*$ and $y_*$  
are outside the shock-wave (see figure \ref{swprpagation}). To arrive at 
eq. (\ref{simplification1}) we have used Eqs. (\ref{defPi}), (\ref{commD}), (\ref{identity1}),
and the fact that $\hat{\alpha}$ commutes with all fields.
Using eq. (\ref{simplification1}) in eq. (\ref{quarksubnoedge}) we arrive at
\begin{eqnarray}
\hspace{-0.2cm}\brax{i\over \hat{\Sp} +i\epsilon}\kety=&&\!\!
\left[\int_0^{+\infty}\!\!{\dhd \alpha\over 2\alpha}\theta(x_*-y_*) - 
\int_{-\infty}^0\!\!{\dhd\alpha\over 2\alpha}\theta(y_*-x_*) \right]e^{-i\alpha(x_\bullet - y_\bullet)}{1\over \alpha s}
\braxp\, e^{-i{\hatp^2_\perp\over \alpha s}x_*}
\nonumber\\
&&\times
\Bigg\{\hat{\ssp}\ssp_2[x_*,y_*]\hat{\ssp}
+{ig\over 2\alpha}\int_{y_*}^{x_*}\!\!\!d{2\over s}z_*
\hat{\ssp}\ssp_2\hat{\ssp}[x_*,z_*]\calb_1[z_*,y_*]
\nonumber\\
&&
+ {1\over 4\alpha^2}{2\over s} \int_{y_*}^{x_*}\!\!\!d{2\over s}z_*\,
\hat{\ssp}\ssp_2\hat{\ssp}
\Big[(i\,\Slash{\mathfrak{D}}_\perp \ssp_2 [x_*,z_*])ig\calb_1[z_*,y_*] 
+ [x_*,z_*]ig\calb_1(i\,\Slash{\mathfrak{D}}_\perp\ssp_2[z_*,y_*])\Big]
\nonumber\\
&&
+ {ig\over 2\alpha}
\int^{x_*}_{y_*}\!\!d{2\over s}\omega_*\,\hat{\ssp}\ssp_2\,\bigg(\big\{p^i,[x_*,\omega_*]\,{2\over s}\,\omega_*\, F_{i\bullet}(\omega_*)\,[\omega_*,y_*]\big\}
\nonumber\\
&&
+ g\!\!\int^{x_*}_{\omega_*}\!\!d{2\over s}\,\omega'_*\,{2\over s}\big(\omega_* - \omega'_*\big)
[x_*,\omega'_*]F^i_{~\bullet}[\omega'_*,\omega_*]\,
\,F_{i\bullet}\,[\omega_*,y_*]\bigg)\hat{\ssp}
\Bigg\}e^{i{\hatp^2_\perp\over \alpha s}y_*}\ketyp\,.
\label{quarksubnoedge2}
\end{eqnarray}
We now use the following results 
\begin{eqnarray}
&&{ig\over 2\alpha}\int_{y_*}^{x_*}\!\!\!d{2\over s}z_*\,\hat{\ssp}\ssp_2\hat{\ssp}[x_*,z_*]\half F_{ij}\sigma^{ij}[z_*,y_*]
\label{simplification2}\\
&&
+ {1\over 4\alpha^2}{2\over s} \int_{y_*}^{x_*}\!\!\!d{2\over s}z_*\,
\hat{\ssp}\ssp_2\hat{\ssp}\Big[(i\,\Slash{\mathfrak{D}}_\perp \ssp_2 [x_*,z_*])ig\half F_{ij}\sigma^{ij}[z_*,y_*] + 
[x_*,z_*]ig\half F_{ij}\sigma^{ij}(i\,\Slash{\mathfrak{D}}_\perp\ssp_2[z_*,y_*])\Big]
 \nonumber\\
&&= {ig\over 2\alpha}\int_{y_*}^{x_*}\!\!\!d{2\over s}z_*\, \hat{\ssp}\ssp_2
\bigg[
[x_*,z_*]\half F_{ij}\sigma^{ij}[z_*,y_*]\hat{\ssp}
+ [x_*,z_*]{1\over 4}(i\,\mathfrak{D}_k F_{ij})\big\{\sigma^{ij},\gamma^k\big\}[z_*,y_*]
\nonumber\\
&&
~~ + \big\{\hat{p}_k,[x_*,z_*]i F_{kj}\gamma^j[z_*,y_*]\big\}
+ [x_*,z_*]i F_{kj}\gamma^j(i\,\mathfrak{D}^k[z_*,y_*]) 
- (i\,\mathfrak{D}^k[x_*,z_*])i F_{kj}\gamma^j[z_*,y_*] 
\bigg]\,,
\nonumber
\end{eqnarray}
and 
\begin{eqnarray}
&&\hspace{-0.2cm}
{ig\over 2\alpha}\int_{y_*}^{x_*}\!\!\!d{2\over s}z_*\,\hat{\ssp}\ssp_2\hat{\ssp}[x_*,z_*]{4\over s^2} F_{\bullet *}\sigma_{*\bullet}[z_*,y_*]
\nonumber\\
&&\hspace{-0.2cm}
+ {1\over 4\alpha^2}{2\over s} \int_{y_*}^{x_*}\!\!\!d{2\over s}z_*\,
\hat{\ssp}\ssp_2\hat{\ssp}\Big[(i\,\Slash{\mathfrak{D}}_\perp \ssp_2 [x_*,z_*])ig{4\over s^2} F_{\bullet *}\sigma_{*\bullet}[z_*,y_*] + 
[x_*,z_*]ig{4\over s^2} F_{\bullet *}\sigma_{*\bullet}(i\,\Slash{\mathfrak{D}}_\perp\ssp_2[z_*,y_*])\Big]
\nonumber\\
 &&\hspace{-0.2cm}=
{ig\over 2\alpha}\int_{y_*}^{x_*}\!\!\!d{2\over s}z_*
\,i{s\over 2}\hat{\ssp}\ssp_2\Big[(\hat{\alpha}\ssp_1-\hat{\ssp}_\perp)[x_*,z_*]{4\over s^2} F_{\bullet *}[z_*,y_*]
\nonumber\\
&&\hspace{-0.2cm}
~~+ (i\,\Slash{\mathfrak{D}}_\perp [x_*,z_*]){4\over s^2} F_{\bullet *}[z_*,y_*] 
- [x_*,z_*]{4\over s^2} F_{\bullet *}(i\,\Slash{\mathfrak{D}}_\perp[z_*,y_*])\Big]\,,
\label{simplification3}
\end{eqnarray}
and eq. (\ref{quarksubnoedge2}) becomes
\begin{eqnarray}
\hspace{-0.2cm}\brax{i\over \hat{\Sp} +i\epsilon}\kety=&&\!\!
\left[\int_0^{+\infty}\!\!{\dhd \alpha\over 2\alpha}\theta(x_*-y_*) - 
\int_{-\infty}^0\!\!{\dhd\alpha\over 2\alpha}\theta(y_*-x_*) \right]e^{-i\alpha(x_\bullet - y_\bullet)}{1\over \alpha s}
\braxp\, e^{-i{\hatp^2_\perp\over \alpha s}x_*}
\nonumber\\
&&\hspace{-0.2cm}\times
\Bigg\{\hat{\ssp}\ssp_2[x_*,y_*]\hat{\ssp}
+ {ig\over 2\alpha}
\int^{x_*}_{y_*}\!\!d{2\over s}\omega_*\,\hat{\ssp}\ssp_2\,\bigg(\big\{p^i,[x_*,\omega_*]\,{2\over s}\,\omega_*\, F_{i\bullet}(\omega_*)\,[\omega_*,y_*]\big\}
\nonumber\\
&&\hspace{-0.2cm}
+ [x_*,\omega_*]\half F_{ij}\sigma^{ij}[\omega_*,y_*]
+ g\!\!\int^{x_*}_{\omega_*}\!\!d{2\over s}\,\omega'_*\,{2\over s}\big(\omega_* - \omega'_*\big)
[x_*,\omega'_*]F^i_{~\bullet}[\omega'_*,\omega_*]\,
\,F_{i\bullet}\,[\omega_*,y_*]\bigg)\hat{\ssp}
\nonumber\\
&&\hspace{-0.2cm}
+{ig\over 2\alpha}\int_{y_*}^{x_*}\!\!\!d{2\over s}\omega_*\,\hat{\ssp}\ssp_2\bigg[
[x_*,\omega_*]{i\over 4}\big\{(i\Slash{D}_\perp F_{ij}),\gamma^i\gamma^j\big\}[\omega_*,y_*]
+ \big\{\hat{p}^k,[x_*,\omega_*]i F_{kj}\gamma^j[\omega_*,y_*]\big\}
\nonumber\\
&&\hspace{-0.2cm}
+ [x_*,\omega_*]i F_{kj}\gamma^j(i\,\mathfrak{D}^k[\omega_*,y_*]) 
- (i\,\mathfrak{D}^k[x_*,\omega_*])i F_{kj}\gamma^j[\omega_*,y_*] 
\nonumber\\
&&\hspace{-0.2cm}
+(\hat{\alpha}\ssp_1-\hat{\ssp}_\perp)[x_*,\omega_*]\,i\,{2\over s} F_{\bullet *}[\omega_*,y_*]
+ (i\,\Slash{\mathfrak{D}}_\perp [x_*,\omega_*])\,i\,{2\over s} F_{\bullet *}[\omega_*,y_*] 
\nonumber\\
&&\hspace{-0.2cm}
- [x_*,\omega_*]\,i\,{2\over s} F_{\bullet *}(i\,\Slash{\mathfrak{D}}_\perp[\omega_*,y_*])
\bigg]
\Bigg\}
 e^{i{\hatp^2_\perp\over \alpha s}y_*}\ketyp\,.
\label{quarksubnoedge21}
\end{eqnarray}

To underline the structure of result (\ref{quarksubnoedge21}), we define the following two operators
\begin{eqnarray}
\hspace{-0.5cm} 
\hat{\mathcal{O}}_1(x_*,y_*;p_\perp) = \!\!\!&&
{ig\over 2\alpha}\int^{x_*}_{y_*}\!\!\!d{2\over s}\omega_*\bigg(
[x_*,\omega_*]\half \sigma^{ij}F_{ij}[\omega_*,y_*]
+ \big\{\hat{p}^i,[x_*,\omega_*]\,{2\over s}\,\omega_*\, F_{i\bullet}(\omega_*)\,[\omega_*,y_*]\big\}
\nonumber\\
&&+ g\!\!\int^{x_*}_{\omega_*}\!\!\!d{2\over s}\,\omega'_*\,{2\over s}\big(\omega_* - \omega'_*\big)
[x_*,\omega'_*]F^i_{~\bullet}[\omega'_*,\omega_*]\,F_{i\bullet}\,[\omega_*,y_*]\bigg)\,,
\label{O1}
\end{eqnarray}
and
\begin{eqnarray}
\hat{\mathcal{O}}_2(x_*,y_*;p_\perp) = \!\!\!&&
{ig\over 2\alpha}\int^{x_*}_{y_*}\!\!\!d{2\over s}\omega_*
\bigg[
[x_*,\omega_*]{i\over 4}\big\{(i\Slash{D}_\perp F_{ij}),\gamma^i\gamma^j\big\}[\omega_*,y_*]
+ \big\{\hat{p}^k,[x_*,\omega_*]i F_{kj}\gamma^j[\omega_*,y_*]\big\}
\nonumber\\
&&\hspace{-0.2cm}
+ [x_*,\omega_*]i F_{kj}\gamma^j(i\,\mathfrak{D}^k[\omega_*,y_*]) 
- (i\,\mathfrak{D}^k[x_*,\omega_*])i F_{kj}\gamma^j[\omega_*,y_*] 
\nonumber\\
&&\hspace{-0.2cm}
- [x_*,\omega_*]\,i\,{2\over s} F_{\bullet *}(i\,\Slash{\mathfrak{D}}_\perp[\omega_*,y_*])
+ (i\,\Slash{\mathfrak{D}}_\perp [x_*,\omega_*])\,i\,{2\over s} F_{\bullet *}[\omega_*,y_*] 
\nonumber\\
&&\hspace{-0.2cm}
+(\hat{\alpha}\ssp_1-\hat{\ssp}_\perp)[x_*,\omega_*]\,i\,{2\over s} F_{\bullet *}[\omega_*,y_*]
\bigg]\,,
\label{O2}
\end{eqnarray}
where
\begin{eqnarray}
&&{ig\over 2\alpha}\int_{y_*}^{x_*}\!\!\!d{2\over s}\omega_*\Big[[x_*,\omega_*]i F_{kj}\gamma^j(i\,\mathfrak{D}^k[\omega_*,y_*]) 
- (i\,\mathfrak{D}^k[x_*,\omega_*])i F_{kj}\gamma^j[\omega_*,y_*] \Big]
\\
&&= {ig\over 2\alpha}\int_{y_*}^{x_*}\!\!\!d{2\over s}\omega_*
\!\int^{x_*}_{\omega_*}\!\!\!d{2\over s}{\omega'_*}
\Big[
[x_*,\omega'_*]gF^k_{~\bullet}[\omega'_*,\omega_*]iF_{kj}\gamma^j[\omega_*,y_*]
- [x_*,\omega'_*]iF_{kj}\gamma^j[\omega'_*,\omega_*]gF^k_{~\bullet}[\omega_*,y_*]
\Big]\nonumber
\end{eqnarray}
and 
\begin{eqnarray}
&&\hspace{-1cm}{ig\over 2\alpha}\int_{y_*}^{x_*}\!\!\!d{2\over s}\omega_*
\Big[- [x_*,\omega_*]\,i\,{2\over s} F_{\bullet *}(i\,\Slash{\mathfrak{D}}_\perp[\omega_*,y_*])
+ (i\,\Slash{\mathfrak{D}}_\perp [x_*,\omega_*])\,i\,{2\over s} F_{\bullet *}[\omega_*,y_*] \Big]
\\
&&\hspace{-1cm} = {ig\over 2\alpha}\int_{y_*}^{x_*}\!\!\!d{2\over s}\omega_*
\!\int^{x_*}_{\omega_*}\!\!\!d{2\over s}{\omega'_*}
\Big[
[x_*,\omega'_*]i{2\over s}F_{\bullet*}[\omega'_*,\omega_*]\gamma^kgF_{k\bullet}[\omega_*,y_*]
\nonumber\\
&&\hspace{4cm} - [x_*,\omega'_*]\gamma^kgF_{k\bullet}[\omega'_*,\omega_*]i{2\over s}F_{\bullet*}[\omega_*,y_*]
\Big]\,,\nonumber
\end{eqnarray}
and result (\ref{quarksubnoedge2}) becomes 
\begin{eqnarray}
\brax{i\over \hat{\Sp} +i\epsilon}\kety
=\!\!\!&& \left[\int_0^{+\infty}\!\!{\dhd \alpha\over 2\alpha}\theta(x_*-y_*) - 
\int_{-\infty}^0\!\!{\dhd\alpha\over 2\alpha}\theta(y_*-x_*) \right] e^{-i\alpha(x_\bullet - y_\bullet)}
{1\over \alpha s}\,\braxp\,e^{-i{\hatp^2_\perp\over \alpha s}x_*}
\nonumber\\
&&\times\Bigg\{
\hat{\ssp}\ssp_2[x_*,y_*]\hat{\ssp}
+ \hat{\ssp}\ssp_2\,\hat{\mathcal{O}}_1(x_*,y_*;p_\perp)\,\hat{\ssp}
+ \hat{\ssp}\ssp_2 \, \hat{\mathcal{O}}_2(x_*,y_*;p_\perp) \Bigg\}e^{i{\hatp^2_\perp\over \alpha s}y_*}\ketyp
\nonumber\\
&&+ O(\lambda^{-2})\,.
\label{quarksubnoedge3}
\end{eqnarray}
Equation (\ref{quarksubnoedge3}) is the final result for the quark propagator with sub-eikonal corrections in a non symmetric form.
In next section we will rewrite it in a symmetric form with respect to left and right of the shock-wave.


\subsubsection{Symmetrizing the quark-sub-eikonal corrections}


We observe that
the sub-eikonal contribution $\hat{\mathcal{O}}_1$ has, like the leading-eikonal term, 
the operator $\hat{\ssp}$ to its left and to its right,    
while the sub-eikonal contribution $\hat{\mathcal{O}}_2$ has the operator $\hat{\ssp}$ only to its left.
We can eliminate the asymmetry between the propagation to the left and to the right of the shock-wave
considering the following symmetrization
\begin{eqnarray}
\brax{i\over \hat{\Sp} + i\epsilon}\kety = &&
\half\brax\Big[\hat{\Sp}{i\over \hat{\Sp}^2 + i\epsilon} + {i\over \hat{\Sp}^2 + i\epsilon}\hat{\Sp}\Big]\kety\,.
\label{symqprop}
\end{eqnarray}
We have to consider eq. (\ref{simplification2}) and eq. (\ref{simplification3}) with $\hat{\ssp}\ssp_2\hat{\ssp}$
to the right and the final result for the quark propagator with sub-eikonal corrections can be written 
in terms of operators $\hat{\mathcal{O}}_1$ and $\hat{\mathcal{O}}_2$ as
\begin{eqnarray}
\hspace{-1cm}\brax{i\over \hat{\Sp} +i\epsilon}\kety
=\!\!\!&& \left[\int_0^{+\infty}\!\!{\dhd \alpha\over 2\alpha}\theta(x_*-y_*) - 
\int_{-\infty}^0\!\!{\dhd\alpha\over 2\alpha}\theta(y_*-x_*) \right] e^{-i\alpha(x_\bullet - y_\bullet)}
{1\over \alpha s}\,
\nonumber\\
&&\hspace{-0.1cm}
\times\braxp\,e^{-i{\hatp^2_\perp\over \alpha s}x_*}\Bigg\{
\hat{\ssp}\,\ssp_2\,[x_*,y_*]\,\hat{\ssp}
+ \hat{\ssp}\,\ssp_2\,\hat{\mathcal{O}}_1(x_*,y_*;p_\perp)\,\hat{\ssp}
\nonumber\\
&&\hspace{-0.1cm}
+ \hat{\ssp} \,\ssp_2\,\half \hat{\mathcal{O}}_2(x_*,y_*;p_\perp) 
- \half\hat{\mathcal{O}}_2(x_*,y_*;p_\perp)\,\ssp_2\,\hat{\ssp} 
\Bigg\}e^{i{\hatp^2_\perp\over \alpha s}y_*}\ketyp + O(\lambda^{-2})\,.
\label{sym-quarksubnoedge2}
\end{eqnarray}

The operators $\hat{\mathcal{O}}_1$, $\hat{\mathcal{O}}_2$ 
are the sub-eikonal corrections to the quark propagator that \textit{measure} the deviation from the straight-line
due to the finite width of the shock-wave.


\subsection{Quark propagator in the background of quark and anti-quark fields}
\label{sec: qbackground}

\begin{figure}[htb]
	\begin{center}
		\includegraphics[width=2.8in]{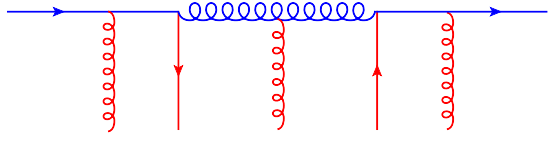}
		\caption{Typical diagram for quark propagator in the background of quark fields. 
			We indicate in blue the quantum field while in red the background one.}
		\label{quarkprop-inq}
	\end{center}
\end{figure}

Up to this point we considered only the background gluon field. In this section we consider the
propagation of a quark in an external field made of quarks and anti-quarks as well (see Fig \ref{quarkprop-inq}). 
Our starting point is
%
\begin{eqnarray}
\hspace{-1cm}\langle \psi(x)\bar{\psi}(y)\rangle_{\psi,\bar{\psi}} 
=\!\!\!&&  g^2\!\!\int\! d^4 z d^4z' \brax{1\over \Sp + i\epsilon}\ketz\gamma^\mu t^a \psi(z)
{\cal G}^{ab}_{\mu\nu}(z,z')\bar{\psi}(z') t^b\gamma^\nu \brazi {1\over \Sp + i\epsilon}\kety\,.
\label{quark-in-q1}
\end{eqnarray}
Propagator (\ref{quark-in-q1}) is made of three terms: two quark propagators in the background of gluon field and 
${\cal G}_{\mu\nu}^{ab}$ the gluon propagator 
in the background of gluon field. Moreover, at point $z$ and $z'$ we have the insertion of the 
background quark fields. The leading eikonal contribution of propagator (\ref{quark-in-q1}) is
\begin{eqnarray}
\langle\psi(x)\bar{\psi}(y)\rangle_{\psi,\bar{\psi}} 
=\!\!\!&& g^2 \!\!\int_{y_*}^{x_*}\!\!\! dz_*\! \int_{y_*}^{z_*}\!\!\!dz'_*
\Big[\int_0^{+\infty}\!{\dhd\alpha\over 2\alpha}\theta(x_*-y_*)
- \int_{-\infty}^0\!{\dhd\alpha\over 2\alpha}\theta(y_* - x_*)\Big]
e^{-i\alpha(x_\bullet - y_\bullet)}\nonumber\\
&&\times {1\over \alpha^4 s^4}\braxp e^{-i{\hatp^2_\perp\over \alpha s}x_*}\ssp\ssp_2\Bigg[
\ssp[x_*,z_*] 
\gamma^\mu t^a\psi(z_*){\cal G}^{ab}_{\mu\nu}(z_*,z'_*)\bar{\psi}(z'_*)t^b\gamma^\nu [z'_*,y_*]\ssp 
\nonumber\\
&&
+ g^2{2\over s}\gamma^i\!
\int_{z_*}^{x_*}\!\!\!d\omega_*[x_*,\omega_*]F_{\bullet i}[\omega_*,z_*]
\gamma^\mu t^a\psi(z_*){\cal G}^{ab}_{\mu\nu}(z_*,z'_*)
\nonumber\\
&&\times\bar{\psi}(z'_*)t^b\gamma^\nu 
{2\over s}\gamma^j\!\!\int_{y_*}^{z'_*}\!\!\!d\omega'_*[z'_*,\omega'_*]
F_{\bullet j}[\omega'_*,y_*]
\nonumber\\
&& - g{2\over s}\gamma^i\!\!
\int_{z_*}^{x_*}\!\!\!d\omega_*[x_*,\omega_*]F_{\bullet i}[\omega_*,z_*]
\gamma^\mu t^a\psi(z_*){\cal G}^{ab}_{\mu\nu}(z_*,z'_*)\bar{\psi}(z'_*)t^b\gamma^\nu [z'_*,y_*]\ssp
\nonumber\\
&& - \ssp[x_*,z_*]\gamma^\mu t^a\psi(z_*){\cal G}^{ab}_{\mu\nu}(z_*,z'_*)\bar{\psi}(z'_*)t^b\gamma^\nu
g{2\over s}\gamma^i\!\!\int_{y_*}^{z'_*}\!\!\!d\omega_*[z'_*,\omega'_*]
F_{\bullet i}[\omega'_*,y_*]
\Bigg]
\nonumber\\
&&\times\ssp_2\ssp \,e^{i{\hatp^2_\perp\over \alpha s}y_*}\ketyp
+ O(\lambda^{-2})\,,
\label{qbackground}
\end{eqnarray}
where ${\cal G}^{ab}_{\mu\nu}$ can either be the eikonal term of the gluon propagator in the background-Feynman gauge eq. (\ref{gluonprofbtotal}) 
or the eikonal term of the light-cone gauge, $p_{2\mu}A^\mu = 0$, propagator eq. (\ref{gluonaxialtotal}). 
To obtain eq. (\ref{qbackground}) we have used 
$e^{-i{\hatp^2_\perp\over\alpha s}z_*}\,\psi\,e^{-i{\hatp^2_\perp\over \alpha s}z_*}$ $=\psi + O(\lambda^{-1})$ and similarly for $\bar{\psi}$. 

We can simplify further eq. (\ref{qbackground}). Indeed, $D^\mu F^a_{\mu\nu} = - g\bar{\psi}t^a\gamma_\nu\psi$
so we can use the following scaling under the boost
\begin{eqnarray}
\bar{\psi}t^a\ssp_1\psi \to  \lambda \bar{\psi}t^a\ssp_1\psi\,,~~~~
\bar{\psi}t^a\gamma^\perp_\nu\psi \to  \bar{\psi}t^a\gamma^\perp_\nu\psi\,,~~~~
\bar{\psi}t^a\ssp_2\psi \to  \lambda^{-1} \bar{\psi}t^a\ssp_2\psi\,.
\label{spinorboost}
\end{eqnarray}
With this power counting in mind, the final result for the quark propagator 
in the background of gluon field is
\begin{eqnarray}
\hspace{-1cm}\langle\psi(x)\bar{\psi}(y)\rangle_{\psi,\bar{\psi}} 
=\!\!\!&&
\Big[\int_0^{+\infty}\!{\dhd\alpha\over 2\alpha}\theta(x_*-y_*)
- \int_{-\infty}^0\!{\dhd\alpha\over 2\alpha}\theta(y_* - x_*)\Big]
e^{-i\alpha(x_\bullet - y_\bullet)}
\label{qbackground1}
\\
&&\times
{g^2\over 4} \!\!\int_{y_*}^{x_*}\!\!\! d{2\over s}z_*\! \int_{y_*}^{z_*}\!\!\!d{2\over s}z'_*\,
\braxp e^{-i{\hatp^2_\perp\over \alpha s}x_*}
\nonumber\\
&&\times
[x_*,z_*] \,\ssp_1\,\gamma^\mu_\perp t^a\psi(z_*)[z_*,z'_*]^{ab}\bar{\psi}(z'_*)t^b\gamma^\perp_\mu
\,\ssp_1\,
[z'_*,y_*]e^{i{\hatp^2_\perp\over \alpha s}y_*}\ketyp 
+ O(\lambda^{-2})\,.
\nonumber
\end{eqnarray}
Notice that, the result (\ref{qbackground1}) is independent of the gauge used for the gluon propagator.


\section{Sub-eikonal corrections to the gluon propagator}
\label{sec: gluonpro}

In this section we calculate the sub-eikonal corrections to the 
gluon propagator in the in the Light-Cone gauge  $p_{2\mu}A^\mu=A_*=0$ in the shock-wave limit.
In the Appendix \ref{sec: feynback} we will calculate the gluon propagator in the background-Feynman gauge.

These corrections have been calculated in refs. \cite{Balitsky:2015qba, Balitsky:2016dgz} for an
external field $A^\mu = (A_\bullet, 0,0)$. We will consider, instead, an
external filed with transverse components different than zero, namely, 
$A^\mu= (A_\bullet, 0, A_\perp)$.

\subsection{Gluon propagator in the light-cone gauge}
\label{sec: gluonprolc}
We start considering the gluon propagator in the background of gluon field in functional form
\begin{eqnarray}
\langle A^a_\mu(x)A^b_\nu(y)\rangle =\!\!\!&&\lim_{w\rightarrow 0}N^{-1}\int DA
A^{qa}_{\mu }(x)A^{qb}_{\nu }(y)\label{c1}
\nonumber\\
&&\times e^{i\int dz {\rm Tr}\{ A^{q}_{\alpha}(z)
	(D^{2}g^{\alpha\beta}-D^{\alpha}D^{\beta} - 2iF_{cl}^{\alpha\beta} - {1\over w}p_2^{\alpha}p_2^{\beta})A^{q}_{\beta}(z)\}}\,,
\label{functionalgprop}
\end{eqnarray}
where $D_{\mu }=\partial _{\mu }-igA^{cl}_{\mu }$. We omit the
superscript ``cl'' from the external field from now on. 
The issue of fixing properly the sub-gauge conditions will not be discussed here. For this, we refer the reader to ref. \cite{Chirilli:2015fza}.

In Schwinger notation the gluon propagator (\ref{functionalgprop}) can be written as
\begin{equation}
i\langle A^a_\mu(x)A^b_\nu(y)\rangle\equiv
iG^{ab}_{\mu \nu }(x,y)=\brax
\frac {1}{\Box^{\mu \nu } - P^{\mu }P^{\nu }+\frac {1}{w}p_2^{\mu }p_2^{\nu }}
\kety^{ab}\,,\label{Shwgaxg}
\end{equation}
where we defined $\Box^{\mu \nu }=P^2 g^{\mu \nu } + 2i\,g\,F^{\mu \nu}$. 
It easy to show that (\ref{Shwgaxg}) satisfies a recursion formula (we omit the symbol $\hat{}$ on top of operators, from now on)
\begin{eqnarray}
{1\over \Box^{\mu\nu} - P^\mu P^\nu + {p_2^\mu p_2^\nu\over \omega}}
=\!\!\!&&
\left(\delta_\mu^\xi - P_\mu {p_2^\xi\over P_*}\right) {1\over \Box^{\xi\eta}}
\left(\delta^\eta_\nu - {p_2^\eta\over P_*}P_\nu \right) + P_\mu{\omega\over P^2_*}P_\nu
\nonumber\\
&& - {g\over \Box^{\mu\alpha} - P^\mu P^\alpha + {p_2^\mu p_2^\alpha\over \omega}}
\Big(D_\lambda F^{\lambda \alpha}{p_2^\beta\over P_*} - P^\alpha{1\over P^2}D_\lambda F^{\lambda \beta}\Big)
{1\over \Box^{\beta\eta}}\left(\delta^\eta_\nu - {p_2^\eta\over P_*}P_\nu \right)
\nonumber\\
&& + {g\over \Box^{\mu\alpha} - P^\mu P^\alpha + {p_2^\mu p_2^\alpha\over \omega}}
D_\lambda F^{\lambda \alpha}{\omega\over P^2_*}P_\nu\,.
\label{Shwgaxg1}
\end{eqnarray}
We need only the terms which are linear and square in the source $D^\mu F^a_{\mu\nu}(x) = -g\bar{\psi}(x) \gamma_\nu t^a \psi(x)$
so, from (\ref{Shwgaxg1}) we get (now we can also set $w\to 0$)
\begin{eqnarray}
{1\over \Box^{\mu\nu} - P^\mu P^\nu + {p_2^\mu p_2^\nu\over \omega}}
=\!\!\!&& \Big(\delta_\mu^\xi - P_\mu {p_2^\xi\over P_*}\Big) {1\over \Box^{\xi\eta}}
\Big(\delta^\eta_\nu - {p_2^\eta\over P_*}P_\nu \Big) 
\label{Shwgaxg2}\\
&& - \Big(\delta_\mu^\xi - P_\mu {p_2^\xi\over P_*}\Big) {1\over \Box^{\xi\eta}}
\Big(
D_\lambda F^{\lambda\eta}{p_2^\beta\over P_*} + {p_2^\eta\over P_*}D_\lambda F^{\lambda \beta}
\nonumber\\
&& - {p_2^\eta\over P_*}P_\alpha
D_\lambda F^{\lambda \alpha}{p_2^\beta\over P_*}
 - P^\eta {1\over P^2}D_\lambda F^{\lambda \beta}
\Big) {1\over \Box^{\beta\sigma}}\Big(\delta_\nu^\sigma -  {p_2^\sigma\over P_*}P_\nu\Big)
\nonumber\\
&& - \Big(\delta_\mu^\xi - P_\mu {p_2^\xi\over P_*}\Big) {1\over \Box^{\xi\eta}}
\Big(
D_\lambda F^{\lambda\eta}{p_2^\beta\over P_*} + {p_2^\eta\over P_*}D_\lambda F^{\lambda \beta}
\nonumber\\
&& - {p_2^\eta\over P_*}P_\alpha
D_\lambda F^{\lambda \alpha}{p_2^\beta\over P_*}
- P^\eta {1\over P^2}D_\lambda F^{\lambda \beta}
\Big) {1\over \Box^{\beta\sigma}}
\nonumber\\
&&\times \Big(
D_{\lambda'} F^{\lambda'\sigma}{p_2^{\beta'}\over P_*} + {p_2^\sigma\over P_*}D_{\lambda'} F^{\lambda' \beta'}
 - {p_2^\sigma\over P_*}P_\alpha
D_{\lambda'} F^{\lambda' \alpha}{p_2^{\beta'}\over P_*}
- P^\sigma {1\over P^2}D_{\lambda'} F^{\lambda' \beta'}
\Big) 
\nonumber\\
&&\times{1\over \Box^{\beta'\sigma'}}
\Big(\delta_\nu^{\sigma'} -  {p_2^{\sigma'}\over P_*}P_\nu\Big)
\,.
\nonumber
\end{eqnarray}
We are interested in calculating corrections up to ${1\over \lambda}$. It is easy to check that 
we need to expand $\Box_{\mu\nu}$ in terms of $F_{\mu\nu}$  up to $F^3$ terms (in subsequent algebra we
will omit the $+i\epsilon$ prescription for each of the ${1\over P^2}$ factor)
\begin{equation}
iG_{\mu \nu }^{ab}(x,y)=
\brax(\delta_\mu^\xi  - P_{\mu }{p_2^\xi \over P_*})
\left[{g_{\xi\eta}\over P^2} \!-\! 2ig{1 \over P^2} F_{\xi\eta}{1\over P^2}\! +\! {\cal O}_{\xi\eta}\right]
(\delta_\nu^\eta \!-\! {p_2^\eta \over P_*}P_\nu)\kety\! +\! O(\lambda^{-2})\,,
\label{Shwgaxg3}
\end{equation}
where we defined the operator ${\cal O}_{\mu\nu}$ as 
\begin{eqnarray}
{\cal O}_{\mu\nu} =\!\!\!\!&& - 4g^2 {1\over P^2} F_\mu^{~\xi }
{ 1\over P^2}F_{\xi \nu}{1 \over P^{2}}
+ 8ig^3 {1\over P^2} F_\mu^{~\xi }{ 1\over P^2}F_{\xi \eta}{1 \over P^{2}}F^\eta_{~\nu}{1 \over P^{2}}\nonumber\\
&&- g{1\over P^2}
\left(D^{\alpha}F_{\alpha\mu}{p_{2\nu}  \over p_*}
+ {p_{2\mu} \over p_*}
D^\alpha F_{\alpha\nu } - {p_{2\mu }\over p_*}P^{\beta}D^{\alpha}F_{\alpha\beta} {p_{2\nu }\over p_*}\right)
{1\over P^2} 
\nonumber\\
&& + 2i{1\over P^2}\Big( {p_{2\mu}\over P_*} D_\lambda F^{\lambda\beta}{1\over P^2}F_{\beta\nu}
+ F_{\mu\eta}{1\over P^2}D_\lambda F^{\lambda\eta}{p_{2\nu}\over P_*}\Big)  {1\over P^2}
\nonumber\\
&&+ {1\over P^2}\Big(2D_\lambda F^\lambda_{~\mu}{1\over P^4}D_\rho F^\rho_{~\nu} 
+ {p_{2\mu}\over P_*}D_\lambda F^{\lambda\beta}{1\over P^2}P_\beta{1\over P^2}D_\rho F^\rho_{~\nu}
\nonumber\\
&&
- {p_{2\mu}\over P_*}P_\alpha D_\lambda F^{\lambda\alpha}{1\over P^4}D_\rho F^\rho_{~\nu}\Big){1\over P^2}\,.
\label{Shwgaxg4}
\end{eqnarray}
Keeping only terms up to $\lambda^{-1}$, eq. (\ref{Shwgaxg3}) becomes
\begin{eqnarray}
iG_{\mu \nu }^{ab}(x,y)
 =\!\!\!&& \bigg[{g_{\mu\nu}\over P^2}
-2i{1\over P^2}\Big({2\over s}p_{2\mu}F_{\bullet \nu} + {2\over s}p_{2\nu}F_{\mu\bullet}\Big){1\over P^2}
- {1\over P^2}{p_{2\mu}\over p_*}P_\nu - P_\mu{p_{2\nu}\over p_*}{1\over P^2}
\nonumber\\
&& 
+{16p_{2\mu}p_{2\nu}\over s^2} {1\over P^2}F^i_{~\bullet}{1\over P^2}F_{i\bullet}{1\over P^2}
- 4{p_{2\mu}p_{2\nu}\over \alpha s^2}{1\over P^2}D^i F_{i\bullet}{1\over P^2}
\bigg]
\nonumber\\
&&- {32i\,p_{2\mu}p_{2\nu}\over s^2}\,{1\over P^2} F^i_{~\bullet}
{1\over P^2}F_{ij}{1 \over P^{2}}F^j_{~\bullet}{1 \over P^{2}} 
-2i{1\over P^2}g^\perp_{\alpha \mu}g^\perp_{\beta \nu}F^{\alpha \beta}{1\over P^2}
\nonumber\\
&&
+ {8 p_{2\mu}g^\perp_{\alpha\nu}\over s}{1\over P^2}F^i_{~\bullet}{1\over P^2}F_i^{~\alpha}{1\over P^2}
+ {8 g^\perp_{\alpha\mu}p_{2\nu}\over s}{1\over P^2}F^{i\alpha}{1\over P^2}F_{i\bullet}{1\over P^2}
\nonumber\\
&& - {1\over P^2}D^i F_i^{~\alpha}g^\perp_{\alpha \mu}{p_{2\nu}\over p_*}{1\over P^2}
-{1\over P^2}{p_{2\mu}\over p_*}D^i F_i^{~\alpha}g^\perp_{\alpha \nu}{1\over P^2} + 
{1\over P^2}{p_{2\mu}\over p_*}P^iD^j F_{ji}{p_{2\nu}\over p_*}{1\over P^2} 
\nonumber\\
&& + {4ip_{2\mu}p_{2\nu}\over s P_*}{1\over P^2}\Big(  D_i F^{ij}{1\over P^2}F_{j\bullet}
- F_{j\bullet}{1\over P^2}D_i F^{ij}\Big)  {1\over P^2}
\nonumber\\
&&
+ {8p_{2\mu}p_{2\nu}\over s^2}{1\over P^2}D_i F^i_{~\bullet}{1\over P^4}D_j F^j_{~\bullet} {1\over P^2}
\label{Shwgaxg6}\,.
\end{eqnarray}
Note that the terms in square bracket contain both eikonal and sub-eikonal terms, while
the terms outside the square bracket are only sub-eikonal terms.
After a bit of algebra we can rewrite (\ref{Shwgaxg6}) as
\begin{eqnarray}
iG_{\mu \nu }^{ab}(x,y)
=\!\!\!&&
\brax\Bigg\{\left(\delta_\mu^\xi - {p_{2\mu}\over p_*}P^\xi\right){g_{\xi\eta}\over P^2}
\left(\delta_\nu^\eta - P^\eta{p_{2\nu}\over p_*}\right)
- {p_{2\mu}p_{2\nu}\over p_*^2}\nonumber\\
&& +\bigg[ {4g^2p_{2\mu}p_{2\nu}\over s p_*} {1\over P^2}F^i_{~\bullet}{1\over P^2}\{P^j,F_{ji}\}{1\over P^2} 
+ {4g^2p_{2\mu}p_{2\nu}\over s p_*}{1\over P^2}\{P^j,F_{ji}\}{1\over P^2}F^i_{~\bullet}{1\over P^2}
\nonumber\\
&& + {2gp_{2\mu}p_{2\nu}\over p_*^2}{1\over P^2}P_iD_j F^{ji}{1\over P^2} 
+ {g^2p_{2\mu}p_{2\nu}\over p_*^2}{1\over P^2}F_{ij}F^{ij}{1\over P^2}
\nonumber\\
&&
- {32i\,g^3\,p_{2\mu}p_{2\nu}\over s^2}\,{1\over P^2} F_{i\bullet}
{1\over P^2}F_{ij}{1 \over P^2}F^j_{~\bullet}{1 \over P^2} 
-2gi{1\over P^2}g^\perp_{\alpha \mu}g^\perp_{\beta \nu}F^{\alpha \beta}{1\over P^2}
\nonumber\\
&& + {8g^2 p_{2\mu}g^\perp_{\alpha\nu}\over s}{1\over P^2}F^i_{~\bullet}{1\over P^2}F_i^{~\alpha}{1\over P^2}
+ {8g^2 g^\perp_{\alpha\mu}p_{2\nu}\over s}{1\over P^2}F^{i\alpha}{1\over P^2}F_{i\bullet}{1\over P^2}
\nonumber\\
&& - {1\over P^2}\,gD^i F_i^{~\alpha}g^\perp_{\alpha \mu}{p_{2\nu}\over p_*}{1\over P^2}
-{1\over P^2}{p_{2\mu}\over p_*}\,gD^i F_i^{~\alpha}g^\perp_{\alpha \nu}{1\over P^2} 
\nonumber\\
&& + {4ip_{2\mu}p_{2\nu}\over s P_*}{1\over P^2}\Big(  D_i F^{ij}{1\over P^2}F_{j\bullet}
- F_{j\bullet}{1\over P^2}D_i F^{ij}\Big)  {1\over P^2}
\nonumber\\
&&
+ {8p_{2\mu}p_{2\nu}\over s^2}{1\over P^2}D_i F^i_{~\bullet}{1\over P^4}D_j F^j_{~\bullet} {1\over P^2}\bigg]\Bigg\}\kety\,.
\label{Shwgaxg7}
\end{eqnarray}
In eq. (\ref{Shwgaxg7}) we need to make expansion of the operator $P^2$. We have to use
the leading-eikonal scalar propagator (in the adjoint representation) for the terms that are already
sub-eikonal \textit{i.e.} all terms in the square bracket, while we need to use the
scalar propagator with sub-eikonal corrections, eq. (\ref{scalpropa-nogcv}), for the first term
right after equal sign in eq. (\ref{Shwgaxg7}). Since we are considering the shock-wave case
we can disregard the terms with fields at the edges of the gauge links (\textit{i.e.} at points $x_*$ and $y_*$) 
in the scalar propagator (\ref{scalpropa-nogcv}).
Taking all this into account, eq. (\ref{Shwgaxg7}) becomes
\begin{eqnarray}
\langle A^a_\mu(x)A^b_\nu(y)\rangle_A
=\!\!\!&& \left[-\int_0^{+\infty}\!\!{\dhd \alpha\over 2\alpha}\theta(x_*-y_*) 
+ \int_{-\infty}^0\!\!{\dhd\alpha\over 2\alpha}\theta(y_*-x_*) \right]e^{-i\alpha(x_\bullet - y_\bullet)}
\braxp e^{-i{\hatp^2_\perp\over \alpha s}x_*}
\nonumber\\
&&\times\left(\delta_\mu^\xi - 
{p_{2\mu}\over p_*}p^\xi\right)\!\calo_\alpha(x_*,y_*)\!
\left(g_{\xi\nu} - 
p_\xi
{p_{2\nu}\over p_*}\right)\!e^{i{\hatp^2_\perp\over \alpha s}y_*} \ketyp^{ab}
+ i\brax{p_{2\mu}p_{2\nu}\over p_*^2}\kety^{ab}\nonumber\\
&& + \left[-\int_0^{+\infty}\!\!{\dhd \alpha\over 2\alpha}\theta(x_*-y_*) 
+ \int_{-\infty}^0\!\!{\dhd\alpha\over 2\alpha}\theta(y_*-x_*) \right]e^{-i\alpha(x_\bullet - y_\bullet)}
\braxp e^{-i{\hatp^2_\perp\over \alpha s}x_*}
\nonumber\\
&&\times\Big[
 \mathfrak{G}^{ab}_{1\mu\nu}(x_*,y_*; p_\perp) +
 \mathfrak{G}^{ab}_{2\mu\nu}(x_*,y_*; p_\perp) + \mathfrak{G}^{ab}_{3\mu\nu}(x_*,y_*; p_\perp) 
 + \mathfrak{G}^{ab}_{4\mu\nu}(x_*,y_*; p_\perp)\Big]
\nonumber\\
&&\times e^{i{\hatp^2_\perp\over \alpha s}y_*}\ketyp
+ O(\lambda^{-2})\,,
\label{Shwgaxg9}
\end{eqnarray}
where we defined 
\begin{eqnarray}
{\calo_\alpha(x_*,y_*)} \equiv\!\!\!&& [x_*,y_*] 
+ {ig\over 2\alpha}
\int^{x_*}_{y_*}\!\!d{2\over s}\omega_*\bigg(
\big\{p^i,[x_*,\omega_*]\,{2\over s}\,\omega_*\, F_{i\bullet}(\omega_*)\,[\omega_*,y_*]\big\}
\nonumber\\
&& + g\!\!\int^{x_*}_{\omega_*}\!\!d{2\over s}\,\omega'_*\,{2\over s}\big(\omega_* - \omega'_*\big)
[x_*,\omega'_*]F^i_{~\bullet}[\omega'_*,\omega_*]\,
\,F_{i\bullet}\,[\omega_*,y_*]\bigg)\,.
\label{Oalfa}
\end{eqnarray}
and
\begin{eqnarray}
\hspace{-1cm}\mathfrak{G}^{ab}_{1\mu\nu}(x_*,y_*; p_\perp) =\!\!\!&&
-{g\,p_{2\mu}p_{2\nu}\over s^2\alpha^3}\int_{y_*}^{x_*}\!\!\!d{2\over s}\omega_*\bigg[
4p^i[x_*,\omega_*]F_{ij}[\omega_*,y_*]p^j
\nonumber\\
&& + ig\int_{\omega'_*}^{x_*}\!\!\!d{2\over s}\omega'_*\,{2\over s}(\omega'_* - \omega_*)
[x_*,\omega'_*]iD^iF_{i\bullet}[\omega'_*,\omega_*]iD^jF_{j\bullet}[\omega_*,y_*]
\bigg]^{ab}\,,
\label{G1}\\
\hspace{-1cm}\mathfrak{G}^{ab}_{2\mu\nu}(x_*,y_*; p_\perp) =\!\!\!&&
- {g\over \alpha}\delta^i_\mu \delta^j_\nu\int_{y_*}^{x_*}\!\!\!d{2\over s}\omega_*
\big([x_*,\omega_*]F_{ij}[\omega_*,y_*]\big)^{ab}\,,
\label{G2}\\
\nonumber\\
\hspace{-1cm}\mathfrak{G}^{ab}_{3\mu\nu}(x_*,y_*; p_\perp) =\!\!\!&&
{g\over  \alpha^2 s}\Big(\delta^j_\mu p_{2\nu} + \delta^j_\nu p_{2\mu}\Big)
\int_{y_*}^{x_*}\!\!\!d{2\over s}\omega_* \,\big([x_*,\omega_*]iD^i F_{ij}[\omega_*,y_*]\big)^{ab}\,,
\label{G3}\\
\nonumber\\
\hspace{-1cm}\mathfrak{G}^{ab}_{4\mu\nu}(x_*,y_*; p_\perp) =\!\!\!&& - {2g^2 \over \alpha^2s}
\int_{y_*}^{x_*}\!\!\!d{2\over s}\omega_*\int_{\omega_*}^{x_*}\!\!\!d{2\over s}\omega'_*
\Big(\delta^j_\nu p_{2\mu}[x_*\omega'_*]F^i_{~\bullet}[\omega'_*,\omega_*]
F_{ij}[\omega_*,y_*]
\nonumber\\
\hspace{-1cm}&& +  \delta^j_\mu p_{2\nu}[x_*\omega'_*]F_{ij}[\omega'_*,\omega_*]
F^i_{~\bullet}[\omega_*,y_*]\Big)^{ab}\,.
\label{G4}
\end{eqnarray}

Equation (\ref{Shwgaxg9}) is our final result for the gluon propagator with sub-eikonal corrections in the light-cone gauge.
At this point we can send $x_*\to\infty$ and $y_*\to -\infty$ and obtain the gluon propagator in the light-cone gauge
with sub-eikonal corrections in the shock-wave case.


\subsection{Gluon propagator in the light-cone gauge  in the background of quark and anti-quark fields}
\label{sec: gluoninquark-axial}

	\begin{figure}[htb]
		\begin{center}
		\includegraphics[width=5.3in]{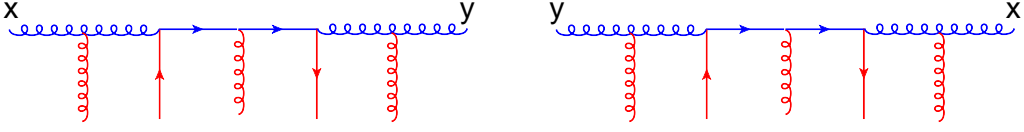}
		\caption{Typical diagram and cross-diagram for gluon propagator in the background of quark fields. As usual, we indicate
			in blue the quantum field while in red the background one.}
		\label{gprop-inquark}
	\end{center}
	\end{figure}

In the previous section we considered the gluon propagator in the axial gauge in the background of gluon fields. 
We now consider the gluon propagator in the background of quarks and anti-quarks in the axial gauge. In section
\ref{sec: gluoninquark-bf} we will consider the background-Feynman gauge.

The Feynman diagram is given in figure (\ref{gprop-inquark})
\begin{eqnarray}
\langle A_\mu^a(x) A_\nu^b(y)\rangle_{\psi,\bar{\psi}} =\!\!\!&&
-g^2\!\!\int\! d^4z_1d^4z_2 \,\langle A_\mu^a(x) A_\rho^c(z_1)\rangle \langle A_\lambda^d(z_2) A_\nu^b(y)\rangle 
\\
&&\times\!\Big[
\bar{\psi}(z_2) t^d \gamma^\lambda\,\bra{z_2}{i\over \Sp \!+\! i\epsilon}\ket{z_1}\,\gamma^\rho t^c\psi(z_1)
\!+\! \bar{\psi}(z_1)t^c\gamma^\rho  \bra{z_1}{i\over \Sp \!+\! i\epsilon}\ket{z_2} \gamma^\lambda t^d \psi(z_2)
\Big]\,.
\nonumber
\end{eqnarray} 

To disentangle the eikonal terms from the sub-eikonal ones, we will use again the scaling in eq. 
(\ref{spinorboost}). The first thing we observe is that the 
gluon propagator in the axial gauge in the background of quark and anti-quark fields starts already with sub-eikonal contribution. 
Therefore, it is sufficient to use the eikonal gluon propagator (\ref{Shwgaxg9}) and the eikonal quark propagator (\ref{leadingscapro}) to obtain
\begin{eqnarray}
&&\hspace{-1cm}\langle A_\mu^a(x) A_\nu^b(y)\rangle_{\psi,\bar{\psi}}
\nonumber\\
~~ =\!\!\!&&
\Big[-\int_0^{+\infty}\!\!{\dhd\alpha \over  2\alpha}\theta(x_* - y_*)
+ \int_{-\infty}^0\!\!{\dhd\alpha\over 2\alpha}\theta(y_* - x_*)\Big]e^{-i\alpha(x_\bullet - y_\bullet)}
\nonumber\\
&&\times g^2\!\int_{y_*}^{x_*}\!\!\! d{2\over s}z_{1*}\int_{y_*}^{z_{1_*}}\!\!\! d{2\over s}z_{2*}
\,{1\over 4\alpha}\Bigg[\braxp
\,e^{-i{\hatp^2_\perp\over \alpha s}x_*}\Big(g_{\perp\mu}^\xi - {p_{2\mu}\over p_*}p^\xi_\perp\Big)
\nonumber\\
&&
\times\!\bar{\psi}(z_{1*}) \gamma^\perp_\xi\,\ssp_1\,[z_{1*},x_*]t^a[x_*,y_*]t^b[y_*,z_{2*}]\gamma^\sigma_\perp\psi(z_{2*})
\Big(g^\perp_{\sigma\nu} - p^\perp_\sigma{p_{2\nu}\over p_*}\Big)
e^{i{\hatp^2_\perp\over \alpha s}y_*}\ketyp 
\nonumber\\
&&
+\,
\brayp
\,e^{-i{\hatp^2_\perp\over \alpha s}y_*}\Big(g_{\perp\nu}^\xi - {p_{2\nu}\over p_*}p^\xi_\perp\Big)
\bar{\psi}(z_{2*})  \gamma^\perp_\xi\,\ssp_1[z_{2*},y_*]t^b[y_*,x_*]t^a[x_*,z_{1*}]\gamma^\sigma_\perp\psi(z_{1*})
\nonumber\\
&&
\times\!\Big(g^\perp_{\sigma\mu} - p^\perp_\sigma{p_{2\mu}\over p_*}\Big)
e^{i{\hatp^2_\perp\over \alpha s}x_*}\ketxp\Bigg] + O(\lambda^{-2}) ~~~~~~~~~~
\label{gluoninqaxial}
\end{eqnarray}
where $p_* = {s\over 2}\alpha$. To arrive at result (\ref{gluoninqaxial}) we have used
$e^{-i{\hatp^2_\perp\over \alpha s}z_*}\psi(z_*) e^{i{\hatp^2_\perp\over \alpha s}z_*} = \psi(z_*) + O(\lambda)$ and similarly for $\bar{\psi}$.

Equation (\ref{gluoninqaxial}) is the gluon propagator in the background of quark and anti-quark fields. 
Contrary to the background-Feynman gauge, as we will see in the Appendix, in the axial gauge the gluon propagator 
in the quark anti-quark external fields does not have any eikonal terms but only sub-eikonal ones. 

\subsection{Summing up the gluon and quark anti-quark external fields contributions}

We can now write the final expression for the gluon propagator in the light-cone gauge in the gluon and quark anti-quark external
fields up to sub-eikonal corrections as
\begin{eqnarray}
&&\hspace{-0.5cm}\langle A_\mu^a(x) A_\nu^b(y)\rangle =
\left[-\int_0^{+\infty}\!\!{\dhd \alpha\over 2\alpha}\theta(x_*-y_*) 
+ \int_{-\infty}^0\!\!{\dhd\alpha\over 2\alpha}\theta(y_*-x_*) \right]e^{-i\alpha(x_\bullet - y_\bullet)}
\braxp e^{-i{\hatp^2_\perp\over \alpha s}x_*}
\nonumber\\
&&\times\left(g_{\perp\mu}^\xi - {p_{2\mu}\over p_*}p^\xi_\perp\right)
\!\calo_\alpha(x_*,y_*)\!
\left(g^\perp_{\xi\nu} - p^\perp_\xi{p_{2\nu}\over p_*}\right)
\!e^{i{\hatp^2_\perp\over \alpha s}y_*} \ketyp^{ab}
+ i\brax{p_{2\mu}p_{2\nu}\over p_*^2}\kety^{ab}\nonumber\\
&& + \left[-\int_0^{+\infty}\!\!{\dhd \alpha\over 2\alpha}\theta(x_*-y_*) 
+ \int_{-\infty}^0\!\!{\dhd\alpha\over 2\alpha}\theta(y_*-x_*) \right]e^{-i\alpha(x_\bullet - y_\bullet)}
\Bigg\{
\braxp\,e^{-i{\hatp^2_\perp\over \alpha s}x_*}
\nonumber\\
&&\times\! 
\bigg(\mathfrak{G}^{ab}_{1\mu\nu}(x_*,y_*; p_\perp) +
\mathfrak{G}^{ab}_{2\mu\nu}(x_*,y_*; p_\perp) + \mathfrak{G}^{ab}_{3\mu\nu}(x_*,y_*; p_\perp) 
+ \mathfrak{G}^{ab}_{4\mu\nu}(x_*,y_*; p_\perp)\bigg)\ketyp\,e^{i{\hatp^2_\perp\over \alpha s}y_*}
\nonumber\\
&&
+ g^2\!\int_{y_*}^{x_*}\!\!\! d{2\over s}z_{1*}\int_{y_*}^{z_{1_*}}\!\!\! d{2\over s}z_{2*}
\,{1\over 4\alpha}\Bigg[
\braxp\,e^{-i{\hatp^2_\perp\over \alpha s}x_*}
\Big(g_{\perp\mu}^\xi - {p_{2\mu}\over p_*}p^\xi_\perp\Big)
\nonumber\\
&&
\times\!\bar{\psi}(z_{1*}) \gamma^\perp_\xi\,\ssp_1\,[z_{1*},x_*]t^a[x_*,y_*]t^b[y_*,z_{2*}]\gamma^\sigma_\perp\psi(z_{2*})
\Big(g^\perp_{\sigma\nu} - p^\perp_\sigma{p_{2\nu}\over p_*}\Big)
e^{i{\hatp^2_\perp\over \alpha s}y_*}\ketyp 
\nonumber\\
&&\
+ \brayp\,e^{-i{\hatp^2_\perp\over \alpha s}y_*}
\Big(g_{\perp\nu}^\xi - {p_{2\nu}\over p_*}p^\xi_\perp\Big)
\bar{\psi}(z_{2*})  \gamma^\perp_\xi\,\ssp_1[z_{2*},y_*]t^b[y_*,x_*]t^a[x_*,z_{1*}]\gamma^\sigma_\perp\psi(z_{1*})
\nonumber\\
&&
\times\!\Big(g^\perp_{\sigma\mu} - p^\perp_\sigma{p_{2\mu}\over p_*}\Big)
e^{i{\hatp^2_\perp\over \alpha s}x_*}\ketxp\Bigg]
\Bigg\}
+ O(\lambda^{-2})\,,
\label{gluonaxialtotal}
\end{eqnarray}
All terms in curly brackets in eq. (\ref{gluonaxialtotal}) are the sub-eikonal corrections: they scale as $\lambda^{-1}$ under the longitudinal boost.

\section{Conclusions and Outlook}


In the high-energy OPE formalism, in the eikonal approximation,
the evolution equations of the relevant operators, the infinite Wilson lines, are the non-linear evolution
BK or B-JIMWLK evolution equations. When sub-eikonal terms are included, new operators appear and, consequently, new non-linear evolution
equations need to be derived. The new evolution equations will describe, for example, the high-energy dynamics of
scattering processes with spin. These sub-eikonal corrections to high-energy OPE are similar 
to higher-twist corrections to the usual light-ray OPE \cite{Balitsky:1987bk} which are now
important part of QCD phenomenology \cite{Braun:2000av, Braun:2008ia, Braun:2009mi, Braun:2017liq}.

The advantage of the operatorial formalism adopted in this paper is provided by
both the gauge invariant representation of the sub-eikonal operators,
and by the systematics provided by the OPE formalism.
 Indeed, given an operatorial definition
of the observable, like, for example, the T-product of two electromagnetic currents in DIS,
the application of the high-energy OPE will
provide a systematic description of the observable in terms of coefficient functions and matrix elements 
of the relevant operators evaluated in the target state. The relevant operators
are provided by the propagation of the projectile-particle
in the background of the target-external field like the propagator in eq. (\ref{sym-quarksubnoedge2}).

The main results obtained in this paper are the sub-eikonal corrections to the 
quark propagator given in eq. (\ref{sym-quarksubnoedge2}), and to the gluon propagator in the light-cone gauge given in eq. (\ref{Shwgaxg9}).

The first step we made was the calculation of the scalar propagator and its sub-eikonal corrections given in eq. (\ref{scalpropa-nogcv}).
To obtained this first result, we showed that the term $\{\hat{P}_\bullet, A_*\}$, which is related to the
possibility for the fields to have dependence also on $x_\bullet$, is actually a sub-sub-eikonal correction. 
We proved it in two different ways. We have also shown that the non gauge-invariant terms at the edges of
the gauge fields in eq. (\ref{scalpropa-nogcv}), which can be put to zero being pure gauges, actually play an important role:
when the scalar propagator is used in intermediate steps, these non gauge-invariant terms
may be in the middle of the shock-wave and therefore cannot be put to zero.

In the quark propagator eq. (\ref{sym-quarksubnoedge2}), the new gauge invariant operators are given in the definition of
$\hat{\mathcal{O}}_1$, $\hat{\mathcal{O}}_2$ 
in eqs. (\ref{O1}) and (\ref{O2}) respectively.

To get the sub-eikonal terms we had to assume that the external field has a finite width. As a consequence one has to
consider also the possibility that the scattering particle starts (or ends) its propagation inside the target-filed. To this end, 
we have derived scalar propagator (\ref{scaprop-subeik-xx}) and the quark 
propagator (\ref{qprop-subeik-xx}) which take into account this possibility.

In section \ref{sec: gluonprolc} we have calculated the sub-eikonal corrections to the gluon propagator in the 
light-cone gauge. In ref. \cite{Balitsky:2016dgz} such correction have been calculated for
a background field $A^\mu(x) = (A_\bullet(x_*,x_\perp),0,0)$. Here we considered an external field where all the components
of the external field are different then zero. The result is given in eq. (\ref{gluonaxialtotal}); it includes the
contribution due to both the gluon and quark anti-quark external fields. The operators
$\mathfrak{G}^{ab}_{1\mu\nu}$, $\mathfrak{G}^{ab}_{2\mu\nu}$, $\mathfrak{G}^{ab}_{3\mu\nu}$ and $\mathfrak{G}^{ab}_{4\mu\nu}$,
given in equations (\ref{G1}), (\ref{G2}), (\ref{G3}) and (\ref{G4}), respectively, are the result of
an external gluon field with all filed components different then zero, while eq. (\ref{gluoninqaxial}) is the sub-eikonal corrections due to 
quark anti-quark external fields.
In background-Feynman gauge, instead, the gluon propagator is given in eq. (\ref{gluonprofbtotal}), and the sub-eikonal correctins
are all included in the operators $\mathfrak{B}^{ab}_{1\mu\nu}$,  $\mathfrak{B}^{ab}_{2\mu\nu}$,  $\mathfrak{B}^{ab}_{3\mu\nu}$,  $\mathfrak{B}^{ab}_{4\mu\nu}$,  
$\mathfrak{B}^{ab}_{5\mu\nu}$,  $\mathfrak{B}^{ab}_{6\mu\nu}$ and $\mathfrak{Q}^{ab}_{1\mu\nu}$,
$\mathfrak{Q}^{ab}_{2\mu\nu}$, $\mathfrak{Q}^{ab}_{3\mu\nu}$, 
$\mathfrak{Q}^{ab}_{4\mu\nu}$, $\mathfrak{Q}^{ab}_{5\mu\nu}$ in Eqs. 
(\ref{B1}), (\ref{B2}), (\ref{B3}),  (\ref{B4}),  (\ref{B5}), and (\ref{B6}) and (\ref{Q1}), (\ref{Q2}), (\ref{Q3}),  (\ref{Q4}),  (\ref{Q5}),  respectively.

In ref. \cite{Bartels:1995iu, Bartels:1996wc} it was shown that 
both the gluon and the quark distribution contribute equally to the spin structure function at low-$x_B$. This
result was obtained within the double-logarithmic approximation (DLA). To study the
quark distribution within the Wilson-line formalism, in section \ref{sec: qbackground}, we considered 
also the sub-eikonal corrections due to quark and anti-quark in the target-external field. 
The quark propagator with such sub-eikonal corrections is given in eq. (\ref{qbackground1}). 
Not only will this result be relevant for high-energy spin-dynamics but also 
to obtain sub-eikonal corrections to the BK equation. Although such corrections to the BK equations
are energy suppressed, one may obtain for the first time the Regge limit of scattering amplitudes
with two-fermions in the t-channel \cite{Gorshkov:1966ht} within the Wilson-line formalism.

The applications of the results derived in this paper does not end with high-energy spin-dynamics. 
The TMD formalism developed so far (for a review see, \cite{fcollins}) 
does not describe data at sufficiently low-$x_B$. If one wants to have at hand a
formalism that can be applicable at a wider range of $x_B$, one has to consider sub-eikonal corrections as a
way to connect to lower energies and thus moderate $x_B$. In the case of gluon TMDs this 
connection has been already made in refs. \cite{Balitsky:2015qba, Balitsky:2016dgz}. In the case 
of quark-TMDs, instead, one my use the results derived in this paper, although here we have not included terms coming from
twist expansion since we considered classical fields and quantum fields with comparable transverse momenta.
Corrections due to the hierarchy between the transverse momenta of the quantum and classical field
is left for future publication.

The author is grateful to I. Balitsky and V.M. Braun for valuable discussions.

\appendix
\section{Notation}
\label{sec: notation}

In this section we explain some of the notations used through out the paper.

Given two light-cone vectors $p_1^\mu$ and $p_2^\mu$, with $p_1^\mu p_{2\mu} = {s\over 2}$,
we can decompose any coordinate as $x^\mu = {2\over s}x_*p_1^\mu + {2\over s}x_\bullet p_2^\mu + x_\perp^\mu$
with $x_* = x_\mu p_2^\mu= \sqrt{{s\over 2}}x^+$, $x_\bullet = x_\mu p_1^\mu= \sqrt{{s\over 2}}x^-$ and 
$x^\pm = {x^0\pm x^3\over \sqrt{2}}$.
We use the notation $x^\mu_\perp = (0,x^1,x^2,0)$ and $x^i = (x^1,x^2)$ such that $x^ix_i = x^\mu_\perp x^\perp_\mu = - x^2_\perp$. 
So, Latin indexes assume values $1, 2$, while Greek indexes run from $0$ to $3$.

We define the gauge link at fixed transverse position as
\begin{eqnarray}
[up_1,vp_2]_z \equiv [up_1 + z_\perp, vp_1 + z_\perp] 
\equiv {\rm Pexp}\Big\{ig\!\!\int_v^u\!\!d t \,A_\bullet(tp_1+z_\perp)\Big\}\,.
\end{eqnarray}
The derivative of the gauge link with respect to the transverse position is
\begin{eqnarray}
{\partial \over \partial z^i} [up_1, vp_1 ]_z
=\!\!\!&& ig A_i(up_1+z_\perp)[up_1,vp_1]_z - ig [up_1,vp_1]_zA_i(vp_1+z_\perp)
\nonumber\\
&&- ig\!\int^u_v\!\!\! ds\,[up_1,sp_1]_zF_{\bullet i}(p_1s+z_\perp)[p_1s,p_1v]_z\,,
\label{deriv-glink}
\end{eqnarray}
with index $i=1,2$.  
From (\ref{deriv-glink}) we may formally define the transverse covariant derivative 
$\mathfrak{D}_i $ that acts on a non-local operator as
\begin{eqnarray}
\hspace{-1cm}i\mathfrak{D}_i\, [up_1, vp_1]_z &\!\equiv\!& 
i{\partial \over \partial z^i} [up_1, vp_1]_z
+g\big[A_i(z_\perp), [up_1,vp_1]_z\big]
\nonumber\\
&\!=\!& g\!\int^u_v\!\!\! ds\,[up_1,sp_1]_zF_{\bullet i}(p_1s+z_\perp)[p_1s,p_1v]_z\,,
\label{coderiv-glink}
\end{eqnarray}
where we have used the implicit notation 
$\big[A_i(z_\perp), [up_1,vp_1]_z\big] = A_i(z_\perp+up_1) [up_1,vp_1]_z - [up_1,vp_1]_z\,A_i(z_\perp + vp_1)$.

Given a gauge link $[x_*,y_*]_z\equiv[{2\over s}x_*p_1 + z_\perp, {2\over s}y_*p_1 + z_\perp]$, in Schwinger notation we have
\begin{eqnarray}
\braxp [x_*,y_*] \ketyp = [x_*,y_*]_x\,\delta^{(2)}(x-y)\,.
\label{schwi-nota}
\end{eqnarray}
The transverse momentum operator $\hat{P}_i = \hat{p}_i + g \hat{A}_i$ acts on the gauge link as
\begin{eqnarray}
&&\hspace{-1.2cm}\braxp \big[\hat{P}_i, [x_*,y_*]\big]\ketyp \!=\! \braxp i\mathfrak{D}_i[x_*,y_*] \ketyp
\!= \!\braxp g{2\over s}\int^{x_*}_{y_*}\!\!\! d\omega_*
\,[x_*,\omega_*]F_{\bullet i}[\omega_*,y_*] \ketyp\,,
\label{defPi}
\end{eqnarray}
where we used again the short-hand notation 
$[x_*,\omega_*]F_{i\bullet}[\omega_*,y_*] = [x_*,\omega_*]F_{i\bullet}(\omega_*)[\omega_*,y_*]$.
The point to make regarding (\ref{defPi}) is that the covariant derivative $i\mathfrak{D}_i$ acts on the gauge link even though the 
transverse coordinate has not been specified yet and, as matter of fact, it does not have to
in order to know how it acts on the gauge link. Therefore, through out the paper we will
make quite a bit of algebra involving the gauge link, the momentum operator $P_i$ and the
covariant derivative $i\mathfrak{D}_i$ without specifying the \textit{bra} $\braxp$ and the \textit{ket} $\ketyp$.

At this point, one is tempted to identify the covariant derivative $i\mathfrak{D}_i$ as the usual
covariant derivative which acts on a local operator. This identification would not be correct. Indeed, one can easily check that
\begin{eqnarray}
\big[i\mathfrak{D}_i,i\mathfrak{D}_j\big][x_*,y_*] = ig F_{ij}[x_*,y_*] - ig[x_*,y_*]F_{ij}\,.
\label{commD}
\end{eqnarray}
To arrive at (\ref{commD}) we have implemented the definition of $i\mathfrak{D}_i$ given in (\ref{coderiv-glink}).

Another identity involving $\mathfrak{D}_i$ that we will often use is
\begin{eqnarray}
\hspace{-0.7cm}i\mathfrak{D}_j([x_*,\omega_*]F_{\bullet i}[\omega_*, y_*])
=\!\!\!&& (i\mathfrak{D}_j[x_*,\omega_*])F_{\bullet i}[\omega_*,y_*] 
+ [x_*,\omega_*]F_{\bullet i}(i\mathfrak{D}_j[\omega_*,y_*])
\nonumber\\
&& + [x_*,\omega_*](iD_jF_{\bullet i})[\omega_*,y_*]
\nonumber\\
=\!\!\!&& g{2\over s}\!\int_{\omega_*}^{x_*}\!\!\!dz_*\,[x_*,z_*]F_{\bullet j}[z_*,\omega_*]F_{\bullet i}[\omega_*,y_*]
\nonumber\\
&&
+g{2\over s}\!\int_{y_*}^{\omega_*}\!\!\!dz_*[x_*,\omega_*]F_{\bullet i}[\omega_*, z_*]F_{\bullet j}[z_*,y_*]
\nonumber\\
&& + [x_*,\omega_*](iD_j F_{\bullet i})[\omega_*,y_*]\,.
\label{identity1}
\end{eqnarray}
Notice that in eq. (\ref{identity1}) derivative $\mathfrak{D}_i$ acts on the gauge link, while the usual covariant derivative $D_j$
acts on the local field operator $F_{\bullet i}$ with respect to the transverse coordinate that will be specified in a second step
according to Schwinger notation (\ref{schwi-nota}). So, as we can see from the first line after the equal sign in eq. 
(\ref{identity1}), $\mathfrak{D}_i$ follows the
usual Leibnitz rule for derivative of product of functions with the exception that when $\mathfrak{D}_i$ acts
on a gauge link it acts as in eq. (\ref{coderiv-glink}), while when it acts on a local filed operator 
it becomes the usual covariant derivative. It is now easy to show that
\begin{eqnarray}
\mathfrak{D}^2_\perp [x_*,y_*] = &&
2 g^2\!\int^{x_*}_{y_*}\!\!d{2\over s}\omega_*\int_{\omega_*}^{x_*}\!\!\! d{2\over s}\omega'_*
\, [x_*,\omega'_*]F_{i\bullet}[\omega'_*,\omega_*]F^i_{~\bullet}[\omega_*,y_*]
\nonumber\\
&& + g\!\int^{x_*}_{y_*}\!\!\!d{2\over s}\omega_*[x_*,\omega_*](iD^iF_{\bullet i})[\omega_*,y_*]\,,
\end{eqnarray}
where $\mathfrak{D}^2_\perp = - \mathfrak{D}^i\mathfrak{D}_i$.

\section{An alternative type of expansion for the quark propagator}
\label{sec: 2type-exp}

\begin{figure}[htb]
	\begin{center}
		\includegraphics[width=4.5in]{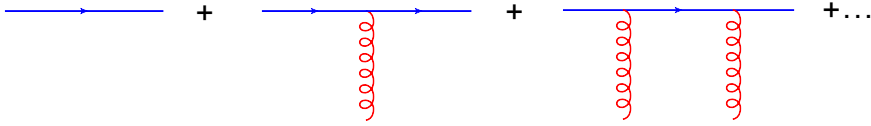}
	\end{center}
	\caption{Diagrammatic expansion of the quark propagator in the external field. Quantum fields are in blue, while 
		classical fields are in red as usual.}
	\label{diagramqexp}
\end{figure}

The procedure we adopted in section \ref{sec: leadinqpro} to obtain the quark propagator in the eikonal approximation is certainly 
not the only one. In this section we consider the following type of expansion   
\begin{eqnarray}
\brax {1\over \Sp+i\epsilon}\kety =\!\!\!&& \brax{1\over \ssp+i\epsilon}\kety - \brax{1\over \ssp+i\epsilon}g\Sa{1\over \ssp+i\epsilon}\kety
\nonumber\\
&&
+ \brax{1\over \ssp+i\epsilon}g\Sa{1\over \ssp+i\epsilon}g\Sa{1\over \ssp+i\epsilon}\kety + \dots\,,
\label{qexp}
\end{eqnarray}
which is diagrammatically shown in figure \ref{diagramqexp}.
Now we are interested only in the eikonal contribution, so we can use, in Schwinger notation, 
 $\brax \Sa \kety = {2\over s}\ssp_2A_{\bullet}(x_*,x_\perp)\delta^{(4)}(x-y)$ and 
eq. (\ref{qexp}) becomes
\begin{eqnarray}
\brax {1\over \Sp+i\epsilon}\kety 
=\!\!\!&& {\ssx - \ssy\over 2\pi^2[(x-y)^2-i\epsilon]^2} 
- {i\over s}\int_{y_*}^{x_*}\!\!\!dz_*
\bigg[\int_0^{+\infty}\!{\dhd\alpha\over 2\alpha^2}\,\theta(x_*-z_*)\theta(z_*-y_*) 
\label{qqexp}\\
&& -
\int^0_{-\infty}\!{\dhd\alpha\over 2\alpha^2}\theta(y_*-z_*)\theta(z_*-x_*)\bigg]e^{-i\alpha(x_\bullet-y_\bullet)}
\nonumber\\
&& \times
\braxp\, \hat{\ssp}\,e^{-i{\hatp^2_\perp\over \alpha s}(x_*-z_*)}\, ig{2\over s}\ssp_2
\hat{A}_\bullet(z_*)\, e^{-i{\hatp^2_\perp\over \alpha s}(z_*-y_*)}\,\hat{\ssp}\,\ketyp
\nonumber\\
&& - {i\over s}\int_{y_*}^{x_*}\!\!\!dz_*\int_{y_*}^{z_*}\!\!\!dz'_*
\Big[\int_0^{+\infty}\!{\dhd\alpha\over 2\alpha^2}\,\theta(x_*-z_*)\theta(z_*-z'_*)\theta(z'_*-y_*) 
\nonumber\\
&& - \int^0_{-\infty}\!{\dhd\alpha\over 2\alpha^2}\theta(y_*-z'_*)\theta(z'_*-z_*)\theta(z_*-x_*)\Big]e^{-i\alpha(x_\bullet-y_\bullet)}
\nonumber\\
&&\times\braxp \,\hat{\ssp}\,e^{-i{\hatp^2_\perp\over \alpha s}(x_*-z_*)}\, ig{2\over s}\ssp_2
\hat{A}_\bullet(z_*)\, e^{-i{\hatp^2_\perp\over \alpha s}(z_*-z'_*)}
\, ig{2\over s}
\hat{A}_\bullet(z'_*)\, e^{-i{\hatp^2_\perp\over \alpha s}(z'_*-y_*)}\,\hat{\ssp}\,\ketyp\,.
\nonumber
\end{eqnarray}
We can rewrite expansion (\ref{qqexp}) as a path ordered exponential
\begin{eqnarray}
\brax {i\over \hat{\Sp}+i\epsilon}\kety 
=\!\!\!&&
{1\over s}\left[\int_0^{+\infty}\!{\dhd\alpha\over 2\alpha^2}\,\theta(x_* - y_*) - 
\int^0_{-\infty}\!{\dhd\alpha\over 2\alpha^2}\theta(y_* - x_*)\right]e^{-i\alpha(x_\bullet-y_\bullet)}
\braxp\,\hat{\ssp}\,e^{-i{\hatp^2_\perp\over \alpha s}x_*}
\nonumber\\
&&\times \ssp_2
{\rm Pexp}\left\{ig\int_{y_*}^{x_*}\!\! d{2\over s}z_*\, e^{-i{\hatp^2_\perp\over \alpha s}z_*}
\hat{A}_\bullet(z_*) e^{i{\hatp^2_\perp\over \alpha s}z_*}\right\} e^{i{\hatp^2_\perp\over \alpha s}y_*}\,
\hat{\ssp}\,\ketyp\,.
\label{qpropa-extA1}
\end{eqnarray}
We may perform a further approximation due to the infinite boost (we are interest only in the eikonal contribution)
\begin{eqnarray}
e^{-i{\hatp^2_\perp\over \alpha s}z_*}
\hat{A}_\bullet(z_*) e^{i{\hatp^2_\perp\over \alpha s}z_*} 
= \hat{A}_\bullet(z_*) + O(\lambda^0)\,,
\end{eqnarray}
where $\lambda$ is the large parameter of the Lorentz boost. With this approximation we arrive at
\begin{eqnarray}
\brax {i\over \hat{\Sp}+i\epsilon}\kety 
=\!\!\!&& 
{1\over s}\left[\int_0^{+\infty}\!{\dhd\alpha\over 2\alpha^2}\,\theta(x_* - y_*) - 
\int^0_{-\infty}\!{\dhd\alpha\over 2\alpha^2}\theta(y_* - x_*)\right]e^{-i\alpha(x_\bullet-y_\bullet)}
\nonumber\\
&&\times \braxp\,\hat{\ssp}\,e^{-i{\hatp^2_\perp\over \alpha s}x_*}\ssp_2
[x_*,y_*] e^{i{\hatp^2_\perp\over \alpha s}y_*}\,
\hat{\ssp}\,\ketyp\,.
\label{qpropa-extA2}
\end{eqnarray}
Notice that to arrive at propagator (\ref{qpropa-extA2}), which is already in the shock-wave form, we did not need
to perform the gauge rotation we performed in section \ref{sec: leadinqpro}, because expansion (\ref{qexp})
assumes from the start a free propagation before and after the interaction with the external field, as can also
be seen from its diagrammatic representation in figure\ref{diagramqexp}. In other words, when we start drawing
Feynman diagrams, like in figure\ref{diagramqexp}, we are implicitly fixing a gauge for the external field.


\section{The $\{\hat{P}_\bullet,A_*\}$ term}
\label{sec: astarterm}

%
In section \ref{sec: subeiko-scalar} we showed that the term $\{\hat{P}_\bullet,A_*\}$ is actually a sub-sub-eikonal correction.
We now demonstrate the same thing following an alternative procedure.
To this end, we notice that (for simplicity we omit the $+i\epsilon$ prescription from each ${1\over \hatp^2 + 2\alpha g A_\bullet}$
and the $\hat{}$ symbol on top of operators)
\begin{eqnarray}
&&{1\over p^2 + 2\alpha g A_\bullet}\{P_\bullet, A_*\}{1\over p^2 + 2\alpha g A_\bullet}
\nonumber\\
&&= {1\over 2\alpha}\{A_*,{1\over \hatp^2 + 2\alpha g A_\bullet}\} +
 {1\over 2\alpha}{1\over p^2 + 2\alpha g A_\bullet}\{p^2_\perp,A_*\}
{1\over p^2 + 2\alpha g A_\bullet}
\nonumber\\
&&= {1\over 2\alpha}\{A_*,{1\over p^2 + 2\alpha g A_\bullet}\} + O(\lambda^{-2})\,.
\label{edgeastar}
\end{eqnarray}
Thus, we are left with sub-eikonal corrections with the $A_*$ (a gauge dependent term) only 
at the edges of the gauge link. When we use the scalar propagator in intermediate steps with the terms (\ref{edgeastar}),
we will have terms like
\begin{eqnarray}
&&{1\over p^2 + 2\alpha g A_\bullet}\{A_*, igF_{i\bullet}\}{1\over p^2 + 2\alpha g A_\bullet}
\nonumber\\
=\!\!\!&& {1\over p^2 + 2\alpha g A_\bullet}\Big(P_i P_\bullet A_* - A_* P_\bullet P_i \Big){1\over p^2 + 2\alpha g A_\bullet}
+ {1\over 2\alpha}{1\over p^2 + 2\alpha g A_\bullet}A_* P_i - {1\over 2\alpha}P_iA_*{1\over p^2 + 2\alpha g A_\bullet}
\nonumber\\
&&
+ {1\over 2\alpha}{1\over p^2 + 2\alpha g A_\bullet}A_*P_ip^2_\perp {1\over p^2 + 2\alpha g A_\bullet}
- {1\over 2\alpha}{1\over p^2 + 2\alpha g A_\bullet}p^2_\perp P_i A_* {1\over p^2 + 2\alpha g A_\bullet}
\nonumber\\
=\!\!\!&& {1\over 2\alpha}{1\over p^2 + 2\alpha g A_\bullet}A_* P_i - {1\over 2\alpha}P_iA_*{1\over p^2 + 2\alpha g A_\bullet}
+ O(\lambda^{-2})\,.
\label{astarinerme}
\end{eqnarray}
So, once again, we arrived at an expression which has the non gauge-invariant terms only at the edges of the gauge links
and they could be eliminated by a proper gauge choice since are outside the shock-wave.
To arrive at eq. (\ref{astarinerme}), we noticed that the term
\begin{eqnarray}
\hspace{-1cm} {1\over 2\alpha}{1\over p^2 + 2\alpha g A_\bullet}A_*P_ip^2_\perp {1\over p^2 + 2\alpha g A_\bullet}
- {1\over 2\alpha}{1\over p^2 + 2\alpha g A_\bullet}p^2_\perp P_i A_* {1\over p^2 + 2\alpha g A_\bullet}
\end{eqnarray}
is sub-sub-eikonal and therefore can be disregarded. Moreover, we used
\begin{eqnarray}
&&\hspace{-2cm}\brax{1\over p^2 + 2\alpha g A_\bullet}A_* P_\bullet P_i {1\over p^2 + 2\alpha g A_\bullet}\kety
\nonumber\\
&&= \int dz \brax {1\over p^2 + 2\alpha g A_\bullet}\ketz A_*(z_*,z_\perp) iD^z_\bullet iD^z_i 
\braz{1\over p^2 + 2\alpha g A_\bullet}\kety
\nonumber\\
&&\hspace{-2cm}=\left[-\int_0^{+\infty}\!\!{\dhd \alpha\over 4\alpha^2}\theta(x_*-y_*) 
+ \int_{-\infty}^0\!\!{\dhd\alpha\over 4\alpha^2}\theta(y_*-x_*) \right]e^{-i\alpha(x_\bullet - y_\bullet)}
\nonumber\\
&&\hspace{-2cm}\times\!
\int_{y_*}^{x_*}\!\!\!d{2\over s}z_{1*}\braxp e^{-i{\hatp^2_\perp\over \alpha s}x_*}
[x_*,z_*]e^{i{\hatp^2_\perp\over \alpha s}z_*} A_*(z_*) iD^z_\bullet iD^z_i 
e^{-i{\hatp^2_\perp\over \alpha s}z_*}[z_*,y_*]e^{i{\hatp^2_\perp\over \alpha s}y_*}\ketyp
\nonumber
\end{eqnarray}
\begin{eqnarray}
=\!\!\!&&  \left[-\int_0^{+\infty}\!\!{\dhd \alpha\over 4\alpha^2}\theta(x_*-y_*) 
+ \int_{-\infty}^0\!\!{\dhd\alpha\over 4\alpha^2}\theta(y_*-x_*) \right]e^{-i\alpha(x_\bullet - y_\bullet)}
\nonumber\\
&&\times
(-g)\int_{y_*}^{x_*}\!\!\!d{2\over s}z_{1*}\braxp e^{-i{\hatp^2_\perp\over \alpha s}x_*}
[x_*,z_*]e^{i{\hatp^2_\perp\over \alpha s}z_*}
A_*(z_*) iD^z_\bullet e^{-i{\hatp^2_\perp\over \alpha s}z_*} 
\nonumber\\
&&\times\!\int_{y_*}^{z_*}\!\!\!d{2\over s}z'_*[z_*,z'_*]F_{i\bullet}[z'_*,y_*]
e^{i{\hatp^2_\perp\over \alpha s}y_*}\ketyp
\nonumber\\
=\!\!\!&&  \left[-\int_0^{+\infty}\!\!{\dhd \alpha\over 4\alpha^2}\theta(x_*-y_*) 
+ \int_{-\infty}^0\!\!{\dhd\alpha\over 4\alpha^2}\theta(y_*-x_*) \right]e^{-i\alpha(x_\bullet - y_\bullet)}
\nonumber\\
&&\times
(-g)\int_{y_*}^{x_*}\!\!\!d{2\over s}z_{1*}\braxp e^{-i{\hatp^2_\perp\over \alpha s}x_*}
[x_*,z_*]
\Big(A_*(z_*)+ {iz_*\over \alpha s}[\hatp^2,A_*]\Big)
\Big({iz_*\over \alpha s}[\hatp^2_\perp,A_\bullet] + {\hatp^2_\perp\over 2\alpha} + iD^z_\bullet  \Big)
\nonumber\\
&&\times\!
\int_{y_*}^{z_*}\!\!\!d{2\over s}z'_*[z_*,z'_*]F_{i\bullet}[z'_*,y_*]
e^{i{\hatp^2_\perp\over \alpha s}y_*}\ketyp
= O(\lambda^{-2})\,,
\end{eqnarray}
where we used the identity $\big(i\partial^x_\bullet + gA_\bullet(x_*)\big)[x_*,z_*] = 0$.
Clearly, the advantage of the procedure adopted in section \ref{sec: subeiko-scalar} is that one does not have to deal with
non gauge-invariant terms even though they are at the edges of the gauge links.

\section{Some calculation details of the sub-eikonal corrections to the quark propagator}
\label{sec: calc-details}

In this section we present some details of the calculation to obtain eq. (\ref{sum4lines}).
Our starting point is eq. (\ref{4terms-quapro}) which we report here
\begin{eqnarray}
&&\hspace{-0.3cm}{\rm Pexp}\left\{ig\!\int_{y_*}^{x_*}\!\!\!d{2\over s}\omega_*
\,e^{i{\hatp^2_\perp\over \alpha s}\omega_*}\left(A_\bullet(\omega_*) + {B(\omega_*)\over 2\alpha}+
{i\over 2\alpha}  {2\over s}F_{\bullet i}(\omega_*)\,\ssp_2\gamma^i\right)\,e^{-i{\hatp^2_\perp\over \alpha s}\omega_*}\right\}
\nonumber\\
&&\hspace{-0.3cm}= [x_*,y_*]
+ ig\!\int_{y_*}^{x_*}\!\!\!d{2\over s}\omega_*
\,[x_*,\omega_*]\left({O(\omega_*)\over 2\alpha} + i{\omega_*\over \alpha s}[p^2_\perp, A_\bullet]\right)[\omega_*,y_*]
\nonumber\\
&&
+  {ig\over 2\alpha}\!\int_{y_*}^{x_*}\!\!\!d{2\over s}\omega_*\,[x_*,\omega_*]
\left({4\over s^2}F_{\bullet*}\sigma_{*\bullet} + {1\over 2}F_{ij}\sigma^{ij}\right)[\omega_*,y_*]
\nonumber\\
&& + ig\!\int_{y_*}^{x_*}\!\!\!d{2\over s}\omega_*\,[x_*,\omega_*]
\left({i\over 2\alpha}  {2\over s}F_{\bullet i}\,\ssp_2\gamma^i 
+ i{\omega_*\over \alpha s}[p^2_\perp, {i\over 2\alpha}  {2\over s}F_{\bullet i}\,\ssp_2\gamma^i]\right)[\omega_*,y_*]
\nonumber\\
&& + (ig)^2\!\int_{y_*}^{x_*}\!\!\!d{2\over s}\omega_*\!\int_{y_*}^{\omega_*}\!\!\!d{2\over s}\omega'_*\,
[x_*,\omega_*]\left({B(\omega_*)\over 2\alpha}
+ i{\omega_*\over \alpha s}[p^2_\perp, A_\bullet] \right)
[\omega_*,\omega'_*]\,
{i\over 2\alpha} {2\over s}F_{\bullet i}(\omega'_*)\,\ssp_2\gamma^i \,[\omega'_*,y_*]
\nonumber\\
&& + (ig)^2\!\int_{y_*}^{x_*}\!\!\!d{2\over s}\omega_*\!\int_{y_*}^{\omega_*}\!\!\!d{2\over s}\omega'_*\,
[x_*,\omega_*]\,{i\over 2\alpha} {2\over s}F_{\bullet i}(\omega_*)\,\ssp_2\gamma^i 
[\omega_*,\omega'_*]\left({B(\omega'_*)\over 2\alpha} + i{\omega'_*\over \alpha s}[p^2_\perp, A_\bullet] \right)[\omega'_*,y_*]
\nonumber\\
&& + O(\lambda^{-2})\,.
\label{4terms-quaproA}
\end{eqnarray}
Our goal is to rewrite (\ref{4terms-quaproA}) in a gauge invariant form.
To this end, we divide the right-hand-side of eq. (\ref{4terms-quaproA}) in four pieces, simplify them separately, and then 
sum them up. 

Using result (\ref{scalpropa-nogcv}), the first piece in (\ref{4terms-quapro}) reduces to
\begin{eqnarray}
&&[x_*,y_*] + ig\!\int_{y_*}^{x_*}\!\!\!d{2\over s}\omega_*
\,[x_*,\omega_*]\left({O(\omega_*)\over 2\alpha} + i{\omega_*\over \alpha s}[p^2_\perp, A_\bullet]\right)[\omega_*,y_*]
\nonumber\\
&&~~ + {ig\over 2\alpha}\!\int_{y_*}^{x_*}\!\!\!d{2\over s}\omega_*\,[x_*,\omega_*]
\left({4\over s^2}F_{\bullet*}\sigma_{*\bullet} + {1\over 2}F_{ij}\sigma^{ij}\right)[\omega_*,y_*]
\nonumber\\
&&= [x_*,y_*] + {ig\over 2\alpha}\Bigg[{2\over s}x_*\Big(\{P_i,A^i(x_*)\} - gA_i(x_*)A^i(x_*)\Big)[x_*,y_*]
\nonumber\\
&&~~ - [x_*,y_*]{2\over s}y_*\Big(\{P_i,A^i(y_*)\} - gA_i(y_*)A^i(y_*)\Big)
\nonumber\\
&&~~+ \int^{x_*}_{y_*}\!\!\!d{2\over s}\omega_*\bigg(
\big\{P^i,[x_*,\omega_*]\,{2\over s}\,\omega_*\, F_{i\bullet}(\omega_*)\,[\omega_*,y_*]\big\}
\nonumber\\
&&~~+ g\!\!\int^{x_*}_{\omega_*}\!\!\!d{2\over s}\,\omega'_*\,{2\over s}\big(\omega_* - \omega'_*\big)
[x_*,\omega'_*]F^i_{~\bullet}[\omega'_*,\omega_*]\,
\,F_{i\bullet}\,[\omega_*,y_*]\bigg)\Bigg]
\nonumber\\
&&~~+  {ig\over 2\alpha}\!\int_{y_*}^{x_*}\!\!\!d{2\over s}\omega_*\,[x_*,\omega_*]
\left({4\over s^2}F_{\bullet*}\sigma_{*\bullet} + {1\over 2}F_{ij}\sigma^{ij}\right)[\omega_*,y_*]\,.
\label{term1}
\end{eqnarray}

Next, making use of the following identity
\begin{eqnarray}
[p^2_\perp,F_{j\bullet}] = \!\!\!&& g\left(\{P^i,A_i\} - gA^iA_i\right)F_{j\bullet} 
\nonumber\\
&& - gF_{j\bullet}\left(\{P^i,A_i\} - gA^iA_i\right)
- \{P^i, (iD_iF_{j\bullet})\}\,,
\label{usefulident}
\end{eqnarray}
the second piece reduces to
\begin{eqnarray}
&&\hspace{-0.5cm}ig\!\int_{y_*}^{x_*}\!\!\!d{2\over s}\omega_*\,[x_*,\omega_*]
\left({i\over 2\alpha}  {2\over s}F_{\bullet i}\,\ssp_2\gamma^i 
+ i{\omega_*\over \alpha s}[p^2_\perp, {i\over 2\alpha}  {2\over s}F_{\bullet i}\,\ssp_2\gamma^i]\right)[\omega_*,y_*]
\nonumber\\
=\!\! && - {i\over 2\alpha}{2\over s}\ssp_2(\Slash{\mathfrak{D}}_\perp[x_*,y_*])  
-  {ig\over 4\alpha^2}\gamma^j{2\over s}\ssp_2\int_{y_*}^{x_*}\!\!\!d{2\over s}\omega_*\Bigg[{2\over s}\omega_*\,
[x_*,\omega_*]
\nonumber\\
&&\times\bigg(g\left(\{P^i,A_i\} - gA^iA_i\right)F_{j\bullet} 
- gF_{j\bullet}\left(\{P^i,A_i\} - gA^iA_i\right) \bigg)[\omega_*,y_*] 
\nonumber\\
&&-  {2\over s}\omega_*\{P^i,[x_*,\omega_*] (iD_iF_{j\bullet})[\omega_*,y_*]\}
- {2\over s}\omega_*[x_*,\omega_*](iD^iF_{j\bullet})(i\mathfrak{D}_i[\omega_*,y_*]) 
\nonumber\\
&& + {2\over s}\omega_*(i\mathfrak{D}^i[x_*,\omega_*])(iD_iF_{j\bullet})[\omega_*,y_*]
\Bigg]\,.
\label{term2}
\end{eqnarray}
Note that identity (\ref{usefulident}) highlights the importance of the transverse fields at the edges of the gauge-link
in the scalar propagator (\ref{scalpropa-nogcv}). These terms can be set to zero only when they are at the point $x_*$ and 
$y_*$, but when we use the scalar propagator as an intermediate step, these terms will not always be at the edges. So, 
setting them to zero once and for all, will lead to a wrong result.

We now turn our attention to the third piece. After some algebra it reduces to
\begin{eqnarray}
&&(ig)^2\!\int_{y_*}^{x_*}\!\!\!d{2\over s}\omega_*\!\int_{y_*}^{\omega_*}\!\!\!d{2\over s}\omega'_*\,
[x_*,\omega_*]\left({B(\omega_*)\over 2\alpha}
+ i{\omega_*\over \alpha s}[p^2_\perp, A_\bullet] \right)
[\omega_*,\omega'_*]\,
{i\over 2\alpha} {2\over s}F_{\bullet i}(\omega'_*)\,\ssp_2\gamma^i \,[\omega'_*,y_*]
\nonumber\\
&&=\left({ig\over 2\alpha}\right)^2\int_{y_*}^{x_*}\!\!\!d{2\over s}\omega'_*\Bigg(
{2\over s}x_*\left(\{P_i,A^i(x_*)\} - gA_iA^i\right)[x_*,\omega'_*]
\nonumber\\
&&~~ - [x_*,\omega'_*]{2\over s}\omega'_*\left(\{P_i,A^i(\omega'_*)\} - gA_iA^i\right)\Bigg)
\,i{2\over s}F_{i\bullet}\gamma^i\ssp_2[\omega'_*,y_*]
\nonumber\\
&&~~ + {g\over 4\alpha^2}{2\over s}\ssp_2\int_{y_*}^{x_*}\!\!\!d{2\over s}\omega_*[x_*,\omega_*]\calb_1
(\Slash{\mathfrak{D}}_\perp[\omega_*,y_*])
\nonumber\\
&&~~+ {g^2\over 4\alpha^2}{2\over s}\ssp_2\int_{y_*}^{x_*}\!\!\!d{2\over s}\omega'_*\int^{\omega'_*}_{y_*}\!\!\!d{2\over s}\omega_*
{2\over s}\omega_*
[x_*,\omega'_*]F^i_{~\bullet}[\omega'_*,\omega_*]F_{i\bullet}\,(\Slash{\mathfrak{D}}_\perp[\omega_*,y_*])
\nonumber\\
&&~~ + {g\over 4\alpha^2}{2\over s}\ssp_2\int_{y_*}^{x_*}\!\!\!d{2\over s}\omega_*\bigg[
\big\{P^i,[x_*,\omega_*]{2\over s}\omega_*F_{i\bullet}(\Slash{\mathfrak{D}}_\perp[\omega_*,y_*])\big\}
\nonumber\\
&&
+ [x_*,\omega_*]{2\over s}\omega_*F_{i\bullet}\big(i\mathfrak{D}^i(\Slash{\mathfrak{D}}_\perp[\omega_*,y_*])\big)
\bigg]\,.
\label{term3}
\end{eqnarray}

Finally, let us consider the forth piece
\begin{eqnarray}
&&(ig)^2\!\int_{y_*}^{x_*}\!\!\!d{2\over s}\omega_*\!\int_{y_*}^{\omega_*}\!\!\!d{2\over s}\omega'_*\,
[x_*,\omega_*]\,{i\over 2\alpha} {2\over s}F_{\bullet i}(\omega_*)\,\ssp_2\gamma^i 
[\omega_*,\omega'_*]\left({B(\omega'_*)\over 2\alpha} + i{\omega'_*\over \alpha s}
[p^2_\perp, A_\bullet] \right)[\omega'_*,y_*]
\nonumber\\
&&=\left({ig\over 2\alpha}\right)^{\!2}\!\int_{y_*}^{x_*}\!\!\!d{2\over s}\omega_*\bigg[
[x_*,\omega_*]\,i{2\over s}F_{i\bullet}\gamma^i\ssp_2
\bigg({2\over s}\omega_*\left(\{P_i,A^i\} - gA_iA^i\right)[\omega_*,y_*] 
\nonumber\\
&&~~ - {2\over s}y_*[\omega_*,y_*]\left(\{P_i,A^i\} - gA_iA^i\right)\bigg)
+ {2\over s}\ssp_2 (i\mathfrak{D}^i(\Slash{\mathfrak{D}}_\perp[x_*,\omega_*])){2\over s}\omega_*F_{i\bullet}[\omega_*,y_*]
\nonumber\\
&&~~
-\{P^i,(\Slash{\mathfrak{D}}_\perp[x_*,\omega_*]){2\over s}\omega_*F_{i\bullet}[\omega_*,y_*]\}
- (\Slash{\mathfrak{D}}_\perp[x_*,\omega_*])\calb_1[\omega_*,y_*]
\nonumber\\
&&~~
+g\!\int_{\omega_*}^{x_*}\!\!\!d{2\over s}\omega'_*\,{2\over s}\omega'_*
(\Slash{\mathfrak{D}}_\perp[x_*,\omega'_*])F^i_{~\bullet}[\omega'_*,\omega_*]F_{i\bullet}[\omega_*,y_*]
\bigg]\,.
\label{term4}
\end{eqnarray}
We can now sum the four terms (\ref{term1}), (\ref{term2}), (\ref{term3}) and (\ref{term4}) and,
since we are interested in the shock-wave limit, we can disregard the fields 
that are at the edges of the gauge-links, that is at points $x_*$ and $y_*$. After some algebra, we obtain
\begin{eqnarray}
&&\hspace{-0.8cm}{\rm Pexp}\left\{ig\!\int_{y_*}^{x_*}\!\!\!d{2\over s}\omega_*
\,e^{i{\hatp^2_\perp\over \alpha s}\omega_*}\left(A_\bullet(\omega_*) + {B(\omega_*)\over 2\alpha}+
{i\over 2\alpha}  {2\over s}F_{\bullet i}(\omega_*)\,\ssp_2\gamma^i\right)\,e^{-i{\hatp^2_\perp\over \alpha s}\omega_*}\right\}
\nonumber\\
&&\hspace{-0.8cm}=
\Bigg(\big(1 - {1\over 2\alpha}{2\over s}\ssp_2\,i\,\Slash{\mathfrak{D}}_\perp\big)[x_*,y_*] 
+ {ig\over 2\alpha}\!\int_{y_*}^{x_*}\!\!\!d{2\over s}\omega_*\,[x_*,\omega_*]
\calb_1[\omega_*,y_*]\Bigg) 
\nonumber\\
&&\hspace{-0.8cm}
+ {1\over 4\alpha^2} \int_{y_*}^{x_*}\!\!\!d{2\over s}z_*
\Big[(i\,\Slash{\mathfrak{D}}_\perp {2\over s}\ssp_2 [x_*,z_*])ig\calb_1[z_*,y_*] 
+ [x_*,z_*]ig\calb_1(i\,\Slash{\mathfrak{D}}_\perp{2\over s}\ssp_2[z_*,y_*])\Big]
\nonumber\\
&&\hspace{-0.8cm}
+ {ig\over 2\alpha}\Bigg[
\int^{x_*}_{y_*}\!\!\!d{2\over s}\omega_*\bigg(
\big\{p^i,[x_*,\omega_*]\,{2\over s}\,\omega_*\, F_{i\bullet}(\omega_*)\,[\omega_*,y_*]\big\}
\nonumber\\
&&\hspace{-0.8cm}
+ g\!\!\int^{x_*}_{\omega_*}\!\!\!d{2\over s}\,\omega'_*\,{2\over s}\big(\omega_* - \omega'_*\big)
[x_*,\omega'_*]F^i_{~\bullet}[\omega'_*,\omega_*]
\,F_{i\bullet}\,[\omega_*,y_*]\bigg)\Bigg]
- {ig\over 4\alpha^2}\gamma^j{2\over s}\ssp_2\int_{y_*}^{x_*}\!\!\!d{2\over s}\omega_*\,{2\over s}\omega_*
\nonumber
\end{eqnarray}
\vspace{-1.2cm}
\begin{eqnarray}
&&\hspace{-0.2cm}
\times\bigg[
(i\mathfrak{D}^i[x_*,\omega_*])(iD_iF_{j\bullet})[\omega_*,y_*]
- [x_*,\omega_*](iD^iF_{j\bullet})(i\mathfrak{D}_i[\omega_*,y_*]) 
- \{p^i,[x_*,\omega_*] (iD_iF_{j\bullet})[\omega_*,y_*]\}
\bigg]
\nonumber\\
&&\hspace{-0.2cm} 
+ {g\over 4\alpha^2}{2\over s}\ssp_2\int_{y_*}^{x_*}\!\!\!d{2\over s}\omega_*
\bigg[
{2\over s}\omega_*(\Slash{\mathfrak{D}}_\perp[x_*,\omega_*])F_{i\bullet}(i\mathfrak{D}^i[\omega_*,y_*])
-{2\over s}\omega_*
(i\mathfrak{D}^i[x_*,\omega_*])F_{i\bullet}(\Slash{\mathfrak{D}}_\perp[\omega_*,y_*])
\nonumber\\
&&\hspace{-0.2cm}
+ \{p^i,[x_*,\omega_*]{2\over s}\omega_*F_{i\bullet}(\Slash{\mathfrak{D}}_\perp[\omega_*,y_*])\}
+ \{p^i,(\Slash{\mathfrak{D}}_\perp[x_*,\omega_*]){2\over s}\omega_*F_{i\bullet}[\omega_*,y_*]\}
\nonumber\\
&&\hspace{-0.2cm} + [x_*,\omega_*]{2\over s}\omega_*F_{i\bullet}\,i\mathfrak{D}^i(\Slash{\mathfrak{D}}_\perp[\omega_*,y_*])
-(i\mathfrak{D}^i(\Slash{\mathfrak{D}}_\perp[x_*,\omega_*])){2\over s}\omega_*F_{i\bullet}[\omega_*,y_*]\bigg]
+ O(\lambda^{-2})\,,
\label{sum4linesA}
\end{eqnarray}
where we have defined $\calb_1\equiv{4\over s^2}F_{\bullet*}\sigma_{*\bullet} + {1\over 2}F_{ij}\sigma^{ij}$.

Note that, to get to eq. (\ref{sum4linesA}) we could have started, alternatively, from the following type of expansion
\begin{eqnarray}
&&\brax \Sp{i\over P^2 + {g\over 2}F_{\mu\nu}\sigma^{\mu\nu}}\kety
\nonumber\\
&&=\brax\Sp\bigg[{i\over P^2} - {i\over P^2}\big(
ig{2\over s}F_{i\bullet}\gamma^i\ssp_2+ g\calb_1\big){1\over P^2}
\nonumber\\
&&~~
+ {i\over P^2}\big(
ig{2\over s}F_{i\bullet}\gamma^i\ssp_2 + g\calb_1\big){1\over P^2}
\big( 
ig{2\over s}F_{i\bullet}\gamma^i\ssp_2+ g\calb_1\big){1\over P^2}
\bigg]\kety
\nonumber\\
&&=  i\Slash{D}\brax\bigg[{i\over p^2 + 2g\alpha A_\bullet + gO} - 
{i\over p^2 + 2g\alpha A_\bullet + gO}ig{2\over s}F_{i\bullet}\gamma^i\ssp_2{1\over p^2 + 2g\alpha A_\bullet + gO}
\nonumber\\
&&
 ~~ - {i\over p^2 + 2g\alpha A_\bullet }\,g\calb_1\,{1\over p^2 + 2g\alpha A_\bullet }
+ {i\over p^2 + 2g\alpha A_\bullet }ig{2\over s}F_{i\bullet}\gamma^i\ssp_2{1\over p^2 + 2g\alpha A_\bullet }
\calb_1{1\over p^2 + 2g\alpha A_\bullet }
\nonumber\\
&&
~~ + {i\over p^2 + 2g\alpha A_\bullet }\,\calb_1\,{1\over p^2 + 2g\alpha A_\bullet }
ig{2\over s}F_{i\bullet}\gamma^i\ssp_2{1\over p^2 + 2g\alpha A_\bullet }\bigg]\kety\,,
\end{eqnarray}

and, substituting the scalar propagator with sub-eikonal corrections, eq. (\ref{scalpropa-nogcv}),
for each ${i\over p^2 + 2g\alpha A_\bullet + gO}$
factor, and the eikonal scalar propagator, eq. (\ref{leadingscapro}), for each ${i\over p^2 + 2g\alpha A_\bullet}$ factor, we would get to
the same gauge invariant expression, eq. (\ref{sum4linesA}), with the help of steps similar to Eqs. (\ref{term1}),
(\ref{term2}), (\ref{term3}) and (\ref{term4}).

\section{Quark and scalar propagators with one end-point in the external field}
\subsection{Scalar propagator}
In section \ref{sec: subeiko-scalar} we derived the scalar propagator for the shock-wave
case, that is, for the case in which the particle starts and ends its propagation
outside the shock-wave as shown in figure \ref{swprpagation}. We expanded around a point which is 
in the middle of the external field, eq. (\ref{subscaexp}). We are now interested in the scalar propagator 
for a particle that starts or ends its propagation inside the external field. We consider the case in which the point $x_*$
is in the shock-wave, \textit{i.e.} inside the interval in which the filed strength tensor is different then zero. 
In this case, expansion (\ref{scalexpan}) can be written as
\begin{eqnarray}
\hspace{-0.5cm}\brax {i\over P^2+i\epsilon}\kety
=\!\!\!&& \left[\int_0^{+\infty}\!\!{\dhd \alpha\over 2\alpha}\theta(x_*-y_*) - 
\int_{-\infty}^0\!\!{\dhd\alpha\over 2\alpha}\theta(y_*-x_*) \right]e^{-i\alpha(x_\bullet - y_\bullet)}
\nonumber\\
&&\times
\braxp\,
{\rm Pexp} \left\{ig\!\! \int_{y_*}^{x_*}d{2\over s}\omega_*\, 
e^{i{\hatp^2_\perp\over \alpha s}(\omega_*-x_*)}\left(A_\bullet(\omega_*) + {O(\omega_*)\over 2\alpha}\right)
e^{-i{\hatp^2_\perp\over \alpha s}(\omega_*-x_*)} \right\} 
\nonumber\\
&&\times e^{-i{\hatp^2_\perp\over \alpha s}(x_*-y_*)}\ketyp\,.
\label{scalprop-xpoint1}
\end{eqnarray}
In eq. (\ref{scalprop-xpoint1}), we observe that the coordinate $x_*$, which is the end point of the particle's propagation,
is now in the path ordered exponential. This means that both, particle and external field end at $x_*$.
We can now repeat the same steps we performed in the previous section and arrive at
\begin{eqnarray}
\brax {i\over P^2+i\epsilon}\kety 
=\!\!\!&& \left[\int_0^{+\infty}\!\!{\dhd \alpha\over 2\alpha}\theta(x_*-y_*) - 
\int_{-\infty}^0\!\!{\dhd\alpha\over 2\alpha}\theta(y_*-x_*) \right]e^{-i\alpha(x_\bullet - y_\bullet)}
\nonumber\\
&&\times
\braxp\,\bigg\{[x_*,y_*] 
+ {ig\over 2\alpha}\bigg[
[x_*,y_*]{2\over s}(x_*-y_*)\Big(\{P_i,A^i(y_*)\} - gA_i(y_*)A^i(y_*)\Big)
\nonumber\\
&& +\!\int^{x_*}_{y_*}\!\!d{2\over s}\omega_*\bigg(
 \big\{P^i,[x_*,\omega_*]\,{2\over s}\,(\omega_*-x_*)\, F_{i\bullet}(\omega_*)\,[\omega_*,y_*]\big\}
\nonumber\\
&& + g\!\!\int^{x_*}_{\omega_*}\!\!d{2\over s}\,\omega'_*\,{2\over s}\big(\omega_* - \omega'_*\big)
[x_*,\omega'_*]F^i_{~\bullet}[\omega'_*,\omega_*]\,
\,F_{i\bullet}\,[\omega_*,y_*]\bigg)\bigg]\bigg\}
\nonumber\\
&&\times e^{-i{\hatp^2_\perp\over \alpha s}(x_*-y_*)}\ketyp
+ O(\lambda^{-2})\,.
\label{scaprop-subeik-x}
\end{eqnarray}
We will now show that, since the particle ends its propagation inside the external field at point $x_*$, 
the non gauge-invariant terms, the transverse field $A_i$,  will be only at point $y_*$. 
To this end we use
\begin{eqnarray}
&&\int_{y_*}^{x_*}\!\!\!d{2\over s}\omega_*\,
\big\{P^i,[x_*,\omega_*]{2\over s}(\omega_*-x_*)F_{i\bullet}[\omega_*,y_*]\big\}
\nonumber\\
&& = \int_{y_*}^{x_*}\!\!\!d{2\over s}\omega_*
\bigg[[x_*,\omega_*]{2\over s}(\omega_*-x_*)(iD^iF_{i\bullet}(\omega_*))[\omega_*,y_*]
+ 2[x_*,\omega_*]{2\over s}(\omega_*-x_*)F_{i\bullet}(\omega_*)[\omega_*,y_*]P^i\bigg]
\nonumber\\
&&~~ - g\int_{y_*}^{x_*}\!\!\!d{2\over s}\omega_*\int_{\omega_*}^{x_*}\!\!\!d{2\over s}\omega'_*\Big(
[x_*,\omega'_*]F_{i\bullet}(\omega'_*)[\omega'_*,\omega_*]{2\over s}(\omega_*-x_*)F^i_{~\bullet}(\omega_*)[\omega_*,y_*]
\nonumber\\
&&~~
+ [x_*,\omega'_*]{2\over s}(\omega'_*-x_*)F_{i\bullet}(\omega'_*)[\omega'_*,\omega_*]F^i_{~\bullet}(\omega_*)[\omega_*,y_*]
\Big)\,.
\label{pitotheright}
\end{eqnarray}
To arrive at eq. (\ref{pitotheright}) we have \textit{pushed} the operator $P_i$ to the right of the gauge link up to point
$y_*$ which is the point outside the range in which $F_{\mu\nu}^{cl}\neq 0$.
Hence, result (\ref{scaprop-subeik-x}) can be written as
\begin{eqnarray}
\brax {i\over P^2+i\epsilon}\kety 
=\!\!\!&& \left[\int_0^{+\infty}\!\!{\dhd \alpha\over 2\alpha}\theta(x_*-y_*) - 
\int_{-\infty}^0\!\!{\dhd\alpha\over 2\alpha}\theta(y_*-x_*) \right]e^{-i\alpha(x_\bullet - y_\bullet)}
\nonumber\\
&&\times
\braxp\,\Bigg\{[x_*,y_*] 
+ {ig\over 2\alpha}\bigg[
[x_*,y_*]{2\over s}(x_*-y_*)\Big(\{P_i,A^i(y_*)\} - gA_i(y_*)A^i(y_*)\Big)
\nonumber\\
&& +\!\int^{x_*}_{y_*}\!\!d{2\over s}\omega_*\bigg(
 [x_*,\omega_*]{2\over s}(\omega_*-x_*)(iD^iF_{i\bullet}(\omega_*))[\omega_*,y_*]\nonumber\\
&&
-2g\int_{\omega_*}^{x_*}\!\!\!d{2\over s}\omega'_*
[x_*,\omega'_*]{2\over s}(\omega'_*-x_*)F_{i\bullet}(\omega'_*)[\omega'_*,\omega_*]F^i_{~\bullet}(\omega_*)[\omega_*,y_*]
\nonumber\\
&&
+ 2[x_*,\omega_*]{2\over s}(\omega_*-x_*)F_{i\bullet}(\omega_*)[\omega_*,y_*]P^i\bigg)
\bigg]\Bigg\}e^{-i{\hatp^2_\perp\over \alpha s}(x_*-y_*)}\ketyp
\nonumber\\
&& + O(\lambda^{-2})\,.
\label{scaprop-subeik-xx}
\end{eqnarray}

\begin{figure}
	\begin{center}
		\includegraphics[width=3.1in]{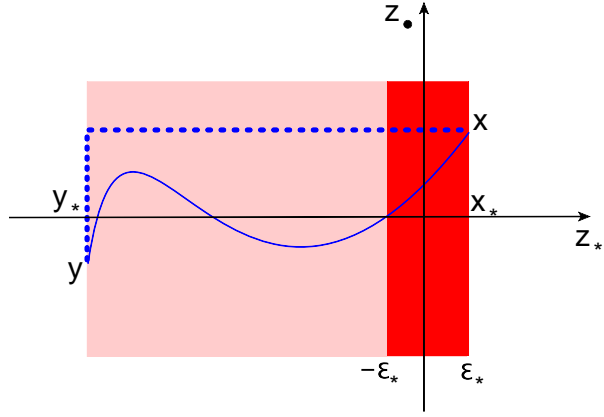}
		\caption{Particle starts its propagation within the shock-wave. 
			In this case the pure gauge is only to the left of the shock-wave.}
		\label{leftswprpagation}
	\end{center}
\end{figure}

All non gauge-invariant terms in (\ref{scaprop-subeik-xx}) are now only
to the right of the gauge links, that is to the point $y_*$. This is opposite to what we have in
eq. (\ref{scalpropa-nogcv}) which has the transverse field $A^i$ on both edges of the gauge link (\textit{i.e.} at points $x_*$ and $y_*$).
In the gauge-rotated field $A^\Omega$, we can set such non gauge-invariant terms to zero.

In a similar way we can expand with respect to 
$y_\perp + {2\over s}p_1 y_*$ and get
\begin{eqnarray}
\hspace{-0.0cm}\brax {i\over P^2+i\epsilon}\kety 
=\!\!\!&& \left[\int_0^{+\infty}\!\!{\dhd \alpha\over 2\alpha}\theta(x_*-y_*) - 
\int_{-\infty}^0\!\!{\dhd\alpha\over 2\alpha}\theta(y_*-x_*) \right]e^{-i\alpha(x_\bullet - y_\bullet)}
\nonumber\\
&&\times
\braxp\,e^{-i{\hatp^2_\perp\over \alpha s}(x_*-y_*)}
\nonumber\\
&&\times\Bigg\{[x_*,y_*] 
+ {ig\over 2\alpha}\bigg[
\Big(\{P_i,A^i(x_*)\} - gA_i(x_*)A^i(x_*)\Big)[x_*,y_*]{2\over s}(x_*-y_*)
\nonumber\\
&&\hspace{0.4cm}
+ \int^{x_*}_{y_*}\!\!d{2\over s}\omega_*\bigg(
- [x_*,\omega_*]{2\over s}(\omega_*-y_*)(iD^iF_{i\bullet}(\omega_*))[\omega_*,y_*]
\nonumber\\
&&\hspace{0.4cm}
+2g\int_{\omega_*}^{x_*}\!\!\!d{2\over s}\omega'_*
[x_*,\omega'_*]F_{i\bullet}(\omega'_*)[\omega'_*,\omega_*]{2\over s}(\omega_*-y_*)F^i_{~\bullet}(\omega_*)[\omega_*,y_*]
\nonumber\\
&&\hspace{0.4cm}
+ 2P^i[x_*,\omega_*]{2\over s}(\omega_*-y_*)F_{i\bullet}(\omega_*)[\omega_*,y_*]\bigg)\bigg]\Bigg\}\ketyp
+ O(\lambda^{-2})\,.
\label{scaprop-subeik-yy}
\end{eqnarray}
In eq. (\ref{scaprop-subeik-yy}) we have the non gauge-invariant terms only at point $x_*$

\subsection{Quark propagator}
\label{sec: qpropsubxx}

In this section we derive the sub-eikonal corrections to the quark propagator with one end-point inside the external filed.
To this end we need to use result (\ref{scaprop-subeik-xx}). 
Let us suppose that $x_\perp + {2\over s}p_1x_*$ is the end-point of the quark propagator in 
the external field as shown in figure \ref{leftswprpagation}. It is convenient to start from the following expression of the quark propagator
\begin{eqnarray}
\brax {i\over \hat{\Sp} + i\epsilon}\kety = \brax 
{i\over \hatp^2 + 2\alpha g A_\bullet + gB + i {2\over s}gF_{\bullet i}\,\ssp_2\gamma^i + i\epsilon}\,\hat{\Sp}\kety\,.
\end{eqnarray}
and we need to include the point $x_*$ in the path-ordered exponential. So, we have
\begin{eqnarray}
\hspace{-1cm}\brax{i\over \hat{\Sp}+i\epsilon}\kety
=\!\!\!&& \left[\int_0^{+\infty}\!\!{\dhd\alpha\over 2\alpha}\theta(x_*-y_*) - 
\int_{-\infty}^0\!\!{\dhd\alpha\over 2\alpha}\theta(y_*-x_*)\right]
e^{-i\alpha(x_\bullet - y_\bullet)}
\braxp\,e^{-i{\hatp^2_\perp\over \alpha s}x_*}
\nonumber\\
&&\times
{\rm Pexp}\Bigg\{ig\!\int_{y_*}^{x_*}\!\!\!d{2\over s}\omega_*
\,e^{i{\hatp^2_\perp\over \alpha s}\omega_*}\left(A_\bullet(\omega_*) + {B(\omega_*)\over 2\alpha} +
{i\over 2\alpha}  {2\over s}F_{\bullet i}(\omega_*)\,\ssp_2\gamma^i\right)
\nonumber\\
&&\times
e^{-i{\hatp^2_\perp\over \alpha s}\omega_*}\Bigg\}
e^{i{\hatp^2_\perp\over \alpha s}y_*}\ketyp
\left(-i\overleftarrow{\slashd}^y + g{2\over s}A_\bullet(y_*,y_\perp)\ssp_2 + g\Sa_\perp(y_*,y_\perp)\right)\,.
\end{eqnarray}
We also need
\begin{eqnarray}
&&
\brazp e^{i{\hatp^2_\perp\over \alpha s}(y_*-x_*)}\ketyp\,
 e^{-i\alpha(x_\bullet - y_\bullet)}\left(i\overleftarrow{\slashd}^y + g{2\over s}\ssp_2 A_\bullet(y_*,y_\perp) 
+ \Sa(y_*)\right)
\nonumber\\
=\!\!\!&&
\brazp\Big({1\over \alpha s}\ssp\ssp_2\ssp - {2\over s}\ssp_2i\overleftarrow{D}^y_\bullet 
- i{(x_*-y_*) \over \alpha s}[\hatp^2_\perp, g{2\over s}\ssp_2A_\bullet(y_*)] + \Sa_\perp(y_*)\Big)
\nonumber\\
&&\times
e^{i{\hatp^2_\perp\over \alpha s}(y_*-x_*)}\ketyp\, e^{-i\alpha(x_\bullet-y_\bullet)}\,.
\end{eqnarray}

At this point the procedure we have to adopt is the same as the one performed in the previous section. 
Therefore, we arrive at
\begin{eqnarray}
&&\hspace{-0.5cm}\brax {i\over \Sp+i\epsilon}\kety
=\left[\int_0^{+\infty}\!\!{\dhd \alpha\over 2\alpha}\theta(x_*-y_*) - 
\int_{-\infty}^0\!\!{\dhd\alpha\over 2\alpha}\theta(y_*-x_*) \right] e^{-i\alpha(x_\bullet-y_\bullet)}
\nonumber\\
&&\times\braxp\Bigg\{
\Big([x_*,y_*]\ssp - g\int_{y_*}^{x_*}\!\!\!d{2\over s}\omega_*\,\gamma^i\,[x_*,\omega_*]F_{i\bullet}
[\omega_*,y_*] \Big){1\over \alpha s}\ssp_2\ssp
\nonumber\\
&&
+ {ig\over 2\alpha}\!\int_{y_*}^{x_*}\!\!\!d{2\over s}\omega_*\Bigg(
[x_*,\omega_*]\Big(\half\sigma^{ij}F_{ij}
 + {4\over s^2}\,F_{\bullet *}\sigma_{*\bullet}\Big)[\omega_*,y_*]
 \nonumber\\
 &&+[x_*,\omega_*]{2\over s}(\omega_*-x_*)(iD^iF_{i\bullet}(\omega_*))[\omega_*,y_*]
\nonumber\\
&&
-2g\int_{\omega_*}^{x_*}\!\!\!d{2\over s}\omega'_*
[x_*,\omega'_*]{2\over s}(\omega'_*-x_*)F_{i\bullet}(\omega'_*)[\omega'_*,\omega_*]F^i_{~\bullet}(\omega_*)[\omega_*,y_*]
\nonumber\\
&&
+ 2[x_*,\omega_*]{2\over s}(\omega_*-x_*)F_{i\bullet}(\omega_*)[\omega_*,y_*]P^i\Bigg) {1\over \alpha s}\ssp\ssp_2\ssp
+ \Bigg[ {ig\over 4\alpha^2}\int_{y_*}^{x_*}\!\!\!d{2\over s}\omega_*\,[x_*,\omega_*]\calb_1(i\,\Slash{\mathfrak{D}}_\perp[\omega_*,y_*])
\nonumber\\
&&
+ {ig\over 4\alpha^2}
\int_{y_*}^{x_*}\!\!\!d{2\over s}\omega_*\,(i\,\Slash{\mathfrak{D}}_\perp[x_*,\omega_*])\calb_1[\omega_*,y_*]
+ {ig\over 4\alpha^2}\int_{y_*}^{x_*}\!\!\!d{2\over s}\omega_*
\,{2\over s}(\omega_*-x_*)\,
\nonumber\\
&&
\times\Bigg(\!\!-2[x_*,\omega_*] (iD_i\gamma^jF_{j\bullet})[\omega_*,y_*]P^i 
- 2[x_*,\omega_*](i\,D^i\gamma^jF_{j\bullet})(i\,\mathfrak{D}_i[\omega_*,y_*]) 
\nonumber\\
&&
+ 2(i\,\Slash{\mathfrak{D}}_\perp[x_*,\omega_*])F^i_{~\bullet}(i\mathfrak{D}_i[\omega_*,y_*]) 
+ 2[x_*,\omega_*]F_{i\bullet}(i\mathfrak{D}^i(i\,\Slash{\mathfrak{D}}_\perp[\omega_*,y_*]))
\nonumber\\
&&
+ [x_*,\omega_*](iD^iF_{i\bullet})(i\,\Slash{\mathfrak{D}}_\perp[\omega_*,y_*]) 
+(i\,\Slash{\mathfrak{D}}_\perp[x_*,\omega_*])(iD^iF_{i\bullet})[\omega_*,y_*] 
\nonumber\\
&&
+ 2[x_*,\omega_*]F_{i\bullet}(i\,\Slash{\mathfrak{D}}_\perp[\omega_*,y_*])P^i
+ 2(i\,\Slash{\mathfrak{D}}_\perp[x_*,\omega_*])F_{i\bullet}[\omega_*,y_*]P^i
\Bigg)\Bigg]{2\over s}\ssp_2\ssp\Bigg\}
\nonumber\\
&&\times e^{i{\hatp^2_\perp\over \alpha s}(y_*-x_*)}\ketyp + O(\lambda^{-2})\,.
\label{qprop-subeik-xx}
\end{eqnarray}
Equation (\ref{qprop-subeik-xx}) is the final result for the quark propagator with sub-eikonal corrections
with point $x_\perp^\mu + {2\over s}x_*p_2^\mu$ in the external field.
Note that, in (\ref{qprop-subeik-xx}), the non gauge-invariant terms are those with gauge field $A_i$ only at point $y_*$ which,
being outside the shock-wave, can be set to zero (see figure (\ref{leftswprpagation})).
In eq. (\ref{qprop-subeik-xx}) the action of the covariant derivative $\mathfrak{D}_i$ can be performed using 
its definition (\ref{coderiv-glink}). 


\section{Gluon propagator in the background-Feynman gauge}

\subsection{In the background of gluon field}
\label{sec: feynback}


Let us consider gluon propagator in the background-Feynman gauge $D^\mu A^q_\mu = 0$
where again $D_\mu = \partial_\mu - ig A^{cl}_\mu$. The superscript ``cl'' will be omitted again from now on.
The propagator in  Schwinger formalism can be written as 
\begin{eqnarray}
i\langle A^a_\mu(x) A^b_\nu(y)\rangle = \brax{1\over \hat{P}^2 + 2ig F + i\epsilon}\kety^{ab}_{\mu\nu} 
\label{gluonpbf}
\end{eqnarray}
where $\hat{P}^2 = \hatp^2 + 2g\alpha \hat{A}_\bullet + g\{\hat{p}^\mu_\perp,\hat{A}^\perp_\mu\} +
{2\over s}g\{\hat{P}_\bullet, \hat{A}_*\} - g^2\hat{A}^2_\perp$. We will omit the symbol
$\,\hat{}\,$ from the operators, for simplicity.
We define again the operator $O \equiv \{p^\mu_\perp,A^\perp_\mu\} +{2\over s}\{P_\bullet, A_*\} - gA^2_\perp$
and get rid of the term $\{P_\bullet, A_*\}$ as we did in section \ref{sec: subeiko-scalar}.

Let us expand eq. (\ref{gluonpbf}) in $F_{\mu\nu}$ up to $F^3$ terms, which are the ones relevant to
get the $\lambda^{-1}$ corrections (we will omit the $+i\epsilon$ prescription for each ${1\over P^2}$ factor)
\begin{eqnarray}
&&\left({1\over P^2 + 2ig F + i\epsilon}\right)_{\mu\nu}
\nonumber\\
&&= {g_{\mu\nu}\over P^2} - 2ig{1\over P^2}F_{\mu\nu} {1\over P^2} 
- 4g^2{1\over P^2}F_{\mu\xi}{1\over P^2}F^\xi_{~\nu}{1\over P^2}
+ 8ig^3{1\over P^2}F_{\mu\xi} {1\over P^2}F^{\xi\eta}{1\over P^2}F_{\eta\nu}{1\over P^2} + \dots
\nonumber\\
&&= {g_{\mu\nu}\over P^2} 
-{4ig\over s}\big(p_{2\nu}g^\perp_{\mu\alpha} - p_{2\mu}g^\perp_{\nu\alpha}\big)
{1\over P^2}F^\alpha_{~\bullet}{1\over P^2}
+ {16g^2p_{2\mu}p_{2\nu}\over s^2}
{1\over P^2}F^i_{~\bullet}{1\over P^2}F_{i\bullet}{1\over P^2}
\nonumber\\
&&~~+ \bigg[ -2ig\, g^\perp_{\alpha \mu}g^\perp_{\beta \nu}{1\over P^2}F^{\alpha \beta}{1\over P^2}
- {8ig\over s^2}\big(p_{1\mu}p_{2\nu} - p_{2\mu}p_{1\nu} \big){1\over P^2}F_{*\bullet}{1\over P^2}
\nonumber\\
&&
~~+ {8g^2\,p_{2\mu}g^\perp_{\alpha\nu}\over s}
{1\over P^2}F^i_{~\bullet}{1\over P^2}F_i^{~\alpha}{1\over P^2}
+ {8g^2\,g^\perp_{\alpha\mu}p_{2\nu}\over s}{1\over P^2}F^{i\alpha}{1\over P^2}F_{i\bullet}{1\over P^2}
\nonumber\\
&&
~~ + {16g^2\,p_{2\mu}g^\perp_{\alpha\nu}\over s^2}{1\over P^2} F_{\bullet*}{1\over P^2}F^\alpha_{~\,\bullet}{1\over P^2}
+ {16g^2\,p_{2\nu}g^\perp_{\alpha\mu}\over s^2}{1\over P^2}F^\alpha_{~\,\bullet}{1\over P^2}F_{\bullet *}{1\over P^2}
\nonumber\\
&&~~ -{64ig^3p_{2\mu2}p_{2\nu}\over s^3}{1\over P^2}F_{\bullet *}{1\over P^2}F_{i\bullet}{1\over P^2}F^i_{~\bullet}{1\over P^2}
+ {64ig^3p_{2\mu2}p_{2\nu}\over s^3}{1\over P^2}F_{i\bullet}{1\over P^2}F^i_{~\bullet}{1\over P^2}F_{\bullet *}{1\over P^2}
\nonumber\\
&&~~ -{32ig^3p_{2\mu2}p_{2\nu}\over s^2}{1\over P^2}F_{i\bullet}{1\over P^2}F^{ij}{1\over P^2}F_{j\bullet}{1\over P^2}
\bigg]
+ O(\lambda^{-2})\,.
\label{expandfb}
\end{eqnarray}
In eq. (\ref{expandfb}) the terms in the square bracket are sub-eikonal terms so, for those terms, for each of the ${1\over P^2}$
factors, we need the leading-eikonal scalar propagator.
The terms outside the square bracket, instead, contain both eikonal and sub-eikonal
corrections so, for those terms,  for each ${1\over P^2}$, we need the scalar propagator with up to 
sub-eikonal corrections, eq. (\ref{scalpropa-nogcv}), in the adjoint representation. Therefore, neglecting the terms with fields
at the edges of the gauge link, that is, at point $x_*$ and $y_*$, we arrive, after some algebra,
\begin{eqnarray}
\hspace{-0.2cm}\langle A_\mu^a(x)A_\nu^b(y)\rangle_A =\!\!\!&&
\left[-\int_0^{+\infty}\!\!{\dhd \alpha\over 2\alpha}\theta(x_*-y_*) +
\int_{-\infty}^0\!\!{\dhd\alpha\over 2\alpha}\theta(y_*-x_*) \right]e^{-i\alpha(x_\bullet - y_\bullet)}
\braxp\, e^{-i{\hatp^2_\perp\over \alpha s}x_*}
\nonumber\\
&&\times
\bigg\{g_{\mu\nu}\,[x_*,y_*]^{ab} 
-{2g\over \alpha s}
\int_{y_*}^{x_*}\!\!\!d{2\over s}\omega_*\bigg[
\big(p_{2\nu}\delta^j_\mu - p_{2\mu}\delta^j_\nu\big)
[x_*,\omega_*]F_{j\bullet} [\omega_*,y_*]
\nonumber\\
&& + {2g\over \alpha s} p_{2\mu}p_{2\nu}
\int_{\omega_*}^{x_*}\!\!\!d{2\over s}\omega'_*\,[x_*,\omega'_*]F^i_{~\bullet}[\omega'_*,\omega_*]F_{i\bullet}[\omega_*,y_*]\bigg]^{ab}
\nonumber\\
&& + \mathfrak{B}^{ab}_{1\mu\nu}(x_*,y_*;p_\perp) + \mathfrak{B}^{ab}_{2\mu\nu}(x_*,y_*;p_\perp)
+ \mathfrak{B}^{ab}_{3\mu\nu}(x_*,y_*;p_\perp)
+ \mathfrak{B}^{ab}_{4\mu\nu}(x_*,y_*;p_\perp)
\nonumber\\
&&  + \mathfrak{B}^{ab}_{5\mu\nu}(x_*,y_*;p_\perp) 
+ \mathfrak{B}^{ab}_{6\mu\nu}(x_*,y_*;p_\perp)+ O(\lambda^{-2})
\bigg\}e^{i{\hatp^2_\perp\over \alpha s}y_*}\ketyp
\label{gpropfeyback}
\end{eqnarray}
where we defined
\begin{eqnarray}
\mathfrak{B}^{ab}_{1\mu\nu}(x_*,y_*;p_\perp) =\!\!\! &&
g_{\mu\nu}\,{ig\over 2\alpha}\Bigg[\int^{x_*}_{y_*}\!\!d{2\over s}\omega_*\bigg(
\big\{p^i,[x_*,\omega_*]\,{2\over s}\,\omega_*\, F_{i\bullet}(\omega_*)\,[\omega_*,y_*]\big\}
\nonumber\\
&& + g\!\!\int^{x_*}_{\omega_*}\!\!d{2\over s}\,\omega'_*\,{2\over s}\big(\omega_* - \omega'_*\big)
[x_*,\omega'_*]F^i_{~\bullet}[\omega'_*,\omega_*]\,
\,F_{i\bullet}\,[\omega_*,y_*]\bigg)\Bigg]^{ab}
\label{B1}
\end{eqnarray}
\begin{eqnarray}
&&\hspace{-0.7cm}\mathfrak{B}^{ab}_{2\mu\nu}(x_*,y_*;p_\perp) =
{1\over \alpha }\big({2\over s}p_{2\nu}\delta^j_\mu - {2\over s}p_{2\mu}\delta^j_\nu\big) {ig\over 2\alpha}
\int_{y_*}^{x_*}\!\!\!d{2\over s}z_{1*}\Bigg\{
{2\over s}z_{1*}\big\{p^i,[x_*,z_{1*}] \,iD_iF_{j\bullet}[z_{1*},y_*]\big\}
\nonumber\\
&& - g\!\int_{z_{1*}}^{x_*}\!\!\!d{2\over s}\omega_*\bigg[
{2\over s}(\omega_* - z_{1*})[x_*,\omega_*]\,iD_iF_{j\bullet}[\omega_*,z_{1*}]F_{j\bullet}[z_{1*},y_*]
\nonumber\\
&& +{2\over s}(\omega_* - z_{1*}) [x_*,\omega_*]F^i_{~\bullet}[\omega_*,z_{1*}]iD_iF_{j\bullet}[z_{1*},y_*]
\nonumber\\
&& + \big\{p^i,[x_*,\omega_*]{2\over s}\omega_*F_{i\bullet}[\omega_*,z_{1*}] F_{j\bullet}[z_{1*},y_*]\big\}
 + \big\{p^i,[x_*,\omega_*]F_{j\bullet}[\omega_*,z_{1*}] {2\over s}z_{1*}F_{i\bullet}[z_{1*},y_*]\big\}
\nonumber\\
&&+ g\!\int^{x_*}_{\omega_*}\!\!\!d{2\over s}\omega'_*
\bigg( {2\over s}(\omega_*-\omega'_*)
[x_*,\omega'_*]F^i_{~\bullet}[\omega'_*,\omega_*]F_{i\bullet}[\omega_*,z_{1*}]F_{j\bullet}[z_{1*},y_*]
\nonumber\\
&&+  {2\over s}(z_{1*}-\omega_*)[x_*,\omega'_*]F_{j\bullet}
[\omega'_*,\omega_*]F^i_{~\bullet}[\omega_*,z_{1*}]F_{i\bullet}[z_{1*},y_*]
\nonumber\\
&&+{2\over s}(z_{1*}-\omega'_*) [x_*,\omega'_*]F_{i\bullet}[\omega'_*,\omega_*] F_{j\bullet}[\omega_*,z_{1*}]
F^i_{~\bullet}[z_{1*},y_*]\bigg)\bigg]
\Bigg\}^{ab}
\label{B2}
\end{eqnarray}
\begin{eqnarray}
&&\hspace{-1.2cm}\mathfrak{B}^{ab}_{3\mu\nu}(x_*,y_*;p_\perp) 
\nonumber\\
&&\hspace{-1.4cm}=
{4g\,p_{2\mu}p_{2\nu}\over \alpha^2 s^2}\int_{y_*}^{x_*}\!\!\!d{2\over s}z_{2*}\Bigg\{
{ig\over 2\alpha}\int_{z_{2*}}^{x*}\!\!\!d{2\over s}z_{1*}
\Bigg[
{2\over s}z_{1*}\big\{p^i,[x_*,z_{1*}]\,iD_iF^j_{~\bullet}[z_{1*},y_*]F_{j\bullet}[z_{2*},y_*]\big\}
\nonumber\\
&&\hspace{-1cm} + {2\over s}z_{2*}\{p^i,[x_*,z_{1*}]F^j_{~\bullet}[z_{1*},z_{2*}]iD_iF_{j\bullet}[z_{2*},y_*]\}
\nonumber\\
&&\hspace{-1cm}
 + {2\over s}(z_{1*}-z_{2*})[x_*,z_{1*}]\,iD_iF^j_{~\bullet}[z_{1*},z_{2*}]iD^iF_{j\bullet}[z_{2*},y_*]
 \nonumber\\
 &&\hspace{-1cm} - g\!\int_{z_{1*}}^{x_*}\!\!\!d{2\over s}\omega_*\bigg(
 \big\{p^i,[x_*,\omega_*]{2\over s}\omega_*F_{i\bullet}[\omega_*,z_{1*}] F^j_{~\,\bullet}[z_{1*},z_{2*}]F_{j\bullet}[z_{2*},y_*]\big\}
\nonumber
\end{eqnarray}
\begin{eqnarray}
&&\hspace{-1cm} + \big\{p^i,[x_*,\omega_*]F^j_{~\,\bullet}[\omega_*,z_{1*}] 
{2\over s}z_{1*}F_{i\bullet}[z_{1*},z_{2*}]F_{j\bullet}[z_{2*},y_*]\big\}
\nonumber\\
&&\hspace{-1cm}
 + \big\{p^i,[x_*,\omega_*]F^j_{~\bullet}[\omega_*,z_{1*}]F_{j\bullet}
[z_{1*},z_{2*}]{2\over s}z_{2*}F_{i\bullet}[z_{2*},y_*]\big\}
\nonumber\\
&&\hspace{-1cm} + {2\over s}(\omega_*-z_{2*})
[x_*,\omega_*]F^i_{~\bullet}[\omega_*,z_{1*}]F^j_{~\bullet}[z_{1*},z_{2*}]iD_iF_{j\bullet}[z_{2*},y_*]
\nonumber\\
&&\hspace{-1cm} + {2\over s}(z_{1*}-z_{2*})
[x_*,\omega_*]F^j_{~\bullet}[\omega_*,z_{1*}]F^i_{~\bullet}[z_{1*},z_{2*}]iD_iF_{j\bullet}[z_{2*},y_*]
\nonumber\\
&&\hspace{-1cm} + {2\over s}(z_{1*}-z_{2*})[x_*,\omega_*]F^j_{~\bullet}[\omega_*,z_{1*}]iD_iF_{j\bullet}[z_{1*},z_{2*}]F^i_{~\bullet}[z_{2*},y_*]
\nonumber\\
&&\hspace{-1cm} + {2\over s}(\omega_*-z_{2*})[x_*,\omega_*]\,iD_iF^j_{~\bullet}[\omega_*,z_{1*}]F_{j\bullet}[z_{1*},z_{2*}]F^i_{~\bullet}[z_{2*},y_*]
\nonumber\\
&&\hspace{-1cm} + {2\over s}(\omega_* - z_{1*})[x_*,\omega_*]\,iD_iF^j_{~\,\bullet}[\omega_*,z_{1*}]F^i_{~\bullet}[z_{1*},z_{2*}]F_{j\bullet}[z_{2*},y_*]
\nonumber\\
&&\hspace{-1cm} + {2\over s}(\omega_* - z_{1*}) [x_*,\omega_*]F^i_{~\bullet}[\omega_*,z_{1*}]iD_iF^j_{~\,\bullet}[z_{1*},z_{2*}]F_{j\bullet}[z_{2*},y_*]
\nonumber\\
&&\hspace{-1cm}
+ 2\,i\,{2\over s}[x_*,\omega_*]F_{\bullet *}[\omega_*,z_{1*}]F_{i\bullet}[z_{1*},z_{2*}]F^i_{~\bullet}[z_{2*},y_*]
\nonumber\\
&&\hspace{-1cm}
- 2\,i\,{2\over s}[x_*,\omega_*]F^i_{~\bullet}[\omega_*,z_{1*}]F_{i\bullet}[z_{1*},z_{2*}]F_{\bullet *}[z_{2*},y_*]
\nonumber\\
&&\hspace{-1cm}
+ 2\,i\,[x_*,\omega_*]F^i_{~\bullet}[\omega_*,z_{1*}]F_{ij}[z_{1*},z_{2*}]F^j_{~\bullet}[z_{2*},y_*]
\nonumber\\
&&\hspace{-1cm} + g\!\int^{x_*}_{\omega_*}\!\!\!d{2\over s}\omega'_*
\bigg( {2\over s}(z_{2*}-z_{1*}+\omega_*-\omega'_*)
[x_*,\omega'_*]F^i_{~\bullet}[\omega'_*,\omega_*]F_{i\bullet}
[\omega_*,z_{1*}]F^j_{~\,\bullet}[z_{1*},z_{2*}]F_{j\bullet}[z_{2*},y_*]
\nonumber\\
&&\hspace{-1cm} + {2\over s}(z_{2*} + z_{1*}-\omega_* - \omega'_*)[x_*,\omega'_*]F^j_{~\bullet}
[\omega'_*,\omega_*]F^i_{~\bullet}
[\omega_*,z_{1*}]F_{i\bullet}[z_{1*},z_{2*}]F_{j\bullet}[z_{2*},y_*]
\nonumber\\
&&\hspace{-1cm} + {2\over s}(z_{2*} + z_{1*} -\omega_* - \omega'_*) 
[x_*,\omega'_*]F_{i\bullet}[\omega'_*,\omega_*] F^j_{~\,\bullet}
[\omega_*,z_{1*}]F^i_{~\bullet}[z_{1*},z_{2*}]F_{j\bullet}[z_{2*},y_*]\bigg)
\Bigg)\Bigg]
\Bigg\}^{ab}
\label{B3}
\end{eqnarray}
\begin{eqnarray}
\hspace{-1cm}\mathfrak{B}^{ab}_{4\mu\nu}(x_*,y_*;p_\perp) =\!\!\!&&
-{2g^2\over \alpha^2 s}\!\int_{y_*}^{x_*}\!\!\!d{2\over s}\omega_*\,\bigg[
p_{2\mu}\delta^j_\nu\int_{\omega_*}^{x_*}\!\!\!d{2\over s}\omega'_*
\,[x_*,\omega'_*]F^i_{~\bullet}[\omega'_*,\omega_*]F_{ij}[\omega_*,y_*]
\nonumber\\
&&
+ p_{2\nu}\delta^j_\mu\int_{\omega_*}^{x_*}\!\!\!d{2\over s}\omega'_*
\,[x_*,\omega'_*]F_{ij}[\omega'_*,\omega_*]F^i_{~\bullet}[\omega_*,y_*]
\nonumber\\
&& + p_{2\mu}\delta^j_\nu\int_{\omega_*}^{x_*}\!\!\!d{2\over s}\omega'_*
\,[x_*,\omega'_*]{2\over s}F_{\bullet*}[\omega'_*,\omega_*]F_{j\bullet}[\omega_*,y_*]
\nonumber\\
&& + p_{2\nu}\delta^j_\mu\int_{\omega_*}^{x_*}\!\!\!d{2\over s}\omega'_*
\,[x_*,\omega'_*]F_{j\bullet}[\omega'_*,\omega_*]{2\over s}F_{\bullet*}[\omega_*,y_*]\bigg]^{ab}
\label{B4}
\\
\mathfrak{B}^{ab}_{5\mu\nu}(x_*,y_*;p_\perp) =\!\!\!&&
-{g\over \alpha}\delta^i_\mu\delta^j_\nu\int_{y_*}^{x_*}\!\!\!d{2\over s}\omega_*\,
\big([x_*,\omega_*]F_{ij}[\omega_*,y_*]\big)^{ab}
\label{B5}
\\
\mathfrak{B}^{ab}_{6\mu\nu}(x_*,y_*;p_\perp) =\!\!\!&&
-{4g\over \alpha s^2}\big(p_{1\mu}p_{2\nu} - p_{2\mu}p_{1\nu} \big)\int_{y_*}^{x_*}\!\!\!d{2\over s}\omega_*\,
\big([x_*,\omega_*]F_{*\bullet}[\omega_*,y_*]\big)^{ab}
\label{B6}
\end{eqnarray}

Equation (\ref{gpropfeyback}) is the final result of the gluon propagator with sub-eikonal corrections in the background-Feynman gauge.
The operators $\mathfrak{B}^{ab}_{1\mu\nu}$, $\mathfrak{B}^{ab}_{2\mu\nu}$, $\mathfrak{B}^{ab}_{3\mu\nu}$, $\mathfrak{B}^{ab}_{4\mu\nu}$, 
$\mathfrak{B}^{ab}_{5\mu\nu}$, and $\mathfrak{B}^{ab}_{6\mu\nu}$ are the sub-eikonal corrections to the gluon propagator
in the gluon external filed in the background-Feynman gauge.

\subsection{In the background of quark and anti-quark fields}
\label{sec: gluoninquark-bf}

Let us consider the gluon propagator in the background of quark and anti-quark fields. 
As we have anticipated in \ref{sec: gluoninquark-axial}, in the background-Feynman gauge the gluon propagator in the background of quark and anti quark fields
receives a eikonal contribution. The diagram and the cross-diagram is given in figure (\ref{gprop-inquark})
\begin{eqnarray}
\langle A_\mu^a(x) A_\nu^b(y)\rangle_{\psi\,\bar{\psi}} =\!\!\!&&
- g^2\int d^4z_1d^4z_2 \, \langle A_\mu^a(x) A_\rho^c(z_1)\rangle \langle A_\lambda^d(z_2) A_\nu^b(y)\rangle 
\label{gpropinq}
\\
&&\times\Big[
\bar{\psi}(z_2) t^d \gamma^\lambda\,\bra{z_2}{i\over \Sp \!+\! i\epsilon}\ket{z_1}\,\gamma^\rho t^c\psi(z_1)
\!+\! \bar{\psi}(z_1)t^c\gamma^\rho  \bra{z_1}{i\over \Sp \!+\! i\epsilon}\ket{z_2} \gamma^\lambda t^d \psi(z_2)
\Big]
\nonumber
\end{eqnarray}

To proceed, we need the eikonal gluon propagator in the gluon external field with the 
transverse $A_i$ fields at the edge of the gauge link (\textit{i.e.} at point $z_{1*}$ and $z_{2*}$)
not set to zero like eq. (\ref{scalpropa-nogcv}) for the scalar propagator. 
The reason is two folds. First, we need to extract eikonal and sub-eikonal contributions
and second, the gluon propagator has points $z_{1*}$ and $z_{2*}$ in the shock-wave external field. Thus, we have
\begin{eqnarray}
\langle A^a_\mu(x)A^c_\rho(z_1)\rangle_{A}
=\!\!\!&&\Big[-\int_0^{+\infty}\!\!{\dhd\alpha\over 2\alpha}\theta(x_*-z_{1*}) 
+ \int_{-\infty}^0\!\!{\dhd\alpha\over 2\alpha}\theta(z_{1*}-x_*)\Big]
e^{-i\alpha(x_\bullet - z_{1\bullet})}\braxp\, e^{-i{\hatp^2_\perp\over\alpha s}x_*}
\nonumber\\
&&
\times\!
\Bigg\{g_{\mu\rho}[x_*,z_{1*}]^{ac}
\!-\! g_{\mu\rho}{ig\over 2\alpha}[x_*,z_{1*}]^{ac}\, {2\over s}z_{1*}\Big(\big\{P_i,A^i(z_{1*})\big\}\! -\! gA_i(z_{1*})A^i(z_{1*})\Big)
\nonumber\\
&&
- {2g\over \alpha s}\int^{x_*}_{z_{1*}}\!\!\!d{2\over s}\omega_*
\Big((p_{2\rho}\delta^j_\mu - p_{2\mu}\delta_\rho^j)[x_*,\omega_*]F_{j\bullet}[\omega_*,z_{1*}]
\nonumber\\
&&
+{2g\over \alpha s}p_{2\mu}p_{2\rho}\int_{\omega_*}^{x_*}\!\!d{2\over s}\omega'_*
[x_*,\omega']F^i_{~\bullet}[\omega'_*,\omega_*]F_{i\bullet}[\omega_*,z_{1*}]
\Big)^{ac}\Bigg\}e^{i{\hatp^2_\perp\over\alpha s}z_{1*}}\ket{z_{1\perp}}
\nonumber\\
\label{gluon-bf-eikonal}
\end{eqnarray}
and
\begin{eqnarray}
\langle A^d_\lambda(z_2)A^b_\nu(y)\rangle_{A}
=\!\!\!&&\Big[-\int_0^{+\infty}\!\!{\dhd\alpha\over 2\alpha}\theta(z_{2*}-y_*) 
+ \int_{-\infty}^0\!\!{\dhd\alpha\over 2\alpha}\theta(y_*-z_{2*})\Big]
e^{-i\alpha(z_{2\bullet} - y_\bullet)}\bra{z_{2\perp}}\, e^{-i{\hatp^2_\perp\over\alpha s}z_{2*}}
\nonumber\\
&&
\times\!
\Bigg\{g_{\lambda\nu}[z_{2*},y_*]^{db}
+ g_{\lambda\nu}{ig\over 2\alpha}{2\over s}z_{2*}\Big(\big\{P_i,A^i(z_{2*})\big\} - gA_i(z_{2*})A^i(z_{2*})\Big)[z_{2*},y_*]^{db}
\nonumber\\
&&
- {2g\over \alpha s}\int^{z_{2*}}_{y_*}\!\!\!d{2\over s}\omega_*
\Big((p_{2\nu}\delta^j_\lambda - p_{2\lambda}\delta_\nu^j)[z_{2*},\omega_*]F_{j\bullet}[\omega_*,y_*]
\nonumber\\
&&
+{2g\over \alpha s}p_{2\lambda}p_{2\nu}\int_{\omega_*}^{z_{2*}}\!\!d{2\over s}\omega'_*
[z_{2*},\omega'_*]F^i_{~\bullet}[\omega'_*,\omega_*]F_{i\bullet}[\omega_*,y_*]
\Big)^{db}\Bigg\}e^{i{\hatp^2_\perp\over\alpha s}y_*}\ketyp\,.
\nonumber\\
\label{gluon-bf-eikonal1}
\end{eqnarray}

It is convenient to start with a symmetric expression with respect to $\Sp$
\begin{eqnarray}
&&\bar{\psi}(z_1)t^c\gamma^\rho\Big(\half\bra{z_1}\Sp{i\over \Sp^2 + i\epsilon}\ket{z_2} + \half\bra{z_1}{i\over \Sp^2 + i\epsilon}\Sp\ket{z_2}
\Big)\gamma^\lambda t^d \psi(z_2)
\nonumber\\
&&=\bra{z_1}\Bigg({8\over s^2} p_2^\rho p_2^\lambda\,\bar{\psi}\,t^c\ssp_1\half\big\{P_\bullet,{i\over  P^2 \!+\! i\epsilon}\big\}\,t^d\psi 
\!+\!  \bar{\psi}\,t^c\ssp_1\half \Big({2\over s}p_2^\rho\gamma^j\gamma^\lambda_\perp \!+\! 
{2\over s}p_2^\lambda \gamma_\perp^\rho\gamma^j\Big)\big\{P_j,{i\over P^2 \!+\! i\epsilon}\big\}\,t^d\psi 
\nonumber\\
&&~
+\!{2\over s}\bar{\psi}t^c\gamma^\rho_\perp P_*\,\ssp_1\gamma^\lambda_\perp\,{i\over P^2 \!+\! i\epsilon}t^d\psi
\!-\! {8\over s^2}\,p_2^\rho p_2^\lambda \, \bar{\psi}\,t^c\ssp_1\,\half\big[\Sp_\perp,{i\over  P^2 \!+\! i\epsilon}gF_{i\bullet}\gamma^i
{i\over P^2 \!+\! i\epsilon}\big]t^d\psi
\nonumber\\
&&~
+ {8\over s^2} \bar{\psi}t^c\,
\half\big(p_2^\lambda\gamma_\perp^\rho\gamma^i \!-\! p_2^\rho\gamma^i\gamma_\perp^\lambda\big)\ssp_1P_*
{i\over P^2 \!+\! i\epsilon}gF_{i\bullet}{i\over P^2 \!+\! i\epsilon}t^d\psi\Bigg)\ket{z_2}\,.
\label{symgproinq}
\end{eqnarray}
Note that, we can use $P_* = p_* = {s\over 2}\alpha$ because $A_*$ will contribute as a sub-sub-eikonal term.
In eq. (\ref{gpropinq}) we have two integrations over the $*$-components $z_{1*}$ and $z_{2*}$, so we can use 
$\Sp^2 = p^2 + 2\alpha g A_\bullet + ig{2\over s}F_{i\bullet}\gamma^i\ssp_2 + O(\lambda^{-1})$ neglecting
all other sub-eikonal terms because they would contribute as sub-sub-eikonal corrections.

The term from which we will extract both eikonal and sub-eikonal contributions is
${8\over s^2}p_2^\rho p_2^\lambda\,\bar{\psi}\,t^c{\ssp_1}P_\bullet\,{i\over  P^2}\,t^d\psi$.
Let us observe that, using scalar propagator (\ref{scalpropa-nogcv}), we have
\begin{eqnarray}
&&\bra{z_1}\,P_\bullet{i\over  p^2 + 2\alpha g A_\bullet +gO}\ket{z_2} = iD^{z_1}_\bullet \bra{z_1}{i\over  p^2 + 2\alpha g A_\bullet +gO}\ket{z_2}
\nonumber\\
&& = \bra{z_1}{i\over 2\alpha}\ket{z_2} 
 + \left[\int_0^{+\infty}\!\!{\dhd \alpha\over 2\alpha}\theta(z_{1*}-z_{2*}) - 
 \int_{-\infty}^0\!\!{\dhd\alpha\over 2\alpha}\theta(z_{2*}-z_{1*}) \right]e^{-i\alpha(z_{1\bullet}- z_{2\bullet})}
\Bigg[
\bra{z_{1\perp}}\, e^{-i{\hatp^2_\perp\over \alpha s}z_{1*}}
\nonumber\\
 &&~~\times
\bigg\{ {\hatp^2_\perp\over 2\alpha} [z_{1*},z_{2*}] - {g\over 2\alpha}\Big(\{P_i,A^i(z_{1*})\} - gA_i(z_{1*})A^i(z_{1*})\Big)[z_{1*},z_{2*}]
\bigg\}e^{i{\hatp^2_\perp\over \alpha s}z_{2*}}\ket{z_{2\perp}}
\nonumber\\
&&= \bra{z_1}{i\over 2\alpha}\ket{z_2} 
+ \left[\int_0^{+\infty}\!\!{\dhd \alpha\over 2\alpha}\theta(z_{1*}-z_{2*}) - 
\int_{-\infty}^0\!\!{\dhd\alpha\over 2\alpha}\theta(z_{2*}-z_{1*}) \right]e^{-i\alpha(z_{1\bullet}- z_{2\bullet})}
\nonumber\\
&&~~\times\!
\bra{z_{1\perp}}\, e^{-i{\hatp^2_\perp\over \alpha s}z_{1*}}
P^2_\perp[z_{1*},z_{2*}]\,e^{i{\hatp^2_\perp\over \alpha s}z_{2*}}\ket{z_{2\perp}}
\end{eqnarray}
and similarly we have
\begin{eqnarray}
&&\bra{z_1}{i\over  p^2 + 2\alpha g A_\bullet + gO}P_\bullet\,\ket{z_2}
\nonumber\\
 && = \bra{z_1}{i\over 2\alpha}\ket{z_2} 
 + \left[\int_0^{+\infty}\!\!{\dhd \alpha\over 2\alpha}\theta(z_{1*}-z_{2*}) - 
 \int_{-\infty}^0\!\!{\dhd\alpha\over 2\alpha}\theta(z_{2*}-z_{1*}) \right]e^{-i\alpha(z_{1\bullet}- z_{2\bullet})}
 \nonumber\\
 &&~~\times\!
 \bra{z_{1\perp}}\, e^{-i{\hatp^2_\perp\over \alpha s}z_{1*}}
[z_{1*},z_{2*}]P^2_\perp\,e^{i{\hatp^2_\perp\over \alpha s}z_{2*}}\ket{z_{2\perp}}\,.
\end{eqnarray}
where $P^2_\perp = - P^iP_i$, and in section \ref{sec: subeiko-scalar} we defined $O = \{p_\mu,A^\mu_\perp\} - gA^2_\perp$.

Let us consider first the term $ \bra{z_1}{i\over 2\alpha}\ket{z_2}$ from which we will have to extract eikonal and sub-eikonal corrections. 
Adding also the cross diagram, we have
\begin{eqnarray}
\hspace{-1.2cm}\langle A_\mu^a(x) A_\nu^b(y)\rangle_{\psi\,\bar{\psi}}\! \ni\!
-\!\!\!&& g^2\!\int d^4z_1d^4z_2 \, \langle A_\mu^a(x) A_\rho^c(z_1)\rangle \langle A_\lambda^d(z_2) A_\nu^b(y)\rangle 
{8\over s^2}p_2^\rho p_2^\lambda
\nonumber\\
\times\!\!\!&&\Big[
\bar{\psi}(z_2) \,\ssp_1\,t^d \bra{z_1} {i\over 2\alpha}\ket{z_2} t^c\psi(z_1)
+ \bar{\psi}(z_1)\,\ssp_1\,t^c\bra{z_1} {i\over 2\alpha}\ket{z_2}  t^d \psi(z_2)
\Big]\,.
\end{eqnarray}
Using propagators (\ref{gluon-bf-eikonal}) and (\ref{gluon-bf-eikonal1}), and the identity 
$g\bar{\psi}\ssp_1[t^c,t^d]\psi = (D^iF_{i\bullet})^{cd} + {2\over s}(D_\bullet F_{*\bullet})^{cd}$, we obtain
\small
\begin{eqnarray}
\langle A_\mu^a(x) A_\nu^b(y)\rangle_{\psi\,\bar{\psi}} \ni
\!\!\!&&- g^2\int\!\! d^4z_1d^4z_2 \, \langle A_\mu^a(x) A_\rho^c(z_1)\rangle \langle A_\lambda^d(z_2) A_\nu^b(y)\rangle 
{8\over s^2}p_2^\rho p_2^\lambda
\nonumber\\
&&\times\Big[
\bar{\psi}(z_2) \,\ssp_1\,t^d \bra{z_1} {i\over 2\alpha}\ket{z_2} t^c\psi(z_1)
\!+\! \bar{\psi}(z_1)\,\ssp_1\,t^c\bra{z_1} {i\over 2\alpha}\ket{z_2}  t^d \psi(z_2)
\Big]
\nonumber\\
= \!\!\!\!&&
 \Big[\!-\!\! \int_0^{+\infty}\!\!{\dhd\alpha\over 8\alpha^3}\theta(x_*-y_*) 
\!+\! \int_{-\infty}^0\!\!{\dhd\alpha\over 8\alpha^3}\theta(y_*-x_*)\Big]
e^{-i\alpha(x_\bullet - y_\bullet)}
\nonumber\\
\!\!\!&&\times{8\over s^2} p_{2\mu} p_{2\nu}
\braxp\, e^{-i{\hatp^2_\perp\over\alpha s}x_*}
i g\!\int_{y_*}^{x_*}\!\!\! d{2\over s}z_{1*}
[x_*,z_{1*}]^{ac}\Big(
\,(D^iF_{i\bullet})^{cd} \!+\! {2\over s}(D_\bullet F_{*\bullet})^{cd}
\nonumber\\
\!\!\!&& -
{iz_{1*}\over \alpha s}\big[\{P^k,(iD_kD^iF_{i\bullet}) \!+\! {2\over s}(iD_kD_\bullet F_{*\bullet})\}\big]^{cd}
\Big)
[z_{1*},y_*]^{db} e^{i{\hatp^2_\perp\over\alpha s}y_*}\ketyp\,.
\label{eikonalingproiq}
\end{eqnarray}
\normalsize

To arrive at result (\ref{eikonalingproiq}) we have used
$e^{-i{\hatp^2_\perp\over \alpha s}z_*}\psi(z_*) e^{i{\hatp^2_\perp\over \alpha s}z_*} = \psi(z_*) + O(\lambda)$ and similarly for $\bar{\psi}$.
The terms $(D^iF_{i\bullet})$ and ${2\over s}(D_\bullet F_{*\bullet})^{cd}$ in eq. (\ref{eikonalingproiq}) are the eikonal contributions to the gluon propagator in the background of quark and anti-quark fields,
while $\{P^k,(iD_kD^iF_{i\bullet}) + {2\over s}(iD_kD_\bullet F_{*\bullet})\}$ is one of the sub-eikonal terms. 
Note that to arrive at eq. (\ref{eikonalingproiq}) we made use of 
\small
\begin{eqnarray}
&&\hspace{-0.2cm}e^{i{\hatp^2_\perp\over \alpha s}z_{1*}}\,
\Big((D^iF_{i\bullet}) + {2\over s}(D_\bullet F_{*\bullet})\Big)\,e^{-i{\hatp^2_\perp\over \alpha s}z_{1*}}
\\
&&\hspace{-0.2cm}= 
(D^iF_{i\bullet}) + {2\over s}(D_\bullet F_{*\bullet}) + 
{iz_{1*}\over \alpha s}\big[\hatp^2_\perp, (D^iF_{i\bullet}) + {2\over s}(D_\bullet F_{*\bullet})\big] + O(\lambda^{-2})
\nonumber\\
&&\hspace{-0.2cm}= (D^iF_{i\bullet}) + {2\over s}(D_\bullet F_{*\bullet}) 
- {iz_{1*}\over \alpha s}\bigg[ g\left(\{P^j,A_j\} - gA^jA_j\right)\Big((D^iF_{i\bullet})+{2\over s}(D_\bullet F_{*\bullet})\Big)
\nonumber\\
&&\hspace{-0.2cm}~~ - g\Big((D^iF_{i\bullet})+{2\over s}(D_\bullet F_{*\bullet})\Big)\left(\{P^j,A_j\} - gA^jA_j\right)
- \big\{P^j, (iD_j(D^iF_{i\bullet})) + {2\over s}(iD_jD_\bullet F_{*\bullet})\big\}\bigg] + O(\lambda^{-2})\,.
\nonumber
\end{eqnarray}
\normalsize
where $g\big(\{P^j,A_j\} - gA^jA_j\big)\Big((D^iF_{i\bullet})+{2\over s}(D_\bullet F_{*\bullet})\Big)$ 
and $- g\Big((D^iF_{i\bullet})+{2\over s}(D_\bullet F_{*\bullet})\Big)\big(\{P^j,A_j\}$ $ - gA^jA_j\big)$ will
cancel out with similar terms coming from the gluon propagators (\ref{gluon-bf-eikonal}) and (\ref{gluon-bf-eikonal1}).
We can rewrite eq. (\ref{eikonalingproiq}), \textit{pushing} the operator $P^j$ all to the left and to the right of the
gauge links, as
\begin{eqnarray}
&&\hspace{-0.8cm}\langle A_\mu^a(x) A_\nu^b(y)\rangle_{\psi\,\bar{\psi}}
\ni
\Big[-\int_0^{+\infty}\!\!{\dhd\alpha\over 8\alpha^3}\theta(x_*-y_*) 
+ \int_{-\infty}^0\!\!{\dhd\alpha\over 8\alpha^3}\theta(y_*-x_*)\Big]
e^{-i\alpha(x_\bullet - y_\bullet)}
\nonumber\\
&&
\times\!{8\over s^2} p_{2\mu} p_{2\nu}
\braxp\, e^{-i{\hatp^2_\perp\over\alpha s}x_*}
i g\!\int_{y_*}^{x_*}\!\!\! d{2\over s}z_{1*}\Bigg[[x_*,z_{1*}]^{ac}
\Big((D^iF_{i\bullet})+{2\over s}(D_\bullet F_{*\bullet})\Big)^{cd}[z_{1*},y_*]^{db}
\nonumber\\
&&
+ {z_{1*}\over \alpha s}\{P^k,[x_*,z_{1*}]^{ac}\Big((D_kD^iF_{i\bullet}) + {2\over s}(D_kD_\bullet F_{*\bullet})\Big)^{cd}[z_{1*},y_*]^{db}\}
\nonumber\\
&&
+\left(z_{1*}\over \alpha s\right) g\!\int_{z_{1*}}^{x_*}\!\!\!d{2\over s}\omega_*\bigg(
\big([x_*,\omega_*]F_{k\bullet}[\omega_*,z_{1*}]\big)^{ac}
\Big((D^kD^iF_{i\bullet}) + {2\over s}(D^kD_\bullet F_{*\bullet})\Big)^{cd}[z_{1*},y_*]^{db} 
\nonumber\\
&&
- [x_*,\omega_*]^{ac}\Big((D^kD^iF_{i\bullet}) + {2\over s}(D^kD_\bullet F_{*\bullet})\Big)^{cd}
([\omega_*,z_{1*}]F_{k\bullet}[z_{1*},y_*])^{db} \bigg)\Bigg] e^{i{\hatp^2_\perp\over\alpha s}y_*}\ketyp
\,.
\label{eikonalsubeikonalbf}
\end{eqnarray}
Note that, to get eq. (\ref{eikonalsubeikonalbf}) we have also added the cross diagram.

Next, we consider the terms $P^2_\perp[z_{1*},z_{2*}]$ and $[z_{1*},z_{2*}]P^2_\perp$ (without the cross diagram)
which are sub-eikonal corrections. We have
\begin{eqnarray}
&&\hspace{-0.5cm}\langle A_\mu^a(x) A_\nu^b(y)\rangle_{\psi\,\bar{\psi}} \ni
 g^2\int d^4z_1d^4z_2 \, \langle A_\mu^a(x) A_\rho^c(z_1)\rangle \langle A_\lambda^d(z_2) A_\nu^b(y)\rangle 
\nonumber\\
&&\hspace{0.3cm}
 \times\!\left[ - \int_0^{+\infty}\!\!{\dhd \alpha\over 2\alpha}\theta(z_{1*}-z_{2*}) 
+ \int_{-\infty}^0\!\!{\dhd\alpha\over 2\alpha}\theta(z_{2*}-z_{1*}) \right]e^{-i\alpha(z_{1\bullet}- z_{2\bullet})}
{4\over s^2}p_{2\mu}p_{2\nu}
\nonumber\\
&&\hspace{0.3cm}\times\!
\bar{\psi}(z_1)t^c\ssp_1 
\bra{z_{1\perp}}\, e^{-i{\hatp^2_\perp\over \alpha s}z_{1*}} 
{1\over 2\alpha}\Big(P^2_\perp[z_{1*},z_{2*}] + [z_{1*},z_{2*}]P^2_\perp\Big)
\,e^{i{\hatp^2_\perp\over \alpha s}z_{2*}}\ket{z_{2\perp}}
t^d \psi(z_2)
\nonumber\\
&&\hspace{-0.5cm}
= 
\left[ - \int_0^{+\infty}\!\!{\dhd \alpha\over 2\alpha}\theta(x_* - y_*) 
+ \int_{-\infty}^0\!\!{\dhd\alpha\over 2\alpha}\theta(y_*-x_*) \right]e^{-i\alpha(x_\bullet-y_\bullet)}
\braxp\, e^{-i{\hatp^2_\perp\over\alpha s}x_*}
\nonumber\\
&&\hspace{0.3cm}
\times\!{g^2\over 8 \, \alpha^3}{4\over s^2}p_{2\mu}p_{2\nu}\int_{x_*}^{y_*}\!\!\!d{2\over s}z_{1*}\!\!\int_{y_*}^{z_{1*}}\!\!\!d{2\over s}z_{2*}
\Bigg(\Big[P^2_\perp\bar{\psi}(z_{1*})
-(\bar{\psi}(z_{1*})\overleftarrow{D}^2_\perp) + 2i P_i(\bar{\psi}(z_{1*})\overleftarrow{D}^i)\Big]
\nonumber\\
&&\hspace{0.3cm}\times\!
[z_{1*},x_*]t^a\ssp_1[x_*,y_*] t^b[y_*,z_{2*}] \psi(z_{2*})[z_{2*},y_*]
+ \bar{\psi}(z_{1*})[z_{1*},x_*]t^a\ssp_1[x_*,y_*]t^b [y_*,z_{2*}]
\nonumber\\
&&\hspace{0.3cm}\times\!
\Big[\psi(z_{2*})P^2_\perp - 2i (D^i\psi(z_{2*}))P_i - (D^2_\perp\psi(z_{2*}))\Big]
\Bigg)e^{i{\hatp^2_\perp\over\alpha s}y_*}\ketyp
\end{eqnarray}
where we used $e^{i{\hatp^2_\perp\over \alpha s}z_{1*}} \psi e^{-i{\hatp^2_\perp\over \alpha s}z_{1*}} = \psi + O(\lambda^{-1})$
and similarly for $\bar{\psi}$.

The other terms in eq. (\ref{symgproinq}) will contribute as sub-eikonal terms so, the final result is
(including also the cross diagram)
\begin{eqnarray}
&&\hspace{-0.5cm}\langle A_\mu^a(x) A_\nu^b(y)\rangle_{\psi\,\bar{\psi}}
=
\Big[-\int_0^{+\infty}\!\!{\dhd\alpha\over 2\alpha}\theta(x_*-y_*) 
+ \int_{-\infty}^0\!\!{\dhd\alpha\over 2\alpha}\theta(y_*-x_*)\Big]
e^{-i\alpha(x_\bullet - y_\bullet)}
\nonumber\\
&&
\times\!\Bigg\{{4\over s^2} p_{2\mu} p_{2\nu}
\braxp\, e^{-i{\hatp^2_\perp\over\alpha s}x_*}
{ig\over 2\alpha^2}\!\int_{y_*}^{x_*}\!\!\! d{2\over s}z_{1*}[x_*,z_{1*}]^{ac}\Big((D^iF_{i\bullet})
+ {2\over s}(D_\bullet F_{*\bullet})\Big)^{cd}[z_{1*},y_*]^{db}
e^{i{\hatp^2_\perp\over\alpha s}y_*}\ketyp
\nonumber\\
&& 
+ \braxp\, e^{-i{\hatp^2_\perp\over\alpha s}x_*}
\Big[
\mathfrak{Q}^{ab}_{1\mu\nu}(x_*,y_*;p_\perp) + \mathfrak{Q}^{ab}_{2\mu\nu}(x_*,y_*;p_\perp) + 
\mathfrak{Q}^{ab}_{3\mu\nu}(x_*,y_*;p_\perp) 
\nonumber\\
&&  \hspace{2.3cm} + \mathfrak{Q}^{ab}_{4\mu\nu}(x_*,y_*;p_\perp) + \mathfrak{Q}^{ab}_{5\mu\nu}(x_*,y_*;p_\perp) 
\Big]
e^{i{\hatp^2_\perp\over\alpha s}y_*}\ketyp
\nonumber\\
&& 
+ \brayp\, e^{-i{\hatp^2_\perp\over\alpha s}y_*}
\Big[ \mathfrak{Q}^{ba}_{2\nu\mu}(y_*,x_*;p_\perp) + 
\mathfrak{Q}^{ba}_{3\nu\mu}(y_*,x_*;p_\perp) 
\nonumber\\
&&  \hspace{2.3cm} + \mathfrak{Q}^{ba}_{4\nu\mu}(y_*,x_*;p_\perp) + \mathfrak{Q}^{ba}_{5\nu\mu}(y_*,x_*;p_\perp) 
\Big]
e^{i{\hatp^2_\perp\over\alpha s}x_*}\ketxp
+ O(\lambda^{-2})\Bigg\}
\label{gluonprpinqfb}
\end{eqnarray}
where we define
\begin{eqnarray}
&&\mathfrak{Q}_{1\mu\nu}^{ab}(x_*,y_*;p_\perp)
\nonumber\\
&&={4\over s^2} p_{2\mu} p_{2\nu}{i g\over 2\alpha^2}\!\int_{y_*}^{x_*}\!\!\! d{2\over s}z_{1*}
\Bigg[ {z_{1*}\over \alpha s}\{P^k,[x_*,z_{1*}]^{ac}\Big((D_kD^iF_{i\bullet}) + {2\over s}(D_kD_\bullet F_{*\bullet})\Big)^{cd}[z_{1*},y_*]^{db}\}
\nonumber\\
&&
~~+\left(g \over 2\alpha \right) \!\int_{z_{1*}}^{x_*}\!\!\!d{2\over s}\omega_*\,{2\over s}z_{1*}\Big(
\big([x_*,\omega_*]F_{k\bullet}[\omega_*,z_{1*}]\big)^{ac}
\Big((D^kD^iF_{i\bullet}) + {2\over s}(D^kD_\bullet F_{*\bullet})\Big)^{cd}[z_{1*},y_*]^{db} 
\nonumber\\
&&\hspace{1cm}  - [x_*,\omega_*]^{ac}\Big((D^kD^iF_{i\bullet}) + {2\over s}(D^kD_\bullet F_{*\bullet})\Big)^{cd}
([\omega_*,z_{1*}]F_{k\bullet}[z_{1*},y_*])^{db} \Big)\Bigg] 
\label{Q1}
\end{eqnarray}

\begin{eqnarray}
&&\mathfrak{Q}_{2\mu\nu}^{ab}(x_*,y_*;p_\perp) 
\nonumber\\
&&=
{g^2\over \alpha}{4\over s^2}p_{2\mu}p_{2\nu}\Bigg\{
{1\over 8\alpha^2}\!\int_{x_*}^{y_*}\!\!\!d{2\over s}z_{1*}\!\!\int_{y_*}^{z_{1*}}\!\!\!d{2\over s}z_{2*}
\Bigg(\Big[P^2_\perp\bar{\psi}(z_{1*})
-(\bar{\psi}(z_{1*})\overleftarrow{D}^2_\perp) + 2i P_i(\bar{\psi}(z_{1*})\overleftarrow{D}^i)\Big]
\nonumber\\
&&\times\!
[z_{1*},x_*]t^a\ssp_1[x_*,y_*] t^b[y_*,z_{2*}] \psi(z_{2*})[z_{2*},y_*]
+ \bar{\psi}(z_{1*})[z_{1*},x_*]t^a\ssp_1[x_*,y_*]t^b [y_*,z_{2*}]
\nonumber\\
&&\times\!
\Big[\psi(z_{2*})P^2_\perp - 2i (D^i\psi(z_{2*}))P_i - (D^2_\perp\psi(z_{2*}))\Big]
\Bigg)
\nonumber\\
&&
+ \Bigg[
\half \!\int^{x_*}_{y_*}\!\!\!d{2\over s}z_{1*}\!\!\int^{z_{1*}}_{y_*}\!\!\!d{2\over s}\omega_{1*}
\!\!\int_{y_*}^{\omega_{1*}}\!\!\!d{2\over s}z_{2*} 
\Big(P_j\bar{\psi}(z_{1*}) + (-i\bar{\psi}(z_{1*})\overleftarrow{D}_j)\Big)
\nonumber\\
&&~~\times\!
[z_{1*},x_*]t^a[x_*,\omega_{1*}]gF_{i\bullet}[\omega_{1*},y_*]t^b[y_*,z_{2*}]
\nonumber\\
&&
+ \int_{y_*}^{x_*}\!\!\!d{2\over s}z_{1*}\!\!\int_{y_*}^{z_{1*}}\!\!\!d{2\over s}z_{2*}\!\!\int_{y_*}^{z_{2*}}\!\!\!d{2\over s}\omega_{1*}
\Big(P_j\bar{\psi}(z_{1*}) + (-i\bar{\psi}(z_{1*})\overleftarrow{D}_j)\Big)
\nonumber\\
&&
\times\!\bigg([z_{1*},x_*]t^a[x_*,\omega_{1*}]gF_{i\bullet}[\omega_{1*},y_*]t^b[y_*,z_{2*}]
\nonumber\\
&&~~~~ - [z_{1*},x_*]t^a[x_*,y_*]t^b[y_*,\omega_{1*}]gF_{i\bullet}[\omega_{1*},z_{2*}]\bigg)
\Bigg]\gamma^j\ssp_1\gamma^i\psi(z_{2*})
\nonumber\\
&&
- \Bigg[\half \!\int^{x_*}_{y_*}\!\!\!d{2\over s}z_{1*}\!\!\int^{z_{1*}}_{y_*}\!\!\!d{2\over s}\omega_{1*}
\!\int_{y_*}^{\omega_{1*}}\!\!\!d{2\over s}z_{2*} \,
\bar{\psi}(z_{1*})[z_{1*},x_*]t^a[x_*,\omega_{1*}]gF_{j\bullet}[\omega_{1*},y_*]t^b[y_*,z_{2*}]
\nonumber\\
&&
+ \int_{y_*}^{x_*}\!\!\!d{2\over s}\omega_{1*}
\!\!\int_{y_*}^{\omega_{1*}}\!\!\!d{2\over s}z_{1*}\!\!\int_{y_*}^{z_{1*}}\!\!\!d{2\over s}z_{2*} \,
\bar{\psi}(z_{1*})\bigg([z_{1*},x_*]t^a[x_*,\omega_{1*}]gF_{j\bullet}[\omega_{1*},y_*]t^b[y_*,z_{2*}]
\nonumber\\
&&
- [z_{1*},\omega_{1*}]gF_{j\bullet}[\omega_{1*},x_*]t^a[x_*,y_*]t^b[y_*,z_{2*}]
\bigg) \Bigg]
\gamma^j \,\ssp_1\gamma^i\Big((iD_i\psi(z_{2*})) + \psi(z_{2*})P_i\Big)
\nonumber\\
&&
+ \Bigg[\int_{y_*}^{x_*}\!\!\!d{2\over s}\omega_{1*}
\!\!\int_{y_*}^{\omega_{1*}}\!\!\!d{2\over s}z_{1*}\!\!\int_{y_*}^{z_{1*}}\!\!\!d{2\over s}z_{2*}\!\!\int_{y_*}^{z_{2*}}\!\!\!d{2\over s}\omega_{2*}
\,\bar{\psi}(z_{1*})
\nonumber\\
&&
\times\!\bigg([z_{1*},x_*]t^a\![x_*,\omega_{1*}]gF_{j\bullet}[\omega_{1*},\omega_{2*}]gF_{i\bullet}[\omega_{2*},y_*]t^b[y_*,z_{2*}]
\nonumber\\
&& - [z_{1*},\omega_{1*}]gF_{j\bullet}[\omega_{1*},x_*]t^a[x_*,\omega_{2*}]gF_{i\bullet}[\omega_{2*},y_*]t^b[y_*,z_{2*}]
\nonumber\\
&& + [z_{1*},\omega_{1*}]gF_{j\bullet}[\omega_{1*},x_*]t^a[x_*,y_*]t^b[y_*,\omega_{2*}]gF_{i\bullet}[\omega_{2*},z_{2*}]
\nonumber\\
&& - [z_{1*},x_*]t^a\![x_*,\omega_{1*}]gF_{j\bullet}[\omega_{1*},y_*]t^b[y_*,\omega_{2*}]gF_{i\bullet}[\omega_{2*},z_{2*}]\bigg)
\nonumber\\
&&
+ \half \!\int_{y_*}^{x_*}\!\!\!d{2\over s}\omega_{1*}\!\!\int^{\omega_{1*}}_{y_*}\!\!\!d{2\over s}z_{1*}\!\!\int^{z_{1*}}_{y_*}\!\!\!d{2\over s}\omega_{2*}
\!\!\int_{y_*}^{\omega_{2*}}\!\!\!d{2\over s}z_{2*}\,\bar{\psi}(z_{1*})
\nonumber\\
&&
\times\!\Big(
[z_{1*},x_*]t^a\,[x_*,\omega_{1*}]gF_{j\bullet}[\omega_{1*},\omega_{2*}]gF_{i\bullet}[\omega_{2*},y_*]t^b[y_*,z_{2*}]
\nonumber\\
&&
- [z_{1*},\omega_{1*}]gF_{j\bullet}[\omega_{1*},x_*]t^a[x_*,\omega_{2*}]gF_{i\bullet}[\omega_{2*},y_*]t^b[y_*,z_{2*}]\Big)
\nonumber\\
&&
+ \half \!\int^{x_*}_{y_*}\!\!\!d{2\over s}z_{1*}\!\!\int^{z_{1*}}_{y_*}\!\!\!d{2\over s}\omega_{1*}
\!\int_{y_*}^{\omega_{1*}}\!\!\!d{2\over s}z_{2*} \!\!\int_{y_*}^{z_{2*}}\!\!\!d{2\over s}\omega_{2*}\,\bar{\psi}(z_{1*})
\nonumber\\
&&
\times\!\Big(
[z_{1*},x_*]t^a[x_*,\omega_{1*}]gF_{j\bullet}[\omega_{1*},\omega_{2*}]gF_{i\bullet}[\omega_{2*},y_*]\,t^b[y_*,z_{2*}] 
\nonumber\\
&&
- [z_{1*},x_*]t^a[x_*,\omega_{1*}]gF_{j\bullet}[\omega_{1*},y_*]t^b[y_*,\omega_{2*}]gF_{i\bullet}[\omega_{2*},z_{2*}]\Big)
\Bigg]\gamma^j\ssp_1\gamma^i\psi(z_{2*})\Bigg\}
\label{Q2}
\end{eqnarray}

\begin{eqnarray}
\hspace{-1.02cm}\mathfrak{Q}_{3\mu\nu}^{ab}(x_*,y_*;p_\perp) = \!\!\!&&
{g^2\over 4\alpha}\!\!\int_{y_*}^{x_*}\!\!\!d{2\over s}z_{1*}\!\!\int_{y_*}^{z_{1*}}\!\!\!d{2\over s}z_{2*}
\,\bar{\psi}(z_{1*})\gamma_\mu^\perp \,\ssp_1\gamma_\nu^\perp[z_{1*},x_*]\,t^a[x_*,y_*]t^b[y_*,z_{2*}]\psi(z_{2*}) 
\nonumber\\
\label{Q3}
\\
\mathfrak{Q}_{4\mu\nu}^{ab}(x_*,y_*;p_\perp)=\!\!\!&&
{g^2\over 8\alpha^2}\!\!\int_{y_*}^{x_*}\!\!\!d{2\over s}z_{1*}\!\!\int_{y_*}^{z_{1*}}\!\!\!d{2\over s}z_{2*}\,
\bar{\psi}(z_{1*}) \big({2\over s}p_{2\mu}\gamma^j\ssp_1\gamma^\perp_\nu - {2\over s}p_{2\nu}\gamma^\perp_\mu\ssp_1\gamma^j \big)
\nonumber\\
&&
\times\Bigg[
\int^{x_*}_{y_*}\!\!\!d{2\over s}\omega_*\,[z_{1*},x_*]\,t^a[x_*,\omega_*]gF_{j\bullet}[\omega_*,y_*]t^b[y_*,z_{2*}]
\nonumber\\
&&
- \int_{z_{1*}}^{x_*}\!\!\!d{2\over s}\omega_*\,[z_{1*},\omega_*]gF_{j\bullet}[\omega_*,x_*]\,t^a[x_*,y_*]t^b[y_*,z_{2*}]
\nonumber\\
&&
- \int_{y_*}^{z_{2*}}\!\!\!d{2\over s}\omega_*\,[z_{1*},x_*]\,t^a[x_*,y_*]t^b[y_*,\omega_*]gF_{j\bullet}[\omega_*,z_{2*}]
\Bigg]\psi(z_{2*})
\label{Q4}
\\
\mathfrak{Q}_{5\mu\nu}^{ab}(x_*,y_*;p_\perp)=\!\!\!&&
{g^2\over 8\alpha^2}\!\int_{y_*}^{x_*}\!\!\!d{2\over s}z_{1*}\!\!\int_{y_*}^{z_{1*}}\!\!\!d{2\over s}z_{2*}
\Bigg[\bar{\psi}(z_{1*})[z_{1*},x_*]\,t^a[x_*,y_*]t^b[y_*,z_{2*}]
\nonumber\\
&&\times\!\Big({2\over s}p_{2\mu}\gamma^j\ssp_1\gamma^\perp_\nu
+ {2\over s}p_{2\nu}\gamma^\perp_\mu\ssp_1\gamma^j\Big)\Big((iD_j\psi(z_{2*})) + \psi(z_{2*})P_j\Big)
\nonumber\\
&&
- \Big(P_j\bar{\psi}(z_{1*}) + (-i\bar{\psi}(z_{1*})\overleftarrow{D}_j)\Big)
\Big({2\over s}p_{2\mu}\gamma^j\ssp_1\gamma^\perp_\nu
+ {2\over s}p_{2\nu}\gamma^\perp_\mu\ssp_1\gamma^j\Big)
\nonumber\\
&&\times\![z_{1*},x_*]\,t^a[x_*,y_*]t^b[y_*,z_{2*}]\psi(z_{2*}) 
\Bigg]
\label{Q5}
\end{eqnarray}
Notice that in $\mathfrak{Q}_{4\mu\nu}^{ab}$ we can use
\begin{eqnarray}
-i\mathfrak{D}_j\big([z_{1*},x_*]\,t^a[x_*,y_*]t^b[y_*,z_{2*}]\big)=\!\!\!&&
\int^{x_*}_{y_*}\!\!\!d{2\over s}\omega_*\,[z_{1*},x_*]\,t^a[x_*,\omega_*]gF_{j\bullet}[\omega_*,y_*]t^b[y_*,z_{2*}]
\nonumber\\
&&
- \int_{z_{1*}}^{x_*}\!\!\!d{2\over s}\omega_*\,[z_{1*},\omega_*]gF_{j\bullet}[\omega_*,x_*]\,t^a[x_*,y_*]t^b[y_*,z_{2*}]
\nonumber\\
&&
- \int_{y_*}^{z_{2*}}\!\!\!d{2\over s}\omega_*\,[z_{1*},x_*]\,t^a[x_*,y_*]t^b[y_*,\omega_*]gF_{j\bullet}[\omega_*,z_{2*}]
\nonumber\\
\end{eqnarray}
and $\mathfrak{Q}_{1\mu\nu}^{ab}(x_*,y_*;p_\perp)$ includes both diagram and cross diagram, so we did not need to include 
$\mathfrak{Q}_{1\nu\mu}^{ba}(y_*,x_*;p_\perp)$ in eq. (\ref{gluonprpinqfb}).

Equation (\ref{gluonprpinqfb}) is the final result for the gluon propagator in the background-Feynman 
gauge in the quark and anti-quark external fields.

\subsection{Summing up the gluon and quark anti-quark external fields contributions}

We can now sum up the contribution due to the gluon external field eq. (\ref{gpropfeyback}) and the quark and anti-quark external field eq. (\ref{gluonprpinqfb}) and
obtain the final expression of the gluon propagator in the Background-Feynman gauge up to sub-eikonal terms 
(\textit{i. e.} terms that scale as $\lambda^{-1}$)

\begin{eqnarray}
\langle A_\mu^a(x) A_\nu^b(y)\rangle=\!\!\!&&
 \left[-\int_0^{+\infty}\!\!{\dhd \alpha\over 2\alpha}\theta(x_*-y_*) +
\int_{-\infty}^0\!\!{\dhd\alpha\over 2\alpha}\theta(y_*-x_*) \right]e^{-i\alpha(x_\bullet - y_\bullet)}
\nonumber\\
\times\!\!\!&&\!\!\Bigg(\braxp\, e^{-i{\hatp^2_\perp\over \alpha s}x_*}
\bigg\{g_{\mu\nu}\,[x_*,y_*]^{ab} 
+ {2g\over \alpha s}
\int_{y_*}^{x_*}\!\!\!d{2\over s}\omega_*\bigg[
\big(p_{2\mu}\delta^j_\nu - p_{2\nu}\delta^j_\mu \big)
[x_*,\omega_*]F_{j\bullet} [\omega_*,y_*]
\nonumber\\
&&+ {1\over \alpha s} p_{2\mu}p_{2\nu}
\bigg([x_*,\omega_*]\Big((iD^iF_{i\bullet}) + {2\over s}(iD_\bullet F_{*\bullet})\Big)[\omega_*,y_*]
\nonumber\\
&&
- 2\,g\!\int_{\omega_*}^{x_*}\!\!\!d{2\over s}\omega'_*\,[x_*,\omega'_*]F^i_{~\bullet}[\omega'_*,\omega_*]F_{i\bullet}[\omega_*,y_*]\bigg)\bigg]^{ab}
\nonumber\\
&&+\ \mathfrak{B}^{ab}_{1\mu\nu}(x_*,y_*;p_\perp) + \mathfrak{B}^{ab}_{2\mu\nu}(x_*,y_*;p_\perp)
+ \mathfrak{B}^{ab}_{3\mu\nu}(x_*,y_*;p_\perp) + \mathfrak{B}^{ab}_{4\mu\nu}(x_*,y_*;p_\perp)
\nonumber\\
&& + \mathfrak{B}^{ab}_{5\mu\nu}(x_*,y_*;p_\perp) 
+ \mathfrak{B}^{ab}_{6\mu\nu}(x_*,y_*;p_\perp)
+ \mathfrak{Q}^{ab}_{1\mu\nu}(x_*,y_*;p_\perp) + \mathfrak{Q}^{ab}_{2\mu\nu}(x_*,y_*;p_\perp)
\nonumber\\
&&
+ \mathfrak{Q}^{ab}_{3\mu\nu}(x_*,y_*;p_\perp) 
+ \mathfrak{Q}^{ab}_{4\mu\nu}(x_*,y_*;p_\perp) + \mathfrak{Q}^{ab}_{5\mu\nu}(x_*,y_*;p_\perp) 
\bigg\}e^{i{\hatp^2_\perp\over \alpha s}y_*}\ketyp
\nonumber\\
&&+
\brayp\, e^{-i{\hatp^2_\perp\over\alpha s}y_*}
\Big[
 \mathfrak{Q}^{ba}_{2\nu\mu}(y_*,x_*;p_\perp) + 
\mathfrak{Q}^{ba}_{3\nu\mu}(y_*,x_*;p_\perp) 
\nonumber\\
&&+ \mathfrak{Q}^{ba}_{4\nu\mu}(y_*,x_*;p_\perp) + \mathfrak{Q}^{ba}_{5\nu\mu}(y_*,x_*;p_\perp) 
\Big]
e^{i{\hatp^2_\perp\over\alpha s}x_*}\ketxp\Bigg) + O(\lambda^{-2})
\label{gluonprofbtotal}
\end{eqnarray}
Equation (\ref{gluonprofbtotal}) is the gluon propagator in the background-Feynman gauge up to sub-eikonal corrections. 
All the sub-eikonal corrections, which scale as $\lambda^{-1}$ under the longitudinal boost, are enclosed in the 
operators $\mathfrak{B}^{ab}_{1\mu\nu}$,  $\mathfrak{B}^{ab}_{2\mu\nu}$,  $\mathfrak{B}^{ab}_{3\mu\nu}$,  $\mathfrak{B}^{ab}_{4\mu\nu}$,  
$\mathfrak{B}^{ab}_{5\mu\nu}$,  $\mathfrak{B}^{ab}_{6\mu\nu}$ and $\mathfrak{Q}^{ab}_{1\mu\nu}$,
$\mathfrak{Q}^{ab}_{2\mu\nu}$, $\mathfrak{Q}^{ab}_{3\mu\nu}$, 
$\mathfrak{Q}^{ab}_{4\mu\nu}$, $\mathfrak{Q}^{ab}_{5\mu\nu}$.


\begin{thebibliography}{99}

\bibitem{balitskyreview}
I. Balitsky, {\it ``High-Energy QCD and Wilson Lines''}, 
In *Shifman, M. (ed.): At the frontier of particle 
physics, vol. 2*, p. 1237-1342  (World Scientific, Singapore, 2001)
[\href{https://arxiv.org/abs/hep-ph/0101042}{{\ttfamily hep-ph/0101042}}]. 

\bibitem{Balitsky:1995ub}
I.~Balitsky,
Nucl.\ Phys.\ B {\bf 463} (1996) 99
doi:10.1016/0550-3213(95)00638-9
[\href{https://arxiv.org/abs/hep-ph/9509348}{{\ttfamily hep-ph/9509348}}].

\bibitem{Jalilian-Marian:1997dw}
J.~Jalilian-Marian, A.~Kovner and H.~Weigert, \emph{The {Wilson}
	renormalization group for low x physics: Gluon evolution at finite parton
	density}, {\emph{Phys. Rev.} {\bfseries D59} (1998) 014015},
[\href{https://arxiv.org/abs/hep-ph/9709432}{{\ttfamily hep-ph/9709432}}].

\bibitem{Jalilian-Marian:1997gr}
J.~Jalilian-Marian, A.~Kovner, A.~Leonidov and H.~Weigert, \emph{The {Wilson}
	renormalization group for low x physics: Towards the high density regime},
{\emph{Phys. Rev.} {\bfseries D59} (1998) 014014},
[\href{https://arxiv.org/abs/hep-ph/9706377}{{\ttfamily hep-ph/9706377}}].

\bibitem{Iancu:2001ad}
E.~Iancu, A.~Leonidov and L.~D. McLerran, \emph{{The renormalization group
		equation for the color glass condensate}},
\href{https://doi.org/10.1016/S0370-2693(01)00524-X}{\emph{Phys. Lett.}
	{\bfseries B510} (2001) 133--144}.

\bibitem{Iancu:2000hn}
E.~Iancu, A.~Leonidov and L.~D. McLerran, \emph{Nonlinear gluon evolution in
	the color glass condensate. {I}}, {\emph{Nucl. Phys.} {\bfseries A692} (2001)
	583--645}, [\href{https://arxiv.org/abs/hep-ph/0011241}{{\ttfamily
		hep-ph/0011241}}].
	
	\bibitem{Kovchegov:1999yj}
	Y.~V. Kovchegov, \emph{Small-x {$F_2$} structure function of a nucleus
		including multiple pomeron exchanges}, {\emph{Phys. Rev.} {\bfseries D60}
		(1999) 034008}, [\href{https://arxiv.org/abs/hep-ph/9901281}{{\ttfamily
			hep-ph/9901281}}].
	
	\bibitem{Kovchegov:1999ua}
	Y.~V. Kovchegov, \emph{Unitarization of the {BFKL} pomeron on a nucleus},
	{\emph{Phys. Rev.} {\bfseries D61} (2000) 074018},
	[\href{https://arxiv.org/abs/hep-ph/9905214}{{\ttfamily hep-ph/9905214}}].
	
	\bibitem{Kuraev:1977fs}
	E.~A. Kuraev, L.~N. Lipatov and V.~S. Fadin, \emph{{The Pomeranchuk
			singlularity in non-Abelian gauge theories}}, {\emph{Sov. Phys. JETP}
		{\bfseries 45} (1977) 199--204}.
	
	\bibitem{Balitsky:1978ic}
	I.~Balitsky and L.~Lipatov, \emph{{The Pomeranchuk Singularity in Quantum
			Chromodynamics}}, {\emph{Sov.J.Nucl.Phys.} {\bfseries 28} (1978) 822--829}.
	
	\bibitem{Kovchegov:2012mbw}
	Y.~V.~Kovchegov and E.~Levin,
	Camb.\ Monogr.\ Part.\ Phys.\ Nucl.\ Phys.\ Cosmol.\  {\bf 33} (2012).
		
		\bibitem{Balitsky:2009yp} 
		I.~Balitsky and G.~A.~Chirilli,
		Phys.\ Lett.\ B {\bf 687}, 204 (2010)
		doi:10.1016/j.physletb.2010.02.084
		[arXiv:0911.5192 [hep-ph]].
		
		\bibitem{Balitsky:2010ze}
		I.~Balitsky and G.~A.~Chirilli,
		Phys.\ Rev.\ D {\bf 83} (2011) 031502
		doi:10.1103/PhysRevD.83.031502
		[arXiv:1009.4729 [hep-ph]].
	
	\bibitem{Balitsky:2012bs}
	I.~Balitsky and G.~A.~Chirilli,
	Phys.\ Rev.\ D {\bf 87} (2013) no.1,  014013
	doi:10.1103/PhysRevD.87.014013
	[arXiv:1207.3844 [hep-ph]].
	
	\bibitem{Chirilli:2011km}
	G.~A.~Chirilli, B.~W.~Xiao and F.~Yuan,
	Phys.\ Rev.\ Lett.\  {\bf 108} (2012) 122301
	doi:10.1103/PhysRevLett.108.122301
	[arXiv:1112.1061 [hep-ph]].
	
	\bibitem{Chirilli:2012jd}
	G.~A.~Chirilli, B.~W.~Xiao and F.~Yuan,
	Phys.\ Rev.\ D {\bf 86} (2012) 054005
	doi:10.1103/PhysRevD.86.054005
	[arXiv:1203.6139 [hep-ph]].
	
	\bibitem{Chirilli:2014dcb}
	G.~A.~Chirilli and Y.~V.~Kovchegov,
	JHEP {\bf 1405} (2014) 099
	Erratum: [JHEP {\bf 1508} (2015) 075]
	doi:10.1007/JHEP05(2014)099, 10.1007/JHEP08(2015)075
	[arXiv:1403.3384 [hep-ph]].
	
	\bibitem{Boer:2011fh}
	D.~Boer {\it et al.},
	arXiv:1108.1713 [nucl-th].
	
	\bibitem{Accardi:2012qut}
	A.~Accardi {\it et al.},
	Eur.\ Phys.\ J.\ A {\bf 52} (2016) no.9,  268
	doi:10.1140/epja/i2016-16268-9
	[arXiv:1212.1701 [nucl-ex]].
	
	\bibitem{Aschenauer:2015ata}
	E.~C.~Aschenauer, R.~Sassot and M.~Stratmann,
	Phys.\ Rev.\ D {\bf 92} (2015) no.9,  094030
	doi:10.1103/PhysRevD.92.094030
	[arXiv:1509.06489 [hep-ph]].
	
	\bibitem{Bjorken:1970ah}
	J.~D.~Bjorken, J.~B.~Kogut and D.~E.~Soper,
	Phys.\ Rev.\ D {\bf 3} (1971) 1382.
	doi:10.1103/PhysRevD.3.1382
	
	\bibitem{tHooft:1987vrq}
	G.~'t Hooft,
	Phys.\ Lett.\ B {\bf 198} (1987) 61.
	doi:10.1016/0370-2693(87)90159-6
	
	\bibitem{Collins:1985gm}
	J.~C.~Collins, D.~E.~Soper and G.~F.~Sterman,
	Nucl.\ Phys.\ B {\bf 263} (1986) 37.
	doi:10.1016/0550-3213(86)90026-X
	
	\bibitem{Cheng:1987ga}
	H.~Cheng and T.~T.~Wu,
	CAMBRIDGE, USA: MIT-PR. (1987) 285p
	
	\bibitem{Nachtmann:1991ua}
	O.~Nachtmann,
	Annals Phys.\  {\bf 209} (1991) 436.
	doi:10.1016/0003-4916(91)90036-8

	
	\bibitem{Balitsky:2015qba}
	I.~Balitsky and A.~Tarasov,
	JHEP {\bf 1510} (2015) 017
	doi:10.1007/JHEP10(2015)017
	[arXiv:1505.02151 [hep-ph]].
	
	\bibitem{Balitsky:2016dgz}
	I.~Balitsky and A.~Tarasov,
	JHEP {\bf 1606} (2016) 164
	doi:10.1007/JHEP06(2016)164
	[arXiv:1603.06548 [hep-ph]].
	
	\bibitem{Kovchegov:2015pbl}
	Y.~V.~Kovchegov, D.~Pitonyak and M.~D.~Sievert,
	JHEP {\bf 1601} (2016) 072
	Erratum: [JHEP {\bf 1610} (2016) 148]
	doi:10.1007/JHEP01(2016)072, 10.1007/JHEP10(2016)148
	[arXiv:1511.06737 [hep-ph]].
	
	\bibitem{Kovchegov:2016zex}
	Y.~V.~Kovchegov, D.~Pitonyak and M.~D.~Sievert,
	Phys.\ Rev.\ D {\bf 95} (2017) no.1,  014033
	doi:10.1103/PhysRevD.95.014033
	[arXiv:1610.06197 [hep-ph]].
	
	\bibitem{Kovchegov:2017jxc}
	Y.~V.~Kovchegov, D.~Pitonyak and M.~D.~Sievert,
	Phys.\ Lett.\ B {\bf 772} (2017) 136
	doi:10.1016/j.physletb.2017.06.032
	[arXiv:1703.05809 [hep-ph]].
	
	\bibitem{Kovchegov:2017lsr} 
	Y.~V.~Kovchegov, D.~Pitonyak and M.~D.~Sievert,
	JHEP {\bf 1710}, 198 (2017)
	doi:10.1007/JHEP10(2017)198
	[arXiv:1706.04236 [nucl-th]].
	
	\bibitem{Bartels:1995iu}
	J.~Bartels, B.~I.~Ermolaev and M.~G.~Ryskin,
	Z.\ Phys.\ C {\bf 70} (1996) 273
	[hep-ph/9507271].
	
	\bibitem{Bartels:1996wc}
	J.~Bartels, B.~I.~Ermolaev and M.~G.~Ryskin,
	Z.\ Phys.\ C {\bf 72} (1996) 627
	doi:10.1007/s002880050285, 10.1007/BF02909194
	[hep-ph/9603204].
	
	\bibitem{Gorshkov:1966ht}
	V.~G.~Gorshkov, V.~N.~Gribov, L.~N.~Lipatov and G.~V.~Frolov,
	Sov.\ J.\ Nucl.\ Phys.\  {\bf 6} (1968) 95
	[Yad.\ Fiz.\  {\bf 6} (1967) 129].
	
	\bibitem{Altinoluk:2014oxa}
	T.~Altinoluk, N.~Armesto, G.~Beuf, M.~Martínez and C.~A.~Salgado,
	JHEP {\bf 1407} (2014) 068
	doi:10.1007/JHEP07(2014)068
	[arXiv:1404.2219 [hep-ph]].
	
	\bibitem{Laenen:2008gt}
	E.~Laenen, G.~Stavenga and C.~D.~White,
	JHEP {\bf 0903} (2009) 054
	doi:10.1088/1126-6708/2009/03/054
	[arXiv:0811.2067 [hep-ph]].
	
	\bibitem{Laenen:2008ux}
	E.~Laenen, L.~Magnea and G.~Stavenga,
	Phys.\ Lett.\ B {\bf 669} (2008) 173
	doi:10.1016/j.physletb.2008.09.037
	[arXiv:0807.4412 [hep-ph]].
	
	\bibitem{Laenen:2010uz}
	E.~Laenen, L.~Magnea, G.~Stavenga and C.~D.~White,
	JHEP {\bf 1101} (2011) 141
	doi:10.1007/JHEP01(2011)141
	[arXiv:1010.1860 [hep-ph]].


\bibitem{Chirilli:2015fza}
G.~A.~Chirilli, Y.~V.~Kovchegov and D.~E.~Wertepny,
JHEP {\bf 1512} (2015) 138
doi:10.1007/JHEP12(2015)138
[arXiv:1508.07962 [hep-ph]].

\bibitem{Balitsky:1987bk} 
I.~I.~Balitsky and V.~M.~Braun,
Nucl.\ Phys.\ B {\bf 311}, 541 (1989).
doi:10.1016/0550-3213(89)90168-5

\bibitem{Braun:2000av}
V.~M.~Braun, G.~P.~Korchemsky and A.~N.~Manashov,
Phys.\ Lett.\ B {\bf 476} (2000) 455
doi:10.1016/S0370-2693(00)00131-3
[hep-ph/0001130].



\bibitem{Braun:2008ia} 
V.~M.~Braun, A.~N.~Manashov and J.~Rohrwild,
Nucl.\ Phys.\ B {\bf 807}, 89 (2009)
doi:10.1016/j.nuclphysb.2008.08.012
[arXiv:0806.2531 [hep-ph]].

\bibitem{Braun:2009mi}
V.~M.~Braun, A.~N.~Manashov and B.~Pirnay,
Phys.\ Rev.\ D {\bf 80} (2009) 114002
Erratum: [Phys.\ Rev.\ D {\bf 86} (2012) 119902]
doi:10.1103/PhysRevD.80.114002, 10.1103/PhysRevD.86.119902
[arXiv:0909.3410 [hep-ph]].

\bibitem{Braun:2017liq}
V.~M.~Braun, Y.~Ji and A.~N.~Manashov,
JHEP {\bf 1705} (2017) 022
doi:10.1007/JHEP05(2017)022
[arXiv:1703.02446 [hep-ph]].
	\bibitem{fcollins}
	J.C. Collins, Foundations of Perturbative QCD, Cambridge University Press, Cambridge
U.K. (2011).
	
\end{thebibliography}
\end{document}